\documentclass[lettersize,journal]{IEEEtran}
\usepackage{amsmath,amsfonts}
\usepackage{algorithmic}
\usepackage{algorithm}
\usepackage{array}
\usepackage[caption=false,font=normalsize,labelfont=sf,textfont=sf]{subfig}
\usepackage{textcomp}
\usepackage{stfloats}
\usepackage{url}
\usepackage{verbatim}
\usepackage{graphicx}
\usepackage{cite}
\usepackage{acro}
\usepackage{etoolbox}
\preto\section\acresetall

\usepackage{mathtools}
\usepackage{siunitx}
\usepackage{pifont}
\usepackage[x11names]{xcolor}
\usepackage{multirow}
\usepackage{booktabs}
\usepackage{colortbl}
\usepackage{lscape}
\usepackage{rotating}

\usepackage{float} 

\usepackage{nicematrix}
\usepackage{acro}
\acsetup{use-id-as-short}

\DeclareAcronym{3G}{long = Third Generation}
\DeclareAcronym{4G}{long = Fourth Generation}
\DeclareAcronym{5G}{long = Fifth Generation}
\DeclareAcronym{6G}{long = Sixth Generation}
\DeclareAcronym{3GPP}{long = 3rd Generation Partnership Project}
\DeclareAcronym{A2G}{long = Air to Ground}
\DeclareAcronym{ABF}{long = adaptive beam forming}
\DeclareAcronym{ACO}{long = Ant Colony Optimization}
\DeclareAcronym{ACM}{long = Adaptive Coding and Modulation}
\DeclareAcronym{AI}{long = Artificial Intelligence}
\DeclareAcronym{ANN}{long = Artificial Neural Network}
\DeclareAcronym{AoI}{long = Age of Information}
\DeclareAcronym{BB}{long = Base Band}
\DeclareAcronym{BBU}{long = Base Band Unit}
\DeclareAcronym{BC}{long = Blockchain}
\DeclareAcronym{BER}{long = Bit Error Rate}
\DeclareAcronym{BH}{long = Beam Hopping}
\DeclareAcronym{BS}{long = Base Station}
\DeclareAcronym{BW}{long = Bandwidth}
\DeclareAcronym{C-RAN}{long = Cloud Radio Access Networks}
\DeclareAcronym{CAPEX}{long = Capital Expenditure}
\DeclareAcronym{CDMA}{long = Code Division Multiple Access}
\DeclareAcronym{CN}{long = Core Network}
\DeclareAcronym{CNN}{long = Convolutional Neural Network}
\DeclareAcronym{CoMP}{long = Coordinated Multipoint}
\DeclareAcronym{CR}{long = Cognitive Radio}
\DeclareAcronym{CSI}{long = Channel State Information}
\DeclareAcronym{D2D}{long = Device-to-Device}
\DeclareAcronym{DAC}{long = Digital-to-Analog Converter}
\DeclareAcronym{DAS}{long = Distributed Antenna Systems}
\DeclareAcronym{DBA}{long = Dynamic Bandwidth Allocation}
\DeclareAcronym{DC}{long = Duty Cycle}
\DeclareAcronym{DL}{long = Deep Learning}
\DeclareAcronym{DNN}{long = Deep Neural Network}
\DeclareAcronym{DBF}{long = Digital Beam Forming}
\DeclareAcronym{DPD}{long = Digital Predistortion}
\DeclareAcronym{DRAA}{long = Direct Radiating Active Antennas}
\DeclareAcronym{DQL}{long = Deep Q-learning}
\DeclareAcronym{DRL}{long = Deep Reinforcement Learning}
\DeclareAcronym{DRT}{long = Dehop-rehop Transponder}
\DeclareAcronym{DSA}{long = Dynamic Spectrum Access}
\DeclareAcronym{DSS}{long = Dynamic Spectrum Sharing}
\DeclareAcronym{DT}{long = Decision Tree}
\DeclareAcronym{DU}{long = Deep Unfolding}
\DeclareAcronym{EO}{long = Earth Observation}
\DeclareAcronym{ESA}{long = European Space Agency}
\DeclareAcronym{EIRP}{long = Effective Isotropic Radiated Power}
\DeclareAcronym{FBMC}{long = Filterbank Multicarrier}
\DeclareAcronym{FDMA}{long = Frequency Division Multiple Access,short-format = \scshape}
\DeclareAcronym{FEC}{long = Forward Error Correction}
\DeclareAcronym{FEM}{long = Finite element method}
\DeclareAcronym{FH-FDMA}{long = frequency hopping-frequency division multiple access, short-format = \scshape}
\DeclareAcronym{FL}{long = Federated Learning}
\DeclareAcronym{FFR}{long = Fractional Frequency Reuse}
\DeclareAcronym{FDTD}{long =Finite-difference time-domain method}
\DeclareAcronym{GA}{long = Genetic Algorithms}
\DeclareAcronym{GAN}{long = Generative Adversarial Network}
\DeclareAcronym{GCN}{long = Graph Convolutional Network}
\DeclareAcronym{GEO}{long = Geo-Stationary Orbit}
\DeclareAcronym{GNSS}{long = Global Navigation Satellite System}
\DeclareAcronym{GO}{long = Geometric optics}
\DeclareAcronym{GS}{long = Ground Segment}
\DeclareAcronym{GSD}{long = Ground Segment Dimensioning}
\DeclareAcronym{GSO}{long = Geosynchronous Orbit}
\DeclareAcronym{GW}{long = Gateway}
\DeclareAcronym{HAP}{long = High Altitude Platform}
\DeclareAcronym{HL}{long = Higher Layer}
\DeclareAcronym{HARQ}{long = Hybrid-Automatic Repeat Request}
\DeclareAcronym{HCA}{long = Hierarchical Cluster Analysis}
\DeclareAcronym{HO}{long = Handover}
\DeclareAcronym{HTS}{long = High Throughput Satellites}
\DeclareAcronym{KNN}{long = k-nearest neighbors}
\DeclareAcronym{KPI}{long = Key Performance Indicator}
\DeclareAcronym{IoT}{long = Internet of Things}
\DeclareAcronym{JCS}{long = Joint Communication and Sensing}
\DeclareAcronym{ISAC}{long = Integrated Sensing and Communication}
\DeclareAcronym{LA}{long = Link Adaptation}
\DeclareAcronym{LAN}{long = Local Area Network}
\DeclareAcronym{LAP}{long = Low Altitude Platform}
\DeclareAcronym{LDPC}{long = Low-Density Parity-Check} 
\DeclareAcronym{LEO}{long = Low Earth Orbit}
\DeclareAcronym{LL}{long = Lower Layer}
\DeclareAcronym{LO}{long = local oscillator}
\DeclareAcronym{LoS}{long = Line of Sight}
\DeclareAcronym{LP}{long = Linear Precoding}
\DeclareAcronym{LSTM}{long = Long Short-term Memory}
\DeclareAcronym{LTE}{long = Long Term Evolution}
\DeclareAcronym{LTE-A}{long = Long Term Evolution Advanced}
\DeclareAcronym{MAC}{long = Medium Access Control}
\DeclareAcronym{MaD-DPD}{long = Memory Polynomial aided Deep Neural Network Digital Predistortion}
\DeclareAcronym{MAP}{long = Medium Altitude Platform}
\DeclareAcronym{MEO}{long = Medium Earth Orbit}
\DeclareAcronym{MEC}{long = Mobile Edge Computing}
\DeclareAcronym{MDP}{long = Markov Decision Process}
\DeclareAcronym{MILP}{long = Mixed-Integer Linear Programming}
\DeclareAcronym{MODCOD}{long = Modulation and Coding}
\DeclareAcronym{MMSE}{long = Minimum Mean-Square Error}
\DeclareAcronym{MPM}{long = Memory Polynomial}
\DeclareAcronym{MIMO}{long = Multiple-Input Multiple-Output}
\DeclareAcronym{ML}{long = Machine Learning}
\DeclareAcronym{MoM}{long = Method of Moments}
\DeclareAcronym{MLP}{long = Multilayer Perceptron}
\DeclareAcronym{M2M}{long = Machine-to-machine}
\DeclareAcronym{mmWave}{long = millimeter Wave}
\DeclareAcronym{NAS}{long = Neural Architecture Search}
\DeclareAcronym{NCC}{long = Network Control Centre}
\DeclareAcronym{NCO}{long =  Numerically-controlled Oscillator}
\DeclareAcronym{NEC}{long = Neural Episodic Control} 
\DeclareAcronym{NFP}{long = Network Flying Platform}
\DeclareAcronym{NFV}{long = Network Function Virtualization}
\DeclareAcronym{NGSO}{long = Non Geo-Stationary}
\DeclareAcronym{NN}{long = Neural Network}
\DeclareAcronym{NOMA}{long = Non-orthogonal Multiple Access}
\DeclareAcronym{NR}{long = New Radio}
\DeclareAcronym{NTN}{long = Non-Terrestrial Network}
\DeclareAcronym{OFDM}{long = Orthogonal Frequency-division Multiplexing}
\DeclareAcronym{OFDMA}{long = Orthogonal Frequency-Division Multiple Access}
\DeclareAcronym{OGS}{long = Optical Ground Station}
\DeclareAcronym{OMA}{long = Orthogonal Multiple Access}
\DeclareAcronym{OPEX}{long = Operational Expenditure}
\DeclareAcronym{ORRF}{long = Online Random Regression Forest} 
\DeclareAcronym{OBO}{long = Output Back-Off}
\DeclareAcronym{PA}{long = Power Amplifier}
\DeclareAcronym{PAM}{long = Pulse Amplitude Modulation}
\DeclareAcronym{PKC}{long =  Public-key Cryptosystem}
\DeclareAcronym{PLS}{long = physical-layer security}
\DeclareAcronym{PNN}{long = Probabilistic neural network}
\DeclareAcronym{PO}{long = Physical optics}
\DeclareAcronym{PGD}{long = Projection Gradient Descent}
\DeclareAcronym{PSO}{long = Particle Swarm Optimization}
\DeclareAcronym{QoE}{long = Quality of Experience}
\DeclareAcronym{QoS}{long = Quality of Service}
\DeclareAcronym{QKD}{long = quantum key distribution}
\DeclareAcronym{RAN}{long = Radio Access Network}
\DeclareAcronym{RBF}{long = radial basis function neural network}
\DeclareAcronym{RDN}{long = residual dense network} 
\DeclareAcronym{RF}{long = Radio Frequency}
\DeclareAcronym{RL}{long = Reinforcement Learning}
\DeclareAcronym{RNN}{long = Recurrent Neural Network}
\DeclareAcronym{RRM}{long = Radio Resource Management}
\DeclareAcronym{RS}{long = Remote Sensing}
\DeclareAcronym{RTT}{long = Round Trip time}
\DeclareAcronym{RZF}{long = Regularized Zero Forcing}
\DeclareAcronym{SAGIN}{long = Space-Air-Ground Integrated Network}
\DeclareAcronym{SATCOM}{long = Satellite Communication}
\DeclareAcronym{SCBS}{long = Small Cell Base Station}
\DeclareAcronym{SCS}{long = Satellite Communication System}
\DeclareAcronym{SDMA}{long = Spaced Division Multiple Access}
\DeclareAcronym{SDN}{long = Software Defined Network}
\DeclareAcronym{SE}{long = Spectral Efficiency}
\DeclareAcronym{SKC}{long = Secret-key Cryptosystem}
\DeclareAcronym{SIC}{long = Successive Interference Cancellation}
\DeclareAcronym{SIP}{long = Satellite Image Processing} 
\DeclareAcronym{SLL}{long = Side Lobe Level}
\DeclareAcronym{SNR}{long = Signal-to-Noise Ratio}
\DeclareAcronym{SON}{long = Self-organising Network}
\DeclareAcronym{ST}{long = Satellite Terminal}
\DeclareAcronym{SSPA}{long = Solid-State Power Amplifier}
\DeclareAcronym{SV}{long = Satellite Video} 
\DeclareAcronym{SVC}{long = Support Vector Classifier}
\DeclareAcronym{SVM}{long = Support Vector Machine}
\DeclareAcronym{PAE}{long = Power-Added Efficiency}
\DeclareAcronym{PEP}{long = Peak Envelope Power}
\DeclareAcronym{PAPR}{long = Peak-to-Average Power Ratio}
\DeclareAcronym{TCL}{long = timing Control Loop}
\DeclareAcronym{TDM}{long = Time Division Multiplexing}
\DeclareAcronym{TDMA}{long = Time Division Multiple Access}
\DeclareAcronym{TLL}{long = Timing Locked Loop}
\DeclareAcronym{TWTA}{long = Traveling-Wave Tube Amplifiers}
\DeclareAcronym{TN}{long = Transport Network}
\DeclareAcronym{UAV}{long = Unmanned Aerial Vehicle}
\DeclareAcronym{UHTS}{long = Ultra High Throughput Satellites}
\DeclareAcronym{UE}{long = User Equipment}
\DeclareAcronym{VCO}{long = Voltage-Controlled Oscillator}
\DeclareAcronym{PoP}{long = Point of Presence}
\DeclareAcronym{OAI}{long = OpenAirInterface}
\DeclareAcronym{TEG}{long = Time-expanded Graph}

\newcommand{\figref}[1]{\figurename~\ref{#1}}

\NewAcroTemplate[list]{customtable}
  {
    \acronymsmapF
      { \AcroAddRow{\acrowrite {short}&\acrowrite {list}\\} }
      { \AcroRerun }  
    \acroheading
    \acropreamble
    \par \noindent
    \begin{NiceTabularX}{\linewidth}{lX}[hvlines, cell-space-limits=3pt]
      \AcronymTable
    \end{NiceTabularX}
  }

 \usepackage{supertabular}

\acsetup{
  index/use = false,
  list/template = supertabular,
}

\begin{document}
\title{Artificial Intelligence for Satellite Communication and Non-Terrestrial Networks: A Survey}
\author{\IEEEauthorblockN{\textsuperscript{} Gianluca Fontanesi, \IEEEmembership{Member, IEEE}, Flor Ort\'iz, \IEEEmembership{Member, IEEE}, Eva Lagunas, \IEEEmembership{Senior Member, IEEE}, Victor Monzon Baeza, \IEEEmembership{Member, IEEE}, Miguel {\'A}ngel V{\'a}zquez, \IEEEmembership{Member, IEEE}, Juan Andr\'es V\'asquez-Peralvo, \IEEEmembership{Member, IEEE}, Mario Minardi, \IEEEmembership{Student Member, IEEE}, Ha Nguyen Vu, \IEEEmembership{Member, IEEE}, Puneeth Jubba Honnaiah, Clement Lacoste, \IEEEmembership{Student Member, IEEE}, Youssouf Drif, \IEEEmembership{Member, IEEE}, Liz Martinez Marrero, \IEEEmembership{Student Member, IEEE}, Saed Daoud, \IEEEmembership{Member, IEEE}, Tedros Salih Abdu, \IEEEmembership{Member, IEEE}, Geoffrey Eappen,  \IEEEmembership{Member, IEEE}, Junaid ur Rehman,  
Luis Manuel Garc\'es-Socorr\'as, \IEEEmembership{Member, IEEE}, Wallace Alves Martins \IEEEmembership{Senior Member, IEEE}, Pol Henarejos, \IEEEmembership{Member, IEEE}, Hayder Al-Hraishawi \IEEEmembership{Senior Member, IEEE}, Juan Carlos Merlano Duncan, \IEEEmembership{Senior Member, IEEE}, Thang X. Vu, \IEEEmembership{Senior Member, IEEE} and Symeon Chatzinotas, \IEEEmembership{Fellow, IEEE}}
\thanks{Manuscript received XXX, XX, 2021; revised XXX, XX, 2021.}
\thanks{The authors from University of Luxembourg have been supported by the Luxembourg National Research Fund (FNR) under the project SmartSpace (C21/IS/16193290). The authors from CTTC have been supported by the European Union’s Horizon 2020 research and innovation programme under grant agreement No 101004215 (ATRIA) and by the Spanish ministry of science and innovation under project IRENE (PID2020-115323RB-C31/AEI/10.13039/501100011033) and grant from the Spanish ministry of economic affairs and digital transformation and of the European union – NextGenerationEU [UNICO-5GI+D/AROMA3D-Space (TSI-063000-2021-70).}
\thanks{G. Fontanesi, F. Ort\'iz, E. Lagunas, V. Monzon Baeza, J. A. V\'asquez-Peralvo, M. Minardi, H. N. Vu, P. J. Honnaiah, C. Lacoste, Y. Drif, T. S. Abdu, G. Eappen, J. Rehman, L. M. Garc\'es-Socorr\'as, W. A. Martins, H. Al-Hraishawi, J. C. Merlano Duncan, T. X. Vu and S. Chatzinotas are with the Interdisciplinary Centre for Security, Reliability, and Trust (SnT), Luxembourg, L-1855 Luxembourg}
\thanks{M. {\'A}. V{\'a}zquez and P. Henarejos are with the Centre Tecnologic de les Telecommunicacions de Catalunya.}
\thanks{(Corresponding author: Gianluca Fontanesi, gianluca.fontanesi@nokia.com)}
}

\maketitle
\begin{abstract}
This paper surveys the application and development of Artificial Intelligence (AI) in Satellite Communication (SatCom) and Non-Terrestrial Networks (NTN). We first present a comprehensive list of use cases, the relative challenges and the main AI tools capable of addressing those challenges. For each use case, we present the main motivation, a system description, the available non-AI solutions and the potential benefits and available works using AI. We also discuss the pros and cons of an on-board and on-ground AI-based architecture, and we revise the current commercial and research activities relevant to this topic. Next, we describe the state-of-the-art hardware solutions for developing ML in real satellite systems. Finally, we discuss the long-term developments of AI in the SatCom and NTN sectors and potential research directions. This paper provides a comprehensive and up-to-date overview of the opportunities and challenges offered by AI to improve the performance and efficiency of NTNs.
\end{abstract}
\IEEEpeerreviewmaketitle
\section{Introduction}
\ac{SATCOM} and \acp{NTN} are seeing unbounded growth, offering a new world of enthralling possibilities for global coverage, connectivity, and scalability. Space-based communication systems have been typically used to cover vast airspace and sea areas, supplementing existing terrestrial networks for global connectivity. With the increasing demand for broadband connectivity, non-terrestrial operators are investing in new orbits and new constellation designs to provide a cost-efficient response to such a market. In addition, \acp{NTN} technology has been identified as a key component of the upcoming 5G and beyond cellular communication, particularly for backhaul services, backup connectivity, and extending coverage in remote or isolated regions of the globe \cite{geraci2022integrating}.

While the term \acp{NTN} includes a huge variety of space-borne and aerial communication networks, such as \ac{GEO}, \ac{MEO}, \ac{LEO} satellite constellations, \acp{HAP} systems, \acp{LAP} systems, and \ac{A2G} networks, for the sake of space and clarity, this survey has focused on satellite communications systems limited to \ac{GEO}, \ac{MEO}, \ac{LEO}.


\subsection{Need for AI/ML in SATCOM}
The \ac{SATCOM} ecosystem is currently experiencing a revolution in advanced technological solutions and raising a new space era where lower orbits are emerging as a low-latency alternative to conventional \ac{GEO} communication systems. The time and geographical dependency of user demands combined with the dynamic movement of low-orbiting satellites have resulted in new antenna architectures with beamforming capabilities and reconfigurable payloads with the capability to adapt the payload configuration in response to traffic needs.

This new communication era brings fundamental operational challenges. Most deployed \ac{SATCOM} systems largely depend on human expertise, and manual intervention \cite{M_A_Vazquez_AISatellite_SatelliteNewsLetters}. 
This has two main drawbacks. First, human involvement in system control activity leads to high \ac{OPEX} and latency \cite{ortiz2023onboard}. Second, the rapidly changing radio environment of the new space scenarios claims for autonomously adaptive mechanisms that human intervention cannot offer.
Lastly, the multitude of use cases and scenarios served by satellite communication in the following years will produce data in large quantities. Thus, it is beneficial and necessary to make satellites capable of automatically generating reliable actions using the data produced.

\ac{AI} is entering satellite communication with the great promise to solve the abovementioned challenges. Intelligent systems enable automation to process specific input from satellite data and translate them into actions. In addition, data-driven techniques might substantially reduce the \ac{OPEX}.


Given the significant momentum and recent activities promoting \ac{AI} in satellite communication, it is timely to survey this brand-new field.

Certain literature agrees to classify \ac{ML} as a branch of \ac{AI}, whose goal is to allow computers to learn from data. In some way, \ac{ML} is just one way to achieve \ac{AI}. However, most of the time, both terms are used interchangeably, such as the idea of a machine mimicking human intelligence. Therefore, in Section II, we address this dilemma for a better understanding of the reader.

\subsection{Related Surveys and Contribution}
%
\ac{ML} has appeared as a promising alternative for dealing with computationally expensive optimization procedures. For this reason, and with the exponential increase in the number of datasets available, ML has become a fundamental technology in different areas of wireless communications \cite{jiang2016machine}. In this context, ML has proven to be an exciting tool for accelerating complex optimization procedures for general wireless communications \cite{bega2019machine}.
Many surveys cover the development of \ac{ML} models to support wireless networks. 

About the direct application of \ac{ML} models in \ac{SATCOM}, there are only a handful of survey papers summarising it.
For this reason, in the following subsections, we will present the surveys describing SATCOM and the application of \ac{AI}.

\subsection{SATCOM Related Surveys}
Due to the increasing importance and interest in satellite communication, several works survey different aspects of satellite communication.
The work in \cite{azari2022evolutionNTN} deals with the vision of \ac{SATCOM} in conjunction with \acp{UAV} and terrestrial networks to provide a feasible and cost-effective solution for continuous and ubiquitous wireless coverage.
The innovation and transformation phase experienced by
\ac{SATCOM}, driven by on-board processing capabilities, non-terrestrial networks and space-based data collection/processing, is explained in \cite{kodheli2020Survey_IEEE}.
The algorithms and applications of \ac{EO} are covered in \cite{hoeser2020object}.
Cubesats are a special class of miniaturized satellites commonly used as teaching tools, and technology demonstration \cite{walsh2021development}. The work in \cite{saeed2020cubesat} describes all the potential future applications of Cubesats for \ac{SATCOM}.

Other papers survey a variery of \ac{SATCOM} aspects, such as handover schemes \cite{chowdhury2006handover_SATCOM_IEEESurvey}, 
 non Geostationary satellite systems \cite{SnT2022survey_Geo_nonGeo}, syncronization techniques \cite{marrero2022Survey_Synchro}, MIMO techniques \cite{arapoglou2010mimo_IEEESurvey}.

The important case of \ac{SATCOM} as \ac{IoT} enabler is treated in several surveys \cite{kodheli2020Survey_IEEE, 2015satellite_IoT_IoTJourn, iot:SurveyOnIoTTechnologies}. The works
\cite{kodheli2020Survey_IEEE, 2015satellite_IoT_IoTJourn} categorize the use cases in which satellites serve IoT devices used in global transportation and agriculture applications. In local-area services, a satellite serves a specific set of IoT devices in applications such as smart grid systems.

\subsection{SATCOM-AI Related Survey}

While great pieces of work, the above papers neglect to address the important challenge of the application of \ac{ML} in the described use cases.
While not intended as a tutorial or survey papers, \cite{homssi2022artificial,M_A_Vazquez_AISatellite_SatelliteNewsLetters} introduce the application of \ac{AI} techniques for integrated terrestrial satellite networks, specifically focusing on mega satellite network communications \cite{homssi2022artificial} and \ac{SATCOM} operations which are strongly dependent on the human  intervention.
The work in \cite{fourati2021_AI_Satellite} provides a general overview of \ac{AI}, its subfields and algorithms and the potential \ac{AI}-based solutions for some satellite challenges.
Other works address \ac{DL} only \cite{kothari2020final,lofqvist2020accelerating} or \ac{AI} for space missions \cite{russo2022using}, without however addressing the challenges of onboard or on-ground computing, hardware constraints and limitations.

\subsection{Scope and Contribution}
This work complements the above works by surveying all possible \ac{SATCOM} use cases and discussing the practical and implementation aspects of the application of \ac{AI} techniques in \ac{SATCOM}. A detailed survey of this type was still missing, and it is of great importance to take a new look at the appealing \ac{AI}/\ac{ML} approaches for \ac{SATCOM} with an implementation perspective. 
The contributions of this paper can be summarized as follows:
\begin{itemize}
    \item  We present an extensive classification of the most important use cases in \ac{SATCOM} and \acp{NTN}, and we group them based on an OSI-oriented architecture. 
    \item  We describe each presented use case, the main related challenges, and the conventional tools used to address them. We further present the main \ac{ML} techniques applied to each use case, and we classify them by whether the training is onboard or on ground.
    \item We assess the hardware landscape for current and next-generation processors, presenting a comprehensive review of commercial AI chipsets and comparing their specifications in terms of \ac{SATCOM} applicability.
    \item We analyze the main challenges for the \ac{SATCOM} and \acp{NTN} from the perspective of future network deployment. We also present novel and futuristic trends.
\end{itemize}

This paper consists of six main sections. Specifically, we first briefly introduce the main \ac{AI}/\ac{ML} learning frameworks. Our survey does not describe how various ML methods operate. There are numerous books and research papers on this topic; we refer the reader to papers such as \cite{wang2020thirty,sun2019application}, for a detailed (although still wireless networking-related) discussion of these methods.
We then discuss in Section III the role of \ac{AI} in \ac{SATCOM}, highlighting the main differences between an onboard or on-ground implementation. Section IV presents an extensive list of use cases in \ac{SATCOM} and \ac{NTN}. For each use case, we show the benefits of using ML-based techniques.
Section V presents an analysis of the HW solutions for AI/ML implementations for \ac{SATCOM}.
Last, Section VI on the challenges and future directions of \ac{SATCOM} and \acp{NTN} concludes the paper.

\subsection{Methodology}
The following survey of \ac{SATCOM} is the result of a systematic literature review methodology.
First, we identified all the possible use-cases and \ac{SATCOM} research fields and their respective challenges.
For each of them, we searched for conventional and \ac{ML} solutions in IEEE, MDPI, Elsevier and Wiley libraries.
A systematic search was implemented to identify the most important related works. The research is restricted to journal and conference papers only.
A second round of research has been computed via manual research and personal knowledge.
Table \ref{tab:Article_Categorization}, to ease the reader understanding, provides a catchy and clear overview of \ac{SATCOM} use cases and related \ac{ML} papers. 

Additionally, we searched for all relevant surveys.
Table \ref{tab:Surveys} summarizes the past surveys on \ac{SATCOM}, allowing the reader to quickly grasp the primary focus of each of the previously conducted surveys.

\begin{table*}
 \centering
 \setlength{\tabcolsep}{3pt}
 \caption{Relevant surveys and tutorials on artificial intelligence (AI)/machine learning (ML) and/or satellite communications.}\label{tab:Surveys}
    \begin{tabular}{|l|l|l|l|l|}
        \hline
        Network & Source & Main Scope & ML Focused & Implementation Oriented \\
        \hline  
    \multirow{10}{*}{\rotatebox[origin=c]{90}{SATCOM}} & \cite{geraci2022integrating,azari2022evolutionNTN,zhu_integrated_2022} & Survey non-terrestrial networks  & No & No\\
         & \cite{kodheli2020Survey_IEEE,de2022open} & Next Generation Satellite Communications & No & No\\
         & \cite{hoeser2020object}  & Earth Observation & No & No\\
         & \cite{saeed2020cubesat} & Cubesat Applications & No & No\\
          & \cite{chowdhury2006handover_SATCOM_IEEESurvey} &  Handover schemes & No & No\\
         & \cite{SnT2022survey_Geo_nonGeo} & non-GEO satellite comm. & No & No \\
         & \cite{marrero2022Survey_Synchro} & Distributed Satellite Systems &  No& No\\ 
         & \cite{arapoglou2010mimo_IEEESurvey} & MIMO Techniques for Satellite  & No& No\\
         & \cite{2015satellite_IoT_IoTJourn} & Satellite Comm in Internet of Things & No & No\\
        & \cite{iot:SurveyOnIoTTechnologies} & Satellite Internet of Things & No& No\\
         & \cite{russo2022using} & Space mission & Yes& No\\
         \hline
          \multirow{5}{*}{\rotatebox[origin=c]{90}{ML \& SATCOM}} &\cite{fourati2021_AI_Satellite} & Review of ML techniques for satellite comm. & Yes & No\\
          & \cite{M_A_Vazquez_AISatellite_SatelliteNewsLetters} & ML for SATCOM operation centers & Yes & No\\
         & \cite{homssi2022artificial}  &  Survey on AI for Mega Constellation &  Yes& No\\
      & \cite{kothari2020final}  & Deep Learning in Space & Yes& No\\
      & \cite{lofqvist2020accelerating}  & Deep Learning in Space & Yes& No\\
       & \cite{ortiz2022machine} & ML for Radio Resource Management &  Yes& No\\
    & This work  & ML for Satellite Communication &  Yes & Yes\\
        \hline
    \end{tabular}
\end{table*}


\printacronyms
\section{Machine Learning Overview}
\subsection{Artificial Intelligence, Machine Learning and Deep Learning}
A multidisciplinary science that develops theories, techniques, and methods to simulate and extend human intelligence is \ac{AI} \cite{Smith_Eckroth_2017}. \ac{AI} attempts to understand the essence of intelligence and to simulate the human brain's information processing by machines. \ac{ML}, a branch of \ac{AI}, is related to computational statistics and predictions by exploiting experience and knowledge gained from data \cite{mitchell1997machine}. 
\ac{ML} involves learning from data and making decisions or predictions. Essentially, it is based on the assumption that machines can have intelligence that allows them to learn from previous computations and adapt to the environment.

A typical \ac{ML} framework includes a training and testing process. Training enables the ML framework to discover the relationships between input and output data.
While a tutorial on how to create a dataset and divide it between training and testing is outside the scope of this paper, we highlight that there are two phases of ML techniques for assisting SATCOM at any stage of network communication: training and inference \cite{ortiz2022machine}. First, the ML model undergoes the training phase. The goal is to find the optimal model parameters to assist the SATCOM task for the system conditions. Next, a trained model is obtained and used to indicate the best decision in the system.

Available ML models in literature can be classified into classification, regression, and structured learning models \cite{zhao2019survey}. A classification model is used to solve binary or multiple classification problems, and a regression model can be used to make predictions. The structured learning model is widely used in many fields, such as natural language processing. 

\begin{figure*}[tp]
\centering
\includegraphics[width =\textwidth]{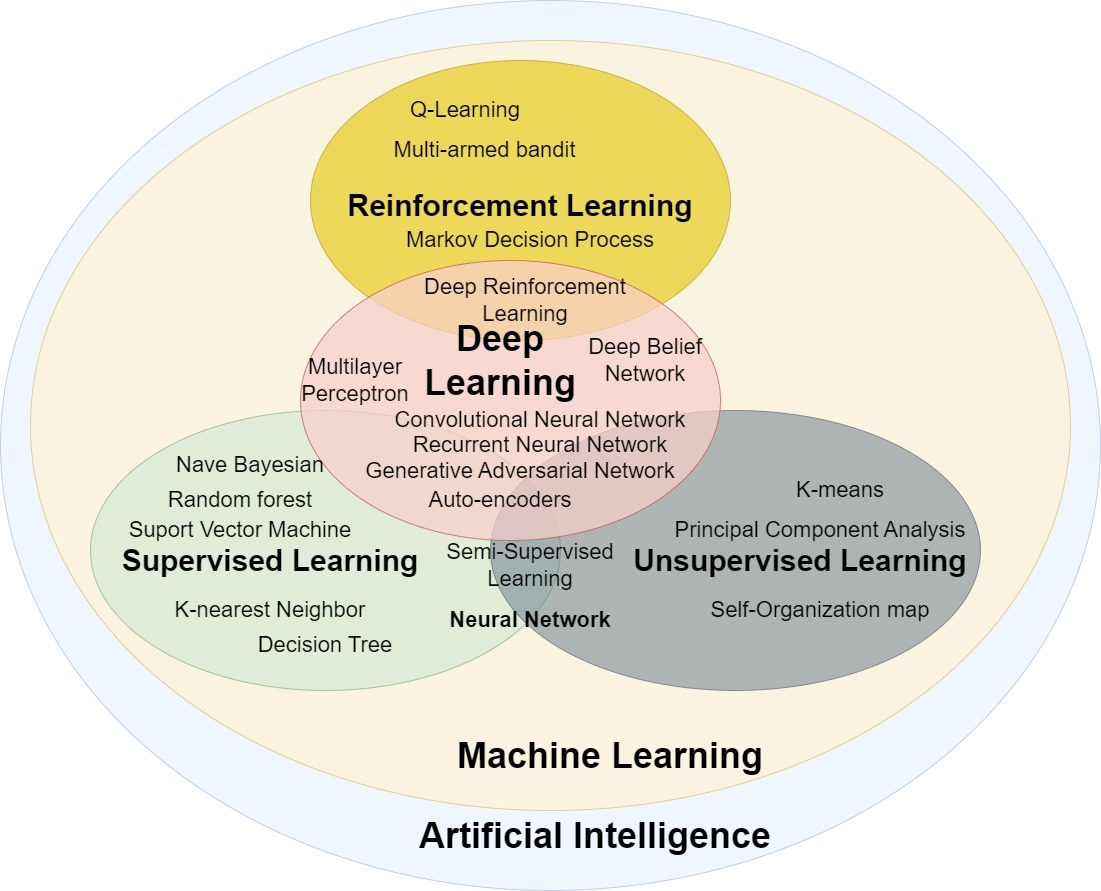}
\caption{Artificial Intelligence, Machine Learning, and Deep Learning relationship, as well as their main methods. }\label{fig:AI_class}
\end{figure*}

\ac{DL} is essentially a branch of ML, which allows a model to make classifications, predictions, or decisions based on large data sets, without being explicitly programmed.
Moreover, ML can be also classified from the training method into supervised, unsupervised, and  \ac{RL} \cite{zhao2019survey}. The classification of the main ML techniques and the intersections between them are shown in Fig. \ref{fig:AI_class}.

Opposite to conventional ML algorithms based on predefined features, DL can extract essential features from the raw data through multiple layers of nonlinear processing units to make predictions or perform target-based actions. DL's main advantages are to extract features automatically. The best-known DL models are neural networks with a sufficient number of hidden layers. Even though the multilayer neural network was proposed several decades ago, unprecedented interest has only recently arisen due to the advancement of backpropagation-based training as well as the success of GPU. \acp{DNN} aims to approximate any complex function by a composition of simple operations on neurons. \acp{DNN} can automatically extract essential features from input data with complex structures. No human-designed learning process is needed, significantly reducing the effort in feature elaboration. DL may learn meaningful patterns from unlabeled data in an unsupervised manner. A key challenge of employing DL in mobile communication systems lies in designing optimal deep neural networks for different scenarios just so that the model can be effectively trained in the offline stage and achieve good performance in the online testing stage. Representative typical DL models include multilayer perceptron (MLP), deep belief network (DBN), autoencoders, \ac{CNN}, recurrent neural network (RNN), and generative adversarial network (GAN) \cite{lecun2015deep,hatcher2018survey,collobert2004links}.

\subsection{Learning Framweworks}
%
\ac{ML} learning approaches can be grouped into supervised learning, unsupervised learning, and reinforcement learning (RL).

\subsubsection{Supervised Learning}
A supervised learning approach needs a supervisor to label the input and output data. During the training step, a learning algorithm is provided with labeled training data containing known input and output data to train a model representing the relationships between the input and output data. A novel test data set is introduced into the learned model during the testing step to obtain the expected output. Usually, supervised learning is employed in scenarios with sufficient labeled data. Many fields, such as object recognition, speech recognition, and spam detection, have used this method extensively. Algorithms typical of this category are Naive Bayes, K-nearest neighbor (KNN), random forest, \acp{NN}), \ac{SVM}, and \ac{DT} \cite{kulkarni2012pruning,buczak2015survey}.

\subsubsection{Unsupervised Learning}
Unsupervised learning algorithms use a set of unlabeled inputs to infer the output correctly. These techniques are frequently used for clustering and aggregation. Some typical unsupervised algorithms include K-mean, self-organizing maps (SOM), hidden Markov model (HMM), and restricted Boltzmann machine (RBM) \cite{na2010research,kohonen1982self,stratonovich1965conditional}.

\subsubsection{Semi-supervised Learing}
Semi-supervised learning involves a combination of supervised learning and unsupervised learning. A large amount of labeled data is expensive and time-consuming to generate in many real-world scenarios, while it is relatively inexpensive and convenient to collect enough unlabeled data. Consequently, semi-supervised learning develops to fully use the unlabeled data to improve the performance of the trained model. Descriptive pseudo labeling, a typical example of the semi-supervised learning algorithm, is simple and efficient. The constrained labeled data are used to train a model. Next, the trained model generates the pseudo labels for the unlabeled data. And finally, the labeled data and the pseudo-labeled data are used to retrain the model. To fully use the unlabeled data, semi-supervised learning algorithms usually require certain assumptions on the dataset, such as manifold assumption, low-density assumption, clustering assumption, and smoothness assumption. For example, expectation maximization is based on the clustering assumption, while transductive SVM requires the low-density separation assumption \cite{mallapragada2008semiboost,wu2017semi,xie2018survey}.
\subsubsection{Reinforcement Learning}
Reinforcement learning (RL) mimics the brain's trial-and-error learning process \cite{kiumarsi2017optimal}. As opposed to learning the structure of the training data set, RL attempts to explore the best actions during a dynamic process. Being able to understand the environment through actions and feedback makes it suitable for solving decision-making problems. RL framework can be classified into two types: model-based and model-free. The framework of model-based RL includes an agent, a state space, and an action space. Interacting with the environment, the agent tries to represent the model of the environment and learn the best action to maximize its long-term reward, which is a cumulative discounted reward and is related to current and future rewards. On each step, the agent monitors a state and takes action from the action space, then receives an immediate reward that indicates the effect of the action, followed by the system moving to another state. In model-based approaches, the agent attempts to learn a policy.

In the state transition process, the agent learns the best policy, mapping from the state to the action space to maximize the long-term reward. When determining the long-term reward of the action in a state space, the value function is applied. A value function most commonly used is the Q function, used by the Q-learning method to learn a table for maintaining state-action pairs and associating long-term rewards. RL based on models is more suitable for dynamic systems due to the difficulty of constructing an accurate model of dynamic networks \cite{elsayed2019ai}. Compared to other learning techniques, the main advantage of reinforcement learning is that it does not rely on an accurate mathematical model of the environment.

Furthermore, the approach addresses long-term rewards, including immediate and future rewards, which allows for long-term optimization results. The key challenge in applying RL to wireless communication systems is designing the system state, action, and reward in different scenarios to achieve optimal performance. During recent years, RL has been widely employed to solve decision-making problems in wireless communications, such as user scheduling, spectrum sharing, or radio access technology handover \cite{qiao2018topology,raj2018spectrum,nguyen2017reinforcement,ZHANG2020107556}. Nevertheless, the technique faces some challenges when dealing with problems with a large state or action space since it is challenging to model each state-action pair directly. Consequently, \ac{RL} is hardly used in practice \cite{ZHANG2020107556}.


\section{Machine-Learning Assisted Satellite System}

\subsection{Introduction}
The satellite communications world is also progressing towards having all the ingredients making ML suitable for particular satellite-related use-cases. First, in response to the increasing telecommunication's market demand for flexibility, satellite payloads are being digitalized. In particular, the digital processing of the uplink signals combined with digital routing and filtering flexibility can be highly efficient to address the emerging broadband market \cite{9513517}.

Full digital payloads are paving the way for satellites with a higher level of flexibility to efficiently support peaks in demand. These payloads include capabilities like onboard channelization, interference mitigation techniques, digital beamforming and/or time flexibility. In principle, the upcoming satellite systems should enable to dynamic allocate much higher capacity to designated areas, while other areas do not require as much bandwidth and are just being scanned to
gauge demand. For most satellite operators, automatic dynamic resource allocation algorithms are the key innovation aspect to unlocking the value of the aforementioned technological investments. Therefore, to unveil the full potential of such new added flexibility, ML can help accelerate the complex procedures typically encountered in system optimization and minimize the human decision-making over such procedures. Furthermore, ML is known to be a great tool for network load prediction, which is also a key aspect when deciding on when and how to reconfigure the satellite payload.

A side from the flexibility of next generation satellite systems, satellite payloads are known by its nonlinear effects, which cause significant headaches to engineers in order to countermeasure their impact on the final performance. Non-linear effects and general impairments introduced by hardware components are part of the channel information, which is typically estimated by the on-ground receiver by measuring pilot signals. Although different models for such impairments exist, such models may not accurately present the actual signal. To deal with such complexend-to-end system model, ML can be exploited to either predict the true model directly from experience or to estimate the parameters to fine-tune the already available model.

Focusing on the ground segment, which has been typically represented by few gateways (GWs) for GEO systems, it is currently evolving towards a multi GW environment. In the case of GEO, this is because the congestion of the Ka band is pushing feeder links to operate in higher bands, e.g. Q/V bands. Given the strong impact of weather impairments on higher frequency bands, the multi-location diversity achieve by mulitple distributed GW becomes an attractive solution. In the case of NGSO, the need of multi-GW is dictated by the movement of the satellites and the need to have often a connection to Earth for the uplink / downlink of the signals. The multi-GW environment resembles the promising Centralized Radio Access Network (C-RAN) architecture of cellular systems \cite{6897914}, where the control plane is logically centralized gathering the whole system intelligence providing a global view of the entire system status and enabling cross-layer system-wide optimization. In this context, machine-learning-based automation applied at the central node can autonomously optimize the network while reducing the critical processing time for decision-making.

Embracing cloud-based architectures, satellite operators look forward to managing and coordinating the different parts of the network in a centralized mode. The benefit for operators is the availability of very large data sets from their networks and the possibility to apply a range of data processing, geo-spatial, ML and other tools to manipulate and analyse such data, with the main goal to optimize the system operation.

Last but not least, the emergence of piggy-back launches, nano and microsat technology and new higher-risk accepting development approaches and the availability of venture capital, new innovative mission concepts have been envisaged and implemented. This trend of having space-based mesh network of small satellites opens up a plethora of opportunities for massive self-organized, reconfigurable and resilient NGSO satellite constellations, which can operate as a global network instead of a single relay.

To exploit NGSO satellite signals, frequency and time misalignment caused by the fast system mobility shall be compensated. For this, accurate estimation and prediction of ephemeris is a key aspect and has been the subject of extensive research, e.g. \cite{9438144, Mahdis2022}. On the other hand, the distributed nature of NGSO constellations translates into highly dynamic and overlapped coverage areas on ground, which bring many operation challenges, e.g. the dynamic handover of users from satellite to satellite, the edge-processing in space, the cooperative transmission from multiple satellites, etc. Each of those relatively new challenges can have potential benefits on the application of ML, as we will see in the next parts of this paper. Before that, we present in the following an overview of the conventional satellite architecture and how ML can be integrated into such systems.

\subsection{Communications Architecture}

\begin{figure*}[tp]
\centering
\includegraphics[width =\textwidth]{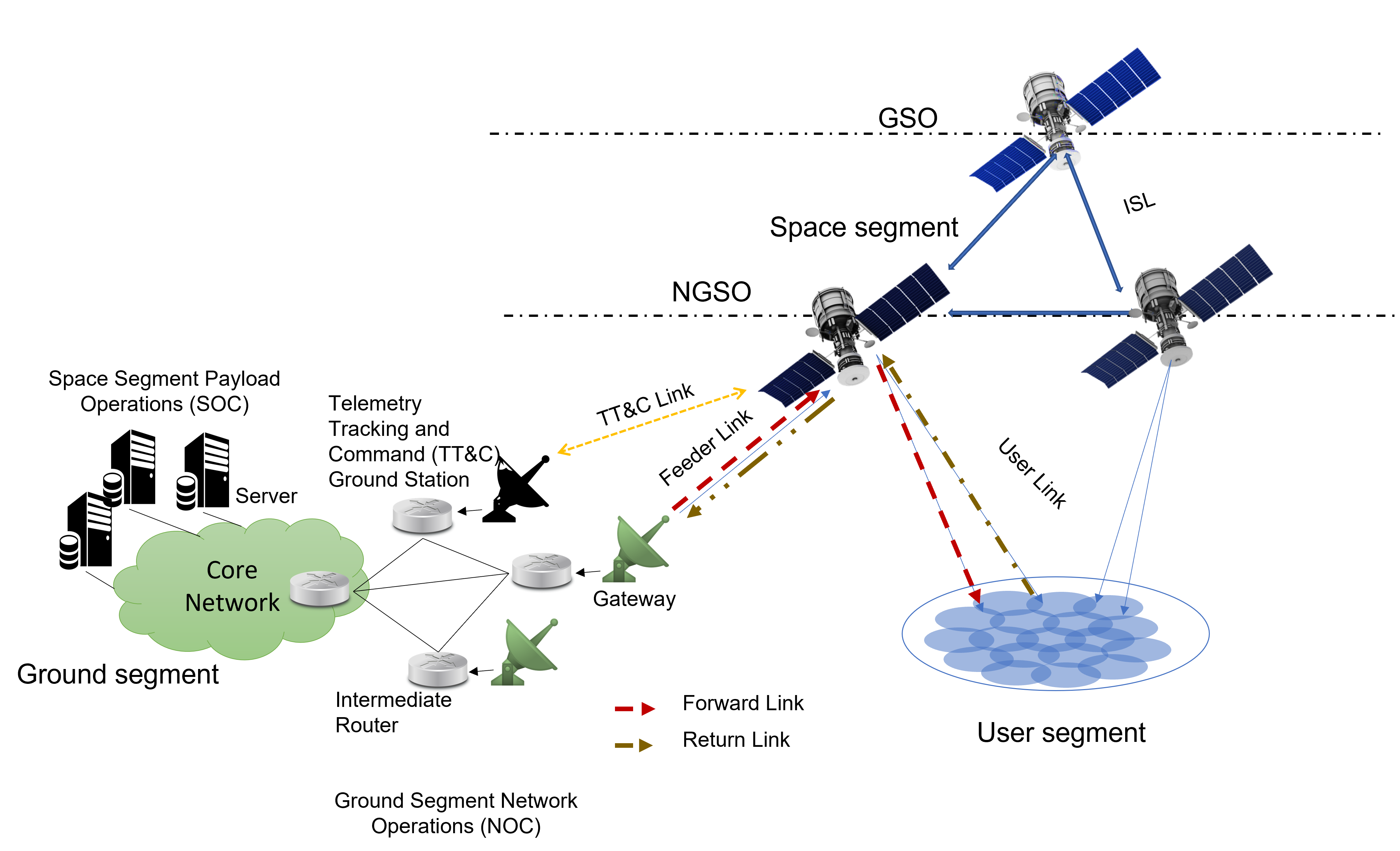}
\caption{Communications Architecture}\label{fig:ComArch}
\end{figure*}

Satellite network architecture comprises space, ground, control, management, and user segments, as shown in Figure \ref{fig:ComArch}. (i) Space segment comprises the satellites organized in the constellation. It supports routing, adaptive access control, and spot beam management. (ii) The ground segment consists of satellite gateways (GWs) interconnected by optical backbones and satellite terminals (STs) that provide connections for end-user devices. The GWs and STs are interconnected through the space segment. (iii) The control and management segment consists of the network operations centers (NOCs) of the ground segment and the payload operations centers (SOCs) of the space segment. The NOCs and SOCs provide real-time management and control functions for the satellite networks. They perform connection establishment, tracking and release, admission control, resource allocation, satellite network element configuration, security, fault, and performance management. Co-located GWs, NOCs, and SOCs are commonly referred to as satellite hubs. (iv) The user segment comprises all end-user devices used by end-users to consume satellite-based services, fixed or mobile. They access satellite networks directly or through terrestrial access points \cite{Xu2018_Arch_Challanges}. 

\subsection{ML for SATCOM Dilemma: Onboard or Onground}\label{sec:Onboard_Onground}
One of the main dilemmas from the space segment architecture perspective is whether to perform operations onboard or on-ground.
Onboard strategies increase the complexity and mass/power of the satellite, which has to be ultra-reliable and robust since there is very little chance of repairing/replacement after the asset is put in orbit \cite{kodheli2020Survey_IEEE}.
Thus, due to the limited satellite process capabilities, onboard applications are challengings in \ac{SATCOM}. On the contrary, on-ground techniques alleviate the onboard complexity.
Recent advances in the efficiency of power generation and the energy efficiency of radio frequency and digital processing components enhanced onboard processing, which can enable innovative communication technologies, such as flexible routing, \ac{RRM}, and beamforming.

As an example of the onboard/on-ground dilemma for the latest mentioned case, \cite{tronc2014_Beamforming:overviewOnBoardOffboard} classifies beamforming methods into onboard and on-ground. While no perfect solution exists, different satellite playmakers have opted for different techniques, onboard (e.g. Inmarsat4, Alphasat or Thuraya) or on-ground techniques, such as US satellite operators.

The same dilemma persists and is emphasized in the application of \ac{ML} in \ac{SATCOM}, where Satellites' ability to capture vast amounts of data from space represents an excellent opportunity to apply ML. 

From the onboard/on-ground perspective, \ac{ML} techniques for \ac{SATCOM} can be first categorized depending on whether the \ac{AI} chipset is onboard the satellite, or at the ground station.
AI algorithms can be very computationally intensive, consume significant power, and are slow when running on standard processors. Thus, running ML models on-ground would be preferable to avoid high payload complexity and reduce cost. However, this would represent a longer delay if the inference waits for the return channel to decide based on the new conditions. 

The industry has developed AI-specific processors capable of executing complex algorithms in real-time and consuming a fraction of the power to enable the practical use of AI for satellite applications in terrestrial and onboard scenarios. These processors are now available as commercial AI chipsets and will be presented in Section \ref{HardwareSolutions}.

In addition, any AI techniques for \ac{SATCOM} must go through a training, where the model parameters are computed, and an inference phase. Thus, an essential decision is where to perform the algorithm's training. 
More specifically, the term online refers to performing the training phase when the \ac{SATCOM} system is active regardless of whether the \ac{ML} hardware can be on board satellites in a space environment or on the ground. However, due to the computational complexity, it is often performed on the ground to avoid the limitations of power, mass and cost of the payload when training is performed online. Still, thanks to the optimized hardware explained in section \ref{HardwareSolutions}, it is hoped that it can help mitigate this challenge to make onboard training more feasible. In contrast, offline training of the ML algorithm is done when the system is not active in the ground segment with a training database and large servers with powerful GPUs. Nevertheless, the AI chipset could go on board the satellite using the model for inference \cite{ortiz2023onboard}.

Based on these criteria, in this survey we categorize \ac{ML} techniques in \ac{SATCOM} as offline or online \ac{ML}.
In what follows, we are interested in showing each configuration's main benefits and challenges.

The main advantage of online training is that the system will have greater flexibility to adapt to unforeseen network changes, e.g., a drastic and unexpected change in traffic demand, because the learning is constantly updated.
Additionally, onboard methods avoid downloading data to servers on the ground, leading to lower latency. Consequently, the data to train the model can be collected and the training can be performed without incurring in outages and network disconnections.  
Last, reducing the need for server communication, the data collected and used for \ac{ML} training, that often captures sensitive information, is less exposed to privacy issues and the risk to be leaked. 

However, while onboard training provides several benefits, implementing such training can be challenging due to the relatively high convergence length and the limited computing capacity onboard \cite{fontanesi2022transfer}.
Consequently, \ac{ML} models in an onboard setting need to be efficient and often have to trade off against processing time, memory, number of computational operations and energy consumption. 


In offline training, once the model has been trained, the \ac{AI} chipset could be located onboard the satellite for inference. Consequently, computing/processing times are reduced since inference on a trained ML model requires a low computational cost compared to the training phase. 
The ML model might thus act as an intelligent switch that selects the configuration based on the current system requirements.
The main advantage of this architecture is the reduced processing time.
However, this architecture strongly depends on the training data and models used. An extensive database may be necessary to obtain successful training that considers multiple parameters, and the model may present significant errors in unexpected situations in the system.


%
Recent advances to solve the on-board/off-board dilemma includes hybrid architectures, where the model training is initially performed online/offline and then updated by collecting new data \cite{ortiz2022machine}.
Therefore, this trade-off strongly impacts the AI Chipset implementation in the system. To ensure successful deployability, designing lightweights \ac{NN}, which might arrive to have hundreds of parameters and layers, is an essential step in deploying resource-constrained computing platforms.
Architectures as MobileNets, ShuffleNet, PeleeNet,and techniques as Model compression, Neural Architecture Search are indicated to reach this goal \cite{mateo2021_TechinicalReport_OnBoardML}.
Another interesting solution is to run the \ac{ML} on-board, but adjusting the \ac{ML} algorithm to the satellite limited capacity \cite{dhar2021survey:OnDeviceML}. In this case, due to the lower compute and lower memory resources onboard, the ML models deployed on the satellite are designed for compute and memory efficiency. Typical techniques to reach increase efficiency are compressing high-performing pre-trained models or designing ad-hoc and application-oriented architectures that are parameter and compute-efficient.
Alternatively, hardware acceleration techniques applied in co-design with software can be used to exploit the model compression algorithms at a hardware level.
The reader will find in \ref{tab:Article_Categorization} the indication whether the literature ML models are run online or offline.
We will discuss the more advanced solutions and visions to solve the onboard/on-ground dilemma in Section \ref{Challenges}.

\begin{table*}[ht]
\centering
 \caption{Benefits and CHallenges for an Online or Offline ML training in SatCom}\label{tab:OnBoardOnGroundComparison}
 \setlength\tabcolsep{4pt}
 \begin{tabular}{|l|l|l|}
        \hline
        ML Training in SatCom & Benefits & Challenges   \\
        \hline
         & Lower Latency  & Limited Computing Capacity on-board  \\
         \emph{Online} & Improved Autonomy  & Longer Processing Time  \\
          & Flexibility to adapt to unforeseen changes   & Limited Power   \\
          & Higher Privacy   &   Radiation Tolerance \\
        \hline
         & Reduced processing time & Not flexible to sudden network' changes \\
          \emph{Offline} & Large computational resources   &  Inaccurate in unexpected situations  \\
           &  & Higher Latency \\
        \hline
\end{tabular}
\end{table*}

\subsection{Ongoing Activities}

This section will provide a brief overview of the most relevant activities carrier out by private and public institutions worldwide on the developmet of ML in \ac{SATCOM}, focusing on the most advanced technology readiness level.

\subsubsection{ML Testbeds}

\begin{figure}[tp]
\centering
\includegraphics[width =\columnwidth]{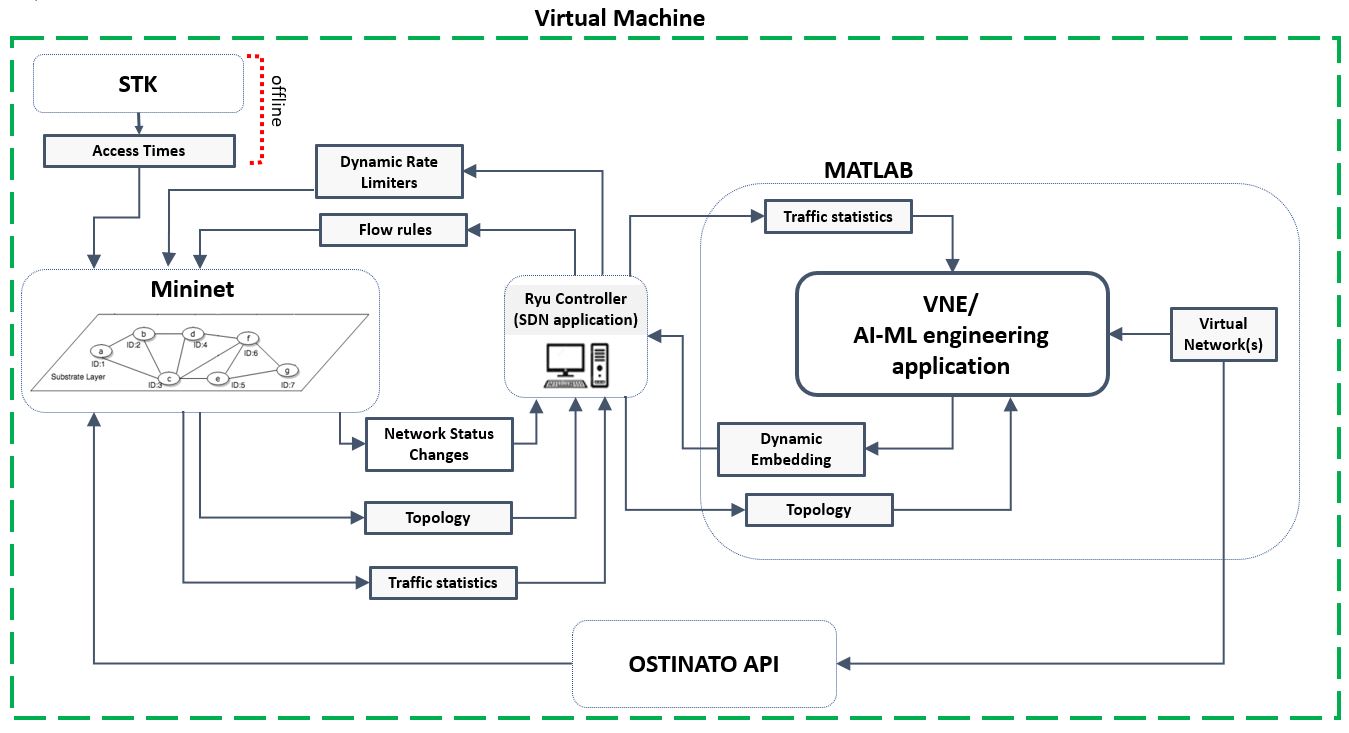}
\caption{Laboratory Testbed }\label{fig:SnT_SDN_testbed}
\end{figure}

MIRSAT testbed \cite{mendoza2021sdn} for Non-Geostationary Orbit (NGSO) constellations proposes an experimentation software platform based on \ac{SDN} for validation of the new autonomous network-slicing algorithms and their performance evaluation in integrated terrestrial-satellite systems. 
The laboratory testbed\footnote{\url{https://wwwen.uni.lu/layout/set/print/snt/research/sigcom/sdn_lab}} has been developed and validated \cite{Minardi_SDN}, consisting of a mininet-based simulator, a Ryu SDN controller with an End-to-End (E2E) Traffic Engineering (TE) application for the Virtual Networks (VNs) establishment, a satellite simulator STK, a real traffic generator OSTINATO and the VNE algorithm implemented in Matlab( \figref{fig:SnT_SDN_testbed}).

Late 2020, SPIRE company (with support from the ESA's Earth Observation Science for Society Programme) launched the so-called Brain in Space project\footnote{\url{https://spire.com/blog/space-services/brain-in-space-new-satellite-testbed/}}. Based on SPIRE's LEMUR 3U platform, Brain in Space provides an on-the-ground simulated testbed including multiple new embedded ML modules for users to schedule, upload and test their ML-powered applications.

On the other hand, INTEL has funded the Intel Neuromorphic Research Community, which includes academic, government, and corporate groups worldwide engaged in advancing the science and application of ML-based neuromorphic computing using Intel's Loihi neuromorphic processor. A discussion on neuromorphic computing is provided in Section VI of this survey. Herein, we highlight the willingness of industrial partners to make the technology available for testing and a better understanding of the ML challenges in the upcoming future.

\subsubsection{ESA}

The \ac{ESA} opened an AI-related call for SATCOMs for the first time in 2019 to investigate the applicability of AI techniques in satellite communications. 
Several potential use cases were shortlisted during these projects. A preliminary evaluation of a small number of them was carried out to provide guidelines for future research; (i) SATAI - Machine Learning and Artificial Intelligence for Satellite Communications 
and (ii) MLSAT - Machine Learning and Artificial Intelligence for Satellite Communication.

In 2020 \ac{ESA} pioneered the deployment of \ac{ML} into space for earth observation onboard a 6U Cubesat called $\Phi$-Sat-1 using an Intel Movidius chip \cite{mateo2021_TechinicalReport_OnBoardML}, with the intent of demonstrating and validating the state-of-the-art \ac{DL} technology applied in-orbit for autonomously processing Earth Observation data.

More recently, ESA is planning a new activity for 2023 named 6G Satellite Precursor, whose objective is to develop an in-orbit laboratory that allows for the R\&D process to be realised early in 6G adoption so that the satellite industry can adapt their technologies products, and use-cases to work side by side with terrestrial communication infrastructure. As part of the in-orbit laboratory, an AI chipset will be considered.

\subsubsection{NASA}

NASA has been actively investigating the \ac{CR} framework in satellite communications within the John H. Glenn Research Center testbed radios aboard the International Space Station.
In addition, a custom edge computing solution for DNA sequencing on the ISS on HPE's Spaceborne Computer-2 eliminates the need to move the massive data produced on the ISS by the DNA Sequencing project by presenting containerized analytical code where the data is being produced \cite{IBM_Space}. By leveraging the local computing available on ISS, the dependence on Earth and the time to get results can be reduced. The custom solution utilizes Red Hat CodeReady Containers, a single-node OpenShift cluster. This solution connects back on the ground with IBM Cloud, where researchers will develop, test and make their code ready to be pushed to the ISS. A ground-based solution on IBM Cloud will permit users to submit jobs, via a VPN connection, to the HPE ground systems that communicate securely with the HPE Spaceborne Computer-2 systems on the ISS, where Red Hat CodeReady Containers are installed.
This solution will not only help expedite the research being performed on the ISS but will also open doors for many new explorations on the ISS and future missions in space.

\subsubsection{European Commission}

From the European Commission side, the Smart Networks and Services Joint Undertaking (SNS JU) recently awarded the first batch of projects to facilitate and develop industrial leadership in Europe in 6G networks and services \cite{SmartNetworks}. Among the awarded projects, some of them included ML-powered optimizations for the integrated 6G and Non-Terrestrial network. The goal of such activities is to improve performance and automation, achieve zero-touch network management and timely monitor and predict overall network behaviour.
 
On the other hand, the European Union’s Horizon 2020 research and innovation programme already granted a few space-related projects, e.g. ATRIA \cite{Atria} or DYNASAT \cite{Dynasat}, both including the investigation of ML for SatCom systems.

\section{Use Cases Analysis}\label{UseCases}
\begin{figure*}[tp]
\centering
\includegraphics[width =\textwidth]{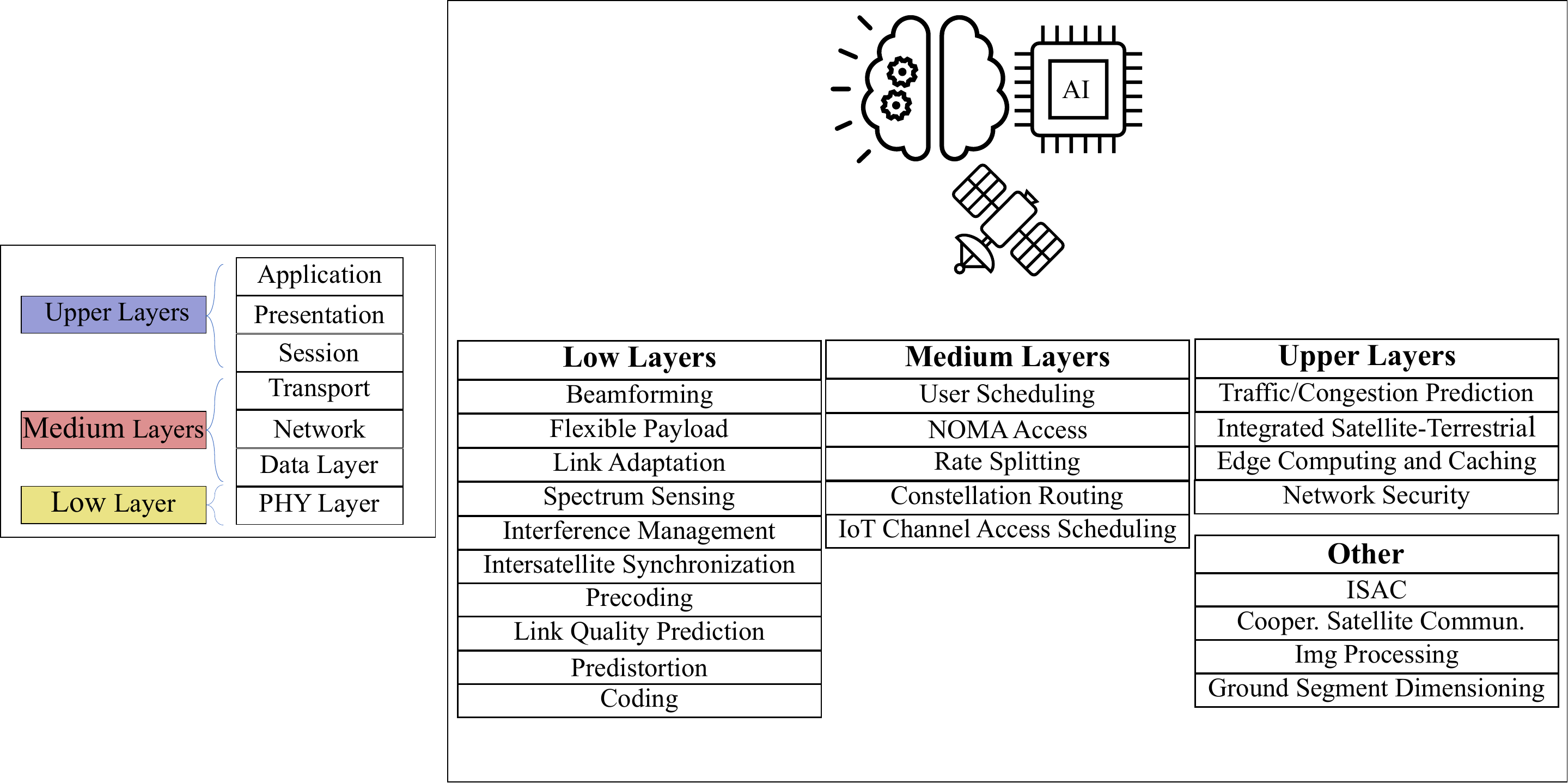}
\caption{Classification OSI-oriented for the use cases presented in Section IV }\label{fig:OSI_classification}
\end{figure*}

This section presents in detail the main \ac{SATCOM} use cases. We have divided the used cases in three layers, that group the conventional OSI layers (\figref{fig:OSI_classification}).
For each use case we present (i) Motivation, (ii) Detailed Description, (iii) Conventional solutions and (iv) ML solutions.
Referenced articles are categorized and summarized in Table \ref{tab:Article_Categorization}. 
Note that \ac{NN} approaches have a dedicated column to include literature that do not refer to supervised learning.
\begin{landscape}
\begin{table}[]
\centering
\caption{Use Cases Analysis for ML and Satellite Communication: in Magenta we highlight an offline ML training, in Cyan an online ML training} 
\begin{tabular}{|l|l|l|l|l|l|l|l|l|}
\hline
  &  \multicolumn{2}{|c|}{Use Case} &   \multicolumn{6}{|c|}{ML Framework} \\ 
  \cline{1-9}
  \multicolumn{3}{|l|}{}  & \rotatebox[origin=c]{90}{Supervised}  & \rotatebox[origin=c]{90}{Unsupervised} & \rotatebox[origin=c]{90}{NN $\&$ DL} &  \rotatebox[origin=c]{90}{RL} & \rotatebox[origin=c]{90}{DRL} &  \rotatebox[origin=c]{90}{FL}  \\ 
  \cline{1-9}
  &  \multicolumn{2}{|l|}{\textit{Low Layers}} &  &  &  &  &   &   \\ 
  \cline{2-9}
  \multirow{25}{*}{\rotatebox[origin=c]{90}{ML for Satellite Communication} }  & \multirow{2}{*}{Beamforming} & Antenna Design & \textcolor{magenta}{\cite{chen2019VTC_DL_faultDetection}} & \textcolor{magenta}{\cite{Nan2007}}& \textcolor{magenta}{\cite{ayestaran2006near_NN,vakula2010using}} &  &  &   \\
  \cline{3-9}
  & &  Antenna Beamforming &  &  \textcolor{magenta}{\cite{bianco2020RadarConf:AdaptivebeamformingDL,singh2020machine}} &\textcolor{magenta}{\cite{bianco2020RadarConf:AdaptivebeamformingDL,mallioras2022_adaptiveBeamforming_DNN,kumar2022dnn}}, \textcolor{cyan}{\cite{li2004performanceDigitalBeamforming}}&  & &   \\
  \cline{2-9}
  &  \multirow{4}{*}{Flexible Payload} & Beam Hopping   &   &  &  \textcolor{magenta}{\cite{lei2020deepL_BeamHopping}}, \textcolor{cyan}{\cite{Leilei2020_Access}} & &  \textcolor{cyan}{\cite{Hu2020_TB,Lin2022_TVT}} &    \\
  \cline{3-9}
  & &  Power  &  & &\textcolor{magenta}{\cite{Flor2019conf1,Flor2019conf2}} & &\textcolor{cyan}{\cite{Zhang2020, Luis2019con1, Luis2019con2}} &   \\
  \cline{3-9}
  & & Bandwidth  &  & & \textcolor{magenta}{\cite{Abdu2022deepcon}}& & \textcolor{cyan}{\cite{Liu218jo, iot:CA-DRLenergyefficient, He2022jo, Lin2022_TVT}}    &   \\
  \cline{3-9}
  &  & Beamwidth &  & \textcolor{cyan}{\cite{Flor2021jo}} & & \textcolor{cyan}{\cite{Flor2021jo} }& \textcolor{magenta}{\cite{Flor_RL2022}} &    \\
    \cline{2-9}
  & \multirow{1}{*}{Link Adaptation} &  & \textcolor{cyan}{\cite{9268889}} & & \textcolor{magenta}{\cite{7924387,9154263,9631216}} \textcolor{cyan}{\cite{8885616}} & &  \textcolor{cyan}{\cite{9631216}  }  &     \\ 
  \cline{2-9}
  &  \multirow{1}{*}{Spectrum Sensing Classification} &  &  & & & &      &     \\
  \cline{2-9}
  &  \multirow{2}{*}{Interference Management} & Detection  &  & & \textcolor{magenta}{\cite{Henarejos2019, Kulin2018}} & &      &     \\ 
  \cline{3-9}
  & &   Classification &  & & \textcolor{magenta}{\cite{Qin2022}} & &      &      \\ 
  \cline{2-9}
    &  \multirow{1}{*}{Intersatellite Synchronization} &  & \textcolor{magenta}{\cite{Wang2013learning, Wang2017,Lee2019}} & \textcolor{magenta}{\cite{Zibar2015}} & \textcolor{magenta}{\cite{Wu2019DL}} & &      &    \\ 
  \cline{2-9}
    &  \multirow{1}{*}{Precoding} &  & \textcolor{magenta}{\cite{VanPhuc_2022}} &  & \textcolor{magenta}{\cite{VanPhuc_2022}} & & &       \\ 
    \cline{2-9}
  & \multirow{1}{*}{Link Quality Prediction} &  &  &  & \textcolor{magenta}{\cite{9253575,Ventouras2019,cornejo2022method,9682090}}  & &   &  \\
  \cline{2-9}
  &  \multirow{1}{*}{Predistortion} &  &  & &  \textcolor{magenta}{\cite{sun2022navigation_DPD_NN}} & &      &    \\ 
  \cline{2-9}
    &  \multirow{1}{*}{Coding} &  &  & \textcolor{magenta}{\cite{9054192}}&\textcolor{magenta}{\cite{DL_decodingSurvey,7852251,phdthesis,lugosch2018learning,Haroon2013DecodingOE,9054192}} & &      &    \\ 
  \cline{2-9}
     & \multicolumn{2}{|l|}{\textit{Medium Layers}} &  &  &  &  &   &   \\ 
  \cline{2-9}
  & \multirow{1}{*}{User Scheduling} &  &  & \textcolor{magenta}{\cite{clust_Guidotti,9815569}} & & &      &     \\ 
    \cline{2-9}
    &  \multirow{1}{*}{NOMA Access} & & \textcolor{magenta}{\cite{8626195}},\textcolor{cyan}{\cite{9053003}}  & \textcolor{magenta}{\cite{8626195}} & \textcolor{magenta}{\cite{Andiappan2022,9685660,8626195,9328471}} \textcolor{cyan}{\cite{iot:downlinkNOMALongTermPower}} &  &  \textcolor{cyan}{\cite{iot:downlinkNOMAdelayConstrained,9839197} }   &     \\ 
  \cline{2-9}
  &  \multirow{1}{*}{Rate Splitting} &  &  & &  \textcolor{magenta}{ \cite{9851793,9837852}} & &   \textcolor{cyan}{ \cite{9850358} }  &     \\ 
    \cline{2-9}
  &  \multirow{1}{*}{Constellation Routing} &  &  &  &  & &  \textcolor{cyan}{\cite{Kato, Cigliano, Chao}}    &      \\ 
    \cline{2-9} 
    \cline{2-9}
   &  \multirow{1}{*}{IoT Channel Access Scheduling} & Fixed Access  & & & & & \textcolor{cyan}{\cite{iot:DynamicChannelAllocationRL,iot:CA-DRLenergyefficient}} &       \\
  \cline{3-9}
  & & Random Access & & & \textcolor{cyan}{\cite{iot:downlinkNOMALongTermPower}} & \textcolor{magenta}{\cite{iot:QL-NOMA-IoT-STRN,iot:QL-STRN,iot:QL-NOMA-IOT,iot:RA_ML_IoRT}} &  \textcolor{cyan}{\cite{iot:downlinkNOMAdelayConstrained,9839197}} \textcolor{cyan}{\cite{iot:DeepDyna-ReinforcementLearningCongestionControl}} &     \\
    \cline{2-9}
    &  \multicolumn{2}{|l|}{\textit{Upper Layers}} &  &  &  &  &   &   \\ 
    \cline{2-9}
    &   \multirow{1}{*}{Traffic/Congestion Prediction} &  &  & & & &      &     \\ 
    \cline{2-9}
  &   \multirow{1}{*}{Satellite-Terrestrial Integration} & & \textcolor{magenta}{\cite{lei_ai_dynamic}} & \textcolor{magenta}{\cite{lei_ai_dynamic}} & \textcolor{magenta}{\cite{bisio_network_2019}} & \textcolor{cyan}{\cite{rodrigues_network_2022}} & &   \\
  \cline{2-9}
  & \multirow{1}{*}{Edge Computing and Caching} &  &  &  & & & \textcolor{cyan}{\cite{Chao, Zhou, Cheng}}  &  \textcolor{cyan}{\cite{ref46}}\\
  \cline{2-9}
    &  \multirow{2}{*}{Network Security} & Devices Authentication  & \textcolor{magenta}{\cite{SSD:20:Sen, SMS:19:ICL, GY:20:Access}} & \textcolor{magenta}{\cite{WFK:18:EL, SNY:20:IJRFI, JRO:20:IoTM}} & \textcolor{magenta}{\cite{QSG:22:MPLB}} &  &    &    \\ 
  \cline{3-9}
  & &   Quantum Key Distribution &  & \textcolor{magenta}{\cite{CJZ:21:npjQI, RCL:21:IEEE_J_COML}} & \textcolor{magenta}{\cite{LDZ:19:PRApp, WL:19:PRA}} & &      &     \\ 
  \cline{2-9}
  & \multicolumn{2}{|l|}{\textit{Other}} &  &  &  &  &   &   \\
  \cline{2-9}
  &   \multirow{1}{*}{ISAC} &  &  & \textcolor{magenta}{\cite{JSC2}} & \textcolor{magenta}{\cite{RS1,9705087,RS37,JSC1}} & &  \textcolor{cyan}{\cite{JSC2}}    &    \\
 \cline{2-9}
  &  \multirow{1}{*}{Cooperative SATCOM} &  &  & & & &      &      \\
  \cline{2-9}
    &  \multirow{1}{*}{Img Processing} &  &  & & \textcolor{magenta}{\cite{9967537,9861276,imag26,imag27,DLLimageSat,imag28}} & &      &   \\ 
  \cline{2-9}
  &  \multirow{1}{*}{Ground Segment Dimensioning} &  &  & & & &      &    \\
  \cline{1-9}
\end{tabular}
\label{tab:Article_Categorization}
\end{table}
\end{landscape}


\subsection{Low Layers}

\subsubsection{Antenna Design and Fail Detection}
\paragraph{Motivation} The advance in antenna design simulation tools has been notoriously significant in the past 30 years. This advancement has helped the research antenna community to design antennas accurately without deep knowledge of computational electromagnetics. However, this advantage comes with the high price of requiring a long computational time to simulate antenna models that use more than one radiating element, as in the case of antenna arrays for satellite communications. Moreover, suppose the designed and manufactured antenna array has problems due to bad connections or broken RF chains; in that case, more time will be necessary to address such problems, which in satellite communications cannot be easily afforded. Therefore, machine learning can help to overcome these time-consuming procedures, reducing antenna design and troubleshooting time.
\paragraph{Description} 
Antenna design nowadays uses full-wave electromagnetic software that applies different electromagnetic computational approaches to obtain the antenna responses like S-parameters, far-field, near-field, E-field, H-field, and current distribution results. Some solvers are very handy for satellite communications and can offer optimal solution time depending on the antenna type. For instance, to model large reflector antennas like antenna reflectors, it is very convenient to use the asymptotic solvers Geometric Optics \ac{GO} and \ac{PO} \cite{azodi2014fast}. Now, if we consider small-size antennae used in CubeSats, for instance, helicoidal, dipole, or patch antennas, then  solvers based on the \ac{FEM} \cite{Zheng2005}, \ac{FDTD} \cite{tirkas1992finite}, or \ac{MoM} \cite{Johnson1999} are the most suitable ones in terms of computational time and resources. Moreover, for antenna arrays, some resources like the array factor can significantly reduce the computational time without considering mutual coupling, network distribution losses, radome effects, or spillover.
On the other hand, antenna fail detection is crucial for phased-array antenna applications. This is because one or more faulty elements can distort the radiation pattern by modifying the main beam gain, beamwidth, \ac{EIRP}, and scanning direction. Moreover, suppose a defective component is present; in that case, the array has to compensate for this by variating the amplitude of the phase of the rest of the element, which may take time using conventional optimization methods.
\paragraph{Conventional Solutions and Issues} 
Limitations of using a combined approach (full-wave simulation, asymptotic solvers, and array factor formula) for phase-array simulation and reflect-array design include the lack or complete absence of mutual coupling, feeding losses, axial ratio results, and cross-polarization. To overcome this limitation, a full-wave simulation that considers the entire structure will solve the problem. Still, at the same time, it will exponentially increase computational time and resources.

Considering the case of antenna failure detection, multiple approaches require the antenna near-field or far-field to obtain the surface currents and identify the faulty element \cite{alavi2019near}. Other methods use a genetic algorithm to detect the faulty element \cite{Beng1999}. The first method needs the antenna near-field or far-field measurement, which makes it impossible for antennas placed on an in-orbit satellite, whereas the second method takes a lot of computational and processing time to find the failure element.

\paragraph{Proposed ML Solutions}
In the antenna design field, some application of \ac{ML} has been used, especially for reflect-array antennas that can be extrapolated to transmit-array and phased-array antennas. 
In \cite{ayestaran2006near_NN} a \ac{NN} is used as a tool able to perform near-field to far-field transformation, directly or through the application of the Theorem of Equivalence. \ac{SVM} algorithm is used in \cite{prado2018support} in the design of reflect-array antennas for Direct Broadcast Satellite. Specifically, \ac{SVM} is adopted to create a surrogate model and substitute the full-wave simulation.
A very comprehensive and complete compilation has been presented in \cite{Misilmani2019}, where the author gathers some applications of \ac{ML} in antenna design to improve the acceleration design process, error minimization, antenna behavior prediction, and antenna optimization.

On the other hand, there are plenty of \ac{ML} algorithms for detecting faulty antennas. In \cite{vakula2010using}, the authors focus on faulty detection of phase arrays for satellite applications using \ac{ANN}, and compare its efficiency with \ac{RBF} and \ac{PNN}.
Similarly, \cite{chen2019VTC_DL_faultDetection} considers a \ac{DL}-based algorithm for fault detection.
An \ac{SVM} classifier is instead proposed in \cite{Nan2007} for planar array failure diagnosis.

While the above works show some preliminary effort on the use of \ac{ML} in the antenna design, we have not identified relevant works in compensation of faulty elements using \ac{ML}, which represents a gap to be filled by the research community in the next years.

\subsubsection{Antenna Beamforming}
\paragraph{Motivation}
Beamforming directs the transmitting antenna energy in a particular direction. Beamforming techniques have evolved over the years with increasing complexity. Being an excellent solution to a variety of challenging fields, \ac{AI} has already begun to be used by researchers as a possible solution to realistic beamforming problems in \ac{SATCOM}. Implementing \ac{AI} algorithms to beamforming allows recognizing underlying patterns, which would be inefficient to identify manually. This "black-box" characteristic of \ac{AI} methods, especially \ac{NN}, is an advantage since the performance of the deployed model does not depend on the antenna characteristics.
The main disadvantage of \ac{NN} is the lengthy training process to be conducted onboard or on-ground.
However, despite the significant time required for training an \ac{NN} in a multi-beam satellite scenario, the fast temporal response of \acp{NN} can be exploited for adaptive beamforming.
In fact, after training, \ac{NN} can perform cumbersome operations in real-time, representing a promising solution for adaptive beamforming, where weights must be repeatedly calculated and conventional solutions are lengthy.
  
\paragraph{Description}
An antenna array consists of multiple same-characteristic antennas separated by a specific distance, determined by the lattice, which can be square, triangular or irregular. The radiation pattern maxima, beamwidth, \ac{SLL}, and null pattern are  functions of the number of radiating elements, amplitude taper, lattice, and progressive phase shift \cite{balanis2015antenna}. Each antenna element, often called a unit cell, will determine mainly the scanning losses in the overall array, the complexity of the feeding network, associate losses, and production cost. Now, considering a fixed-size array employed for a given service in a certain region on the Earth, the beamforming characteristics control of the array are mainly a function of the complex-valued weight matrix, whose entries' magnitudes control the beamwidth in both planes and the \ac{SLL}; on the other hand, the entries' phases of the matrix control the antenna beam scanning, and nulls \cite{haupt_2010}. 

\begin{figure*}[tp]
\centering
\includegraphics[width =\textwidth]{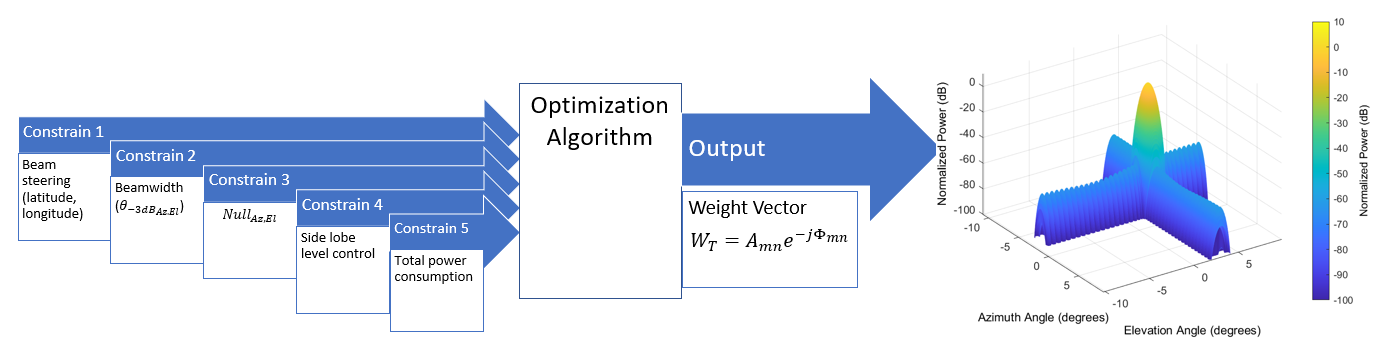}
\caption{Diagram with the inputs and outputs of a beamforming optimization.}\label{fig:bf_optimization}
\end{figure*}

\paragraph{Conventional Solutions and Issues}
The conventional solution applied to beamforming for satellite multi-beam application is an optimization problem, which consists of calculating the beamforming weights accurately to match the desired beam pattern \cite{j_vazquez2023_surrogate}, focusing the available transmit power on locations of interest while placing nulls at other directions, achieving a predetermined 3-dB main-beam width, and minimizing the \ac{SLL}. In the literature, there are multiple solutions to beam synthesize \cite{kai1996, kautz1999, Schippers2008, Sherman2000, chen2006synthesis, Boeringer2004}. Usually, a simple beamforming scenario, which is controlling the gain, can be easily simplified to a sum of the contributions of each antenna element with the required weight matrix as indicated in \ref{eq:antenna_1}
\begin{align}
\label{eq:antenna_1}
    \mathrm{AF}=&\sum_{m=1}^{M_x} \sum_{n=1}^{N_y} W_{mn} {\rm e}^{\mathrm{j}\left(m-1\right)\left(\kappa d_x \sin \left(\theta \right)\cos \left(\phi \right) \right)}\times \nonumber\\  
      &\times {\rm e}^{\mathrm{j}\left(n-1\right)\left(\kappa d_y \sin \left(\theta \right)\sin \left(\phi \right) \right)}\,,
\end{align}
where $M_x$ is the number of elements in the $x$-direction, $N_y$ is the number of elements in the $y$-direction, $\kappa $ is the wave number, $d_x$ is the period in the $x$-direction, $d_y$ is the period in the $y$-direction, $\theta$ and $\phi$ are the evaluating angles, and $W_{mn}$ is the weight matrix entry of index $(m,n)$. Now, if we consider a beam scanning angle ($\theta_0, \phi_0$) scenario, then the computation of the weight matrix can be done via 
\begin{align}
\label{eq:scanning_1}
    W_{mn}=  {\rm e}^{-{\rm j}\kappa (m d_x \sin\theta_0\cos\phi_0+n d_y sin\theta_0\sin\phi_0)}\,.
\end{align}
Moreover, the previous weight matrix can be modified by applying an amplitude tapper like Hamming \cite{webster1978}, Taylor \cite{Taylor1955}, Hann \cite{Blackman1958}, Chebyshev \cite{Dolph1946}, or others to reduce the \ac{SLL} and adjust the beamwidth. Complementarily, to avoid interference from another adjacent beam, it is convenient to add a nulling in a certain direction using, for example, \cite{haupt_2010}. Therefore, combining all the previously analyzed constraints to synthesize the beams in a multi-beam scenario will require proper optimization tool, computational time, computational resources, and power consumption, being limited in an in-orbit satellite. A diagram covering all the beamforming inputs and outputs is presented in Figure \ref{fig:bf_optimization}.

\paragraph{Proposed ML Solutions}
Literature offers several contributions to the implementation of AI techniques and especially deep learning \ac{DL}. Due to their fast response, rapid convergence rates, successful failure detection, and proactive decision capability, most adaptive beamforming \ac{ABF}) techniques are now based on \ac{DL} realizations.
In \cite{li2004performanceDigitalBeamforming}, the neural beamformer utilizes \ac{RNN} to search the optimal between three digital beamforming strategies. The work in \cite{mallioras2022_adaptiveBeamforming_DNN}, even if not explicitly targeting a \ac{SATCOM} scenario, proposes a feedforward \ac{NN} and \ac{RNN} to compute the complex feeding weights for a linear microstrip antenna array.
The work 
\cite{bianco2020RadarConf:AdaptivebeamformingDL} exploits \acp{CNN} with offline training to match a specific radiation pattern and computes the array antenna excitations. \ac{CNN} represents a faster alternative to conventional array synthesis techniques.
\cite{singh2020machine} utilize a linear regression model to steer to predict the directions of arrival of the incoming signal to steer the array beam on the ground in the satellite direction.
\cite{kumar2022dnn} implement a \ac{DNN} to perform beamforming in a \ac{LEO} system at sub-THz, where the atmospheric absorption is compensated by very large antenna arrays, that need to achieve very high gain in the desired direction via very narrow beamforming.
The DNN beamformer is trained to mimic the actual output voltage generated by a TTD beamformer as the angle of arrival (AoA) changes at the receiver due to the rapid movement of the LEO satellite.

\subsubsection{Flexible Payload-Beam Hopping}
\paragraph{Motivation}
The main idea of the \ac{HTS} technology breakthrough is generating multiple small beams and re-using the available satellite spectrum numerous times across the coverage area to increase the total satellite capacity significantly.
To cope with the critical adjacent-beam interference in \ac{HTS} systems, one excellent solution has been introduced is to divide the whole spectrum into various bands so-called ``colors'', which are separated by frequency allocation and polarization.
Recently, in the new \ac{SATCOM} generation of \ac{UHTS}, the current coverage area is foreseen to be divided into a higher number of smaller beams. Herein, the reduction in the beam pattern angle focuses one beam on a smaller spot and enhances a higher capacity gain even at a lower CAPEX cost. 
However, dividing the coverage into smaller spots comes at the expense of a high probability of no demand within some beam spots at a specific time \cite{Anzalchi_2010}.
In such cases, the corresponding no-demand beams can be turned off to reduce the operating cost. 
That is the key idea of beam hopping technology in the new generation of satellite systems \cite{Zhao2022_CostConf}.
\paragraph{Description}
\ac{BH} helps \ac{SATCOM} adapting to the time-varying communication demands which dynamically change due the time, mobility, and weather condition. 
Different from the quasi-static lighting in traditional multi-beam satellite systems, the satellite circularly illuminates a set of specific beams according to the generated schedule.
The \ac{BH} technology refers to the use of less numerous active beams in accordance with the design of the BH pattern to serve the ground terminals within the satellite coverage area, herein, the available satellite resources can be further optimally used to provide services with the heterogeneous demands \cite{Zhao2022_CostConf,Angeletti2006_ICSSConf}. The main difficulty in \ac{BH} pattern design is the large search space for identifying the optimal patterns. That is, in order to find the optimal solution, the number of possible \ac{BH} patterns to be searched increases exponentially with the number of beams \cite{Leilei2020_Access}.
In this context, \ac{ML} appears as a promising technique that offers an alternative to design optimization algorithms for complex resource management in wireless networks.

\paragraph{Conventional Solutions and Issues}
The SATCOM systems realize the BH technology the the flexible payload where the design parameters can flexible configured. 
The existing BH-related works aim to compose the beam switching and jointly optimizing beam pattern illumination mechanism with power/spectrum allocation and precoding/beamforming design.
The work in \cite{Lei2011letter} propose an iterative algorithm for \ac{BH} illumination design while \ac{BH} can also be optimized to minimize the transmitting power, as shown in \cite{alegre2012offered}.
Lin Chen et. al. in \cite{LinChen_TWC2022} have addressed a dynamic precoding-regarded beam-hopping design to minimize the cross interference and also avoid precoding whenever possible while it can satisfy all beam demands. The numerical simulations have illustrated the superior of the dynamic BH strategy to the traditional cluster-based BH mechanisms.
\cite{ha2022geo} aims to jointly determine \ac{LP} vectors, \ac{BH}, and discrete DVB-S2X transmission rates for the GEO satellite communication systems to minimize the payload power consumption and satisfy ground users' demands within a time window. Regarding the constraint on the maximum number of illuminated beams per time slot, the technical requirement is formulated as a sparse optimization problem in which the hardware-related beam illumination energy is modeled in a sparsity form of the LP vectors. To cope with this problem, an iterative compress-sensing-based algorithm is employed to transform the sparsity parts into the quadratic form of precoders. 
Authors in \cite{Tang_Access2021} design a dynamic baem hopping and beam position division for \ac{LEO} \ac{SATCOM} systems to shorten the packet queueing delay. In particular, this work aims to develop a novel mechanism to cover all users with the least number of beam positions in a set of beams which can be illuminated. The problem is first transferred into a $p$-center one, then the beam positions among the footprint of \ac{LEO} satellites are determined dynamically by the user distribution and the traffic distribution.  
The work in \cite{Lin2022_TCom} focuses on developing a novel cooperative beam-hopping mechanism for \ac{NGSO} constellation consisting of multiple satellites. Regarding the load-balancing issue, the proposed algorithm in this work is designed to manage both intra-satellite interference and inter-satellite interference by designing beam-hopping patterns with spatial isolation characteristics.

\paragraph{Recent ML-based BH solutions and Future Directions}
A number of recent works have also regarded employing the \ac{ML} tools to several \ac{BH}-enabled SATCOM schemes.
Lei Lei et. al. investigates in \cite{lei2020deepL_BeamHopping} a \ac{DL} method for \ac{BH} optimization and to predict the number of elements in the beam pattern for \ac{BH} and speed up the process of \ac{BH} pattern selection and allocation. 
The same authors in \cite{Leilei2020_Access} have explored a combined learning-and-optimization approach to provide an efficient, feasible, and near-optimal solution. 
The investigations are from the following aspects: 1) Integration of BH optimization and learning techniques; 2) Features to be learned in BH design; 3) How to address the feasibility issue incurred by machine learning. 
The numerical results have illustrated that the learning feature can enable a very high accuracy in predictions for the size of beam clusters that should be illuminated together to meet users' demands.
In \cite{Hu2020_TB}, the authors investigated the optimal fairness policy for beam hopping in DVB-S2X satellite regarding two main goals: minimizing the delay of real-time services transmission and maximizing the throughput of non-instant services transmission. 
To cope with the time-varying and unpredictable wireless channel issues, and differentiated service arrival rates in the multi-beam satellite environment, this work employed the model-free multi-objective deep reinforcement learning approach to learn and retrieve the optimal policy through interactions with the situation. 
To solve the problem with action dimensional disaster, a novel multi-action selection method based on Double-Loop Learning (DLL) is proposed. Moreover, the multi-dimensional state is reformulated and obtained by the deep neural network. 
Under realistic conditions achieving evaluation results demonstrate that the proposed method can pursue multiple objectives simultaneously, and it can also allocate resources intelligently adapting to the user requirements and channel conditions.
The DRL-based approach is also exploited in \cite{Lin2022_TVT} where a dynamic beam pattern and bandwidth allocation scheme based on DRL, which flexibly uses three degrees of freedom of time, space and frequency. Considering that the joint allocation of bandwidth and beam pattern will lead to an explosion of action space, a cooperative multi-agents deep reinforcement learning (MADRL) framework is presented in this paper, where each agent is only responsible for the illumination allocation or bandwidth allocation of one beam. The agents can learn to collaborate by sharing the same reward to achieve the common goal, which refers to maximize the throughput and minimize the delay fairness between cells.
Regarding future works on applying ML/DL in BH design for the next generation of SATCOM, it can consider developing the payload with high-flexible adjustment and control capabilities at different levels such as beam shape and network topology.
This can be done by employing more sophisticated processing capabilities of satellite digital devices, as well as ML/DL technologies. The work may be also enlarged by linking it some critical technical problem is multi-connectivity SATCOM schemes such as user scheduling, beam placement, and traffic prediction.

\subsubsection{Flexible Payload-Power Allocation}
\paragraph{Motivation}
The next generation of satellite communication systems will incorporate flexible digital payloads, allowing advanced resource management techniques to be implemented. In this case, the digital payload can be configured to change the bandwidth, carrier frequency, and transmit power of the system in response to heterogeneous traffic demands \cite{Miguel2021}. Notably, the payloads of the system are equipped with multiport amplifiers so that the power of the system can be divided across multiple beams. Hence, it is possible to use the remaining power in beams with lower traffic loads to those with higher traffic loads \cite{Elena2018}. However, these flexible payloads require advanced power allocation algorithms to optimize the total transmit power based on each beam's traffic load \cite{ortiz2019use}.
\paragraph{Description}
In satellite systems, power consumption is a major limitation due to its significant impact on satellite mass and operation lifetime. Additionally, the upcoming on-board beamforming technology, which is highly power-hungry, presents a major challenge for on-board power optimization. In order to reduce power consumption, it is necessary to implement innovative techniques, and one feasible approach is to employ advanced power optimization techniques to reduce radiated power. In addition, conventional satellites provide connectivity for users through a multibeam footprint with uniform power allocation per beam. Uniform power allocation, however, is inefficient when there is a distribution of non-uniform demand, which may cause beams with low demands to be assigned too much power while beams with high demands to be assigned too little. Consequently, we may be unable to meet some users' demands. Thus, it is necessary to implement an advanced power allocation algorithm in the next generation of satellite communication systems to ensure efficient power management \cite{ortiz2022supervised, ortiz2020supervised}.
\paragraph{Conventional Solutions and Issues}
Several studies have been conducted to develop power allocation algorithms using classical methods such as analytical optimization and heuristic/metaheuristic methods. With analytical optimization, the algorithm provides an optimal or nearly optimal solution. In this context, in \cite{Kang2018}, an energy-efficient power allocation scheme has been designed to maximize the average rate ratio over the total consumed power. In addition, energy-efficient power allocation over multibeam satellite downlinks with imperfect \ac{CSI} has been proposed in \cite{Qi2015}. Furthermore, an energy-aware power allocation algorithm has been developed to jointly minimize unmet system capacity (USC) and total radiated power in \cite{Efrem2020}. In \cite{Choi2005}, a power optimization approach based on demand and channel quality has been evaluated to minimize the difference between offered capacity and demand while maintaining reasonable fairness for all users. 

A metaheuristic method may not guarantee optimality, but it is well-suited to nonlinear, multi-objective, and NP-hard problems. In this context, \ac{PSO}is used in \cite{Fabio2017} to solve the power allocation problem in the multibeam satellite system, providing the minimum signal-to-noise plus interference ratio that the user terminal requires for reliable communication. Furthermore, a power allocation technique has been proposed in \cite{Aravanis2015} to simultaneously minimize unmet capacity and power consumption in multibeam satellite systems. In \cite{Liu2020}, a game–based dynamic power allocation (AG-DPA) algorithm has been developed for multibeam satellite systems to match the offered capacity and per-beam demand.   

The above-proposed method lacks flexibility in bandwidth allocation, limiting the ability to fully exploit flexible on-board payloads. This is addressed in \cite{Lei2010con,Abdu2021con1,Abdu2021con2,abdu2021jo,Ramirez2022jo} using analytical optimization and in \cite{Cocco2018,Paris2019} using a metaheuristic technique to jointly optimize the power and bandwidth to match the capacity offered with the per-beam demand. Nevertheless, using analytical optimization and metaheuristic techniques for power allocation involves many optimization parameters, so these optimizations are very time-consuming. Furthermore, the algorithm needs to be performed frequently if there are slight changes in traffic requirements or channel conditions. Thus, the proposed power allocation algorithms may not be suitable for real-time applications. In this case, machine learning techniques can be vital to provide a low-complexity power allocation algorithm that adapts to the change of demand and/or channel conditions. 
\paragraph{Proposed ML Solutions}
Several researchers have proposed machine learning-based solutions to overcome the limitations of conventional power optimization techniques. Using a neural network, \cite{Flor2019conf1,Flor2019conf2} proposes a classification algorithm that determines the power required for each beam to minimize the error between the offered capacity and the requested demand. Data is generated from a linear approximation function of spectral efficiency to train this neural network. Since neural network training is performed offline, the main advantage of this method is that it performs resource allocation at a low computational cost. However, the exponential relationship between the number of classes and the number of beams may increase the complexity of the computation and require further investigation. In \cite{Flor2021jo}, a \ac{CNN} network is used to minimize the error difference between the offered capacity and the per-beam demand while saving unused satellite resources. For this, training data is generated from a realistic traffic model. The proposed algorithm in \cite{Flor2021jo} shows better performance on the tradeoff of reducing capacity error and power consumption when compared with traditional resource allocation approaches. However, the algorithm depends on training data which is required to train the neural network if there is a change in the traffic model. 

A power allocation algorithm using \ac{DRL} to maximize the system transmit data has been investigated in \cite{Zhang2020}. In this method, the satellite acts as an agent, while the traffic between beams acts as an environment. Additionally, the DRL state contains a combination of buffered data among beams, power allocations, and beam geographical distributions. Accordingly, the agent adjusts the power among the beams to maximize data transmission. The simulation results in \cite{Zhang2020} demonstrate that the proposed method outperforms heuristic power allocation methods in terms of the system's throughput. However, the power of each beam is adjusted to some discrete values, which may limit the system's performance.

In contrast, DRL for continuous power allocation in flexible high throughput satellites to minimize the overall unmet system demand and power consumption has been proposed in \cite{Luis2019con1}. In this DRL-based power optimization method, the agent takes action (power allocation) given the state of the environment, which contains both the satellite model and demand per beam. For each action, a reward is assigned, modeled as a function of the difference between unmet system demand and power consumption. The proposed method is evaluated using time-series demand data provided by SES, and the authors demonstrate that the algorithm can match beam demand. Furthermore, the same DRL model as \cite{Luis2019con1} has been used in \cite{Luis2019con2} to compare with metaheuristic techniques. This comparison shows that DRL is the most appropriate option when time is critical and a solution is needed in seconds. 
\subsubsection{Flexible Payload-Bandwidth Allocation}
\paragraph{Motivation}
Future satellite networks are expected to offer extremely high data rates, enabling seamless integration with large satellite-terrestrial networks. However, due to spectral limitations, the main challenge will be providing high data rates while reducing the cost (satellite launch and operation) per bit. The key to solving this challenge is to maximize the system's spectral efficiency by utilizing the limited spectrum effectively \cite{Kisseleff2021letter}. Thanks to the emerging on-board digital payloads, the system not only offers flexibility in power optimization but also allows flexibility in bandwidth utilization. Typically, a flexible payload consists of a channelizer, which digitally routes channels within beams by digitizing the signal. It has a variable number of ports, each capable of processing a specific amount of bandwidth (channel bandwidth) while providing switching capability at the subchannel level \cite{Elena2018}. However, an optimal bandwidth allocation technique is required to configure these digital payloads. Consequently, we can manage the limited spectrum to enhance the system's spectral efficiency.
\paragraph{Description}
A satellite with multi-beam technology generates narrow beams, each covering a specific geographical area. As the beams represent a specific area, the same frequency can be reused on non-adjacent beams. Therefore, several beams can share the same frequency band, resulting in a significant increase in spectral efficiency. In traditional satellite systems, bandwidth is divided uniformly among the beams, with each beam sharing the total bandwidth equally. While some beams may have different demands, uniform bandwidth allocation may not be efficient in this case. As a result, bandwidth may be allocated excessively to beams with low demand and insufficiently to beams with high demand. Consequently, we may not satisfy some beam demands. Thus, advanced bandwidth allocation techniques must be developed while leveraging flexible payloads to allocate bandwidth per beam demand. 
\paragraph{Conventional Solutions and Issues}
Based on the characteristics of the beam demand, bandwidth allocation techniques can be categorized into three types:
\begin{enumerate}
    \item 	Orthogonal frequency band assignment across adjacent beams: One example is a four-color scheme in which the total bandwidth is divided into four frequency bands (subchannels). Accordingly, adjacent beams receive orthogonal frequency bands, while non-adjacent beams receive the same frequency band \cite{Destounis2011,Fenech2015,Aravanis2015}. However, this method is inefficient for heterogeneous demand, where some high-demand beams require more bandwidth while others need less. Furthermore, assigning orthogonal frequencies to some adjacent beams may not be necessary if their respective terminals are located near the center of the beams and their demands are low. 
    \item	Semi-orthogonal frequency band assignment across adjacent beams: In this case, adjacent beams may reuse the same frequency band. In \cite{Lei2010con,Lei2011letter}, subchannels (frequency bands) are iteratively assigned to each beam according to the traffic request while minimizing co-channel interference. Similarly, in \cite{Kawamoto2020}, subchannels are allocated to each beam to maximize system throughput. This method uses the same subchannels between adjacent beams within a range where inter-beam interference is minimal, while other ranges use different subchannels. Alternatively, analytical optimization in \cite{Park2012conf,Wang2013conf,abdu2021jo,Ramirez2022jo} and metaheuristics in \cite{Cocco2018,Paris2019,Nils2020conf} have been considered for bandwidth allocation based on the per-beam demand. However, it may be challenging to meet all beams that have a high demand using the techniques mentioned above. 
    \item	Full frequency reuse can be applied to satisfy high beam demand while managing interference among the beams using precoding techniques \cite{Bhavan2021letter,Ana2019jo}. 
\end{enumerate}
Most of the methods mentioned above use combinatorial optimization, resulting in longer computation times as the number of beams or subchannels increases. Furthermore, re-executing the algorithm may be required if traffic requirements or channel conditions change slightly. In this situation, developing an algorithm based on machine learning that can adapt to changing traffic and/or channel conditions could be useful. 

\paragraph{Proposed ML Solutions}
Several researchers have proposed machine learning-based bandwidth allocation techniques. In \cite{Liu218jo}, a dynamic channel allocation algorithm for multibeam satellite systems is presented based on deep reinforcement learning, where the total bandwidth is divided into subchannels and assigned to each beam in order to reduce the blocking probability of the system. A discrete-time event system based on service arrival events is assumed and the satellite checks which channel to assign to the new arrival user terminal. In this case, the satellite serves as the agent, the action determines which channels to allocate, and a positive reward is provided when the new service is satisfied.  On the other hand, a deep reinforcement learning-based approach for energy-efficient channel allocation in the satellite internet of things has been studied in \cite{iot:CA-DRLenergyefficient}. This method involves finding an optimal channel allocation strategy that allocates the limited channels to users on the ground while saving transmitting power in the long term. In \cite{iot:CA-DRLenergyefficient}, the satellite is viewed as an agent, and the user requirement and location represent the environment's state. Based on the environment's state, the agent allocates channels. In this method, the reward is considered the sum of the system's power consumption and service-blocking rate. Based on the numerical results in \cite{iot:CA-DRLenergyefficient}, the proposed method has better energy efficiency than traditional algorithms.

In \cite{He2022jo}, an algorithm for time-frequency resource allocation with deep reinforcement learning has been investigated to maximize the number of users and system throughput. In \cite{He2022jo}, the ground gateway is considered as an agent, as well as the number of time-frequency resource blocks and the user requirements as the environment. Consequently, the agent allocates the resource block while examining the state of the environment. Based on the simulation results presented in \cite{He2022jo}, the proposed deep reinforcement learning method gives better resource utilization with low computational complexity compared to the Genetic Algorithm (GA) and Ant Colony Optimization (ACO). Similarly, cooperative multi-agent deep reinforcement learning to allocate bandwidth with flexibility in time, space, and frequency has been investigated in \cite{Lin2022_TVT}.

Generally, several studies have investigated the use of machine learning applications to maximize system quality of service \cite{Zongwang2021letter}, spectral efficiency \cite{Danhao2022letter}, transmission efficiency \cite{Xianglai2020letter}, and resource utilization \cite{Chen2020letter,Xiaojin2020letter}. While \cite{Ferreira2018jo} applies multi-objective reinforcement learning to manage multiple systems' performance attributes such as throughput, bandwidth, spectral efficiency, bit error rate, power efficiency, and power consumption. On the other hand, in \cite{Abdu2022deepcon}, a demand-aware bandwidth and power allocation algorithm have been proposed based on analytical optimization and Deep Learning (DL). Thus, analytical optimization enables bandwidth and power allocation, while DL can speed up computation.

\subsubsection{Flexible Payload-Beamwidth Allocation}
\paragraph{Motivation}
User demand requirements are generally met in satellite industry through multiple spot beams of high-throughput satellites with fixed multiple beam pattern and footprint planning. 
However, mobile broadband users such as air (airplanes) and water (ships) located in remote geographic locations have dynamic traffic demand \cite{SnT_traffic_emul}.
Additionally, in next generation \ac{SATCOM} the traffic demand is expected to become less heterogeneous due to non-uniformly distributed user terminals with varying traffic demand. This cannot be faced by the current fixed footprint management and requires satellites equipped with beamforming capabilities and beam pattern plan \cite{FlorIrreguBeam2021,Puneeth2021}.

\paragraph{Description}
The main objective of beamwidth allocation is minimizing a proposed primary KPI (Key Performance Indicator), such as the gap between the offered and required capacity or the Coverage Error. The system characteristics, such as service area, satellite orbital position, bandwidth, and power per beam, must be defined. Together with the cost function, technical requirements or constraints must be specified to define the design parameters, such as the possible beamwidth values that can be assigned or the possible antenna orientations that can be provided \cite{FlorIrreguBeam2021}.

Once all the system conditions have been defined, the distribution of the average traffic demand in the service area must be studied and divided into $K$ regions. In addition, the beamwidth of $k-th$ region is assigned to meet the predicted traffic demand distribution. Generally, higher traffic demand require narrower beams, and larger beams should be assigned for lower traffic demand.

The coverage is thus designed for each possible configuration of the beam orientation, defined by the technological constraints. All \ac{SATCOM} link conditions, such as the resulting number of beams, the capacity offered on each beam, the provided capacity of the overall satellite, and the KPIs to calculate the cost function for time instant $t$, must be evaluated.

\paragraph{Conventional Solutions and Issues}
A first contribution for a more efficient beam pattern planning of a multibeam antenna with reflector and overlapping beams has been made by the authors of \cite{rao1994multiple}. To compare conventional contoured beam antennas with the multibeam antenna, the authors of \cite{rao2015advanced} propose a fixed beam for a specified number of beams to coverage using overlapping spot beams with a hexagonal grating arrangement. Additionally, the authors of \cite{schneider2011antennas,balanis2015antenna,jo2011satellite} analyze the antenna requirements for planning a good beam pattern for fixed beam footprint plans. Hence, all the first works consider only the geometrical feasibility of beams in the beam pattern and footprint plan guaranteeing global coverage. Consequently, the coverage region and the offered throughput remain the same for all multiple spot beams, regardless of traffic demand.

In \cite{FlorIrreguBeam2021}, a proposal is presented to optimize the coverage and pattern of irregular beams by minimizing the cost per Gbps in orbit, the Normalized Coverage Error, and the Offered Capacity Error per beam. The analysis and performance of the case study are presented, and a comparison is made with a previous algorithm for a uniform coverage area.

Recent work has focused on adaptive schemes to accommodate fluctuations in traffic demand. Global resource management for dynamic beam steering due to changes in \ac{QoS} requirements and user channel conditions is discussed by the authors of \cite{su2019broadband}. The authors of \cite{takahashi2019adaptive} optimize the beam directivity and transmit power based on the traffic demand to improve the overall system performance. Also, the authors of \cite{tani2016adaptive} adjust the satellite transmit antenna aiming point to maximize the satellite ground station's signal-to-noise power ratio (SNR) and minimize interference with terrestrial networks. An optimization tool is proposed by the authors of \cite{kisseleff2019new} to jointly optimize the number of beams, beamwidth, power, and bandwidth allocation to match the provided data rate to the expected traffic demand. Till they optimize the beam positions and beam shape. A mathematical model to derive multipoint beam arrangements is proposed by the authors of \cite{takahashi2020adaptive}, and they discuss the relationship between multipoint beam placement and user throughput.
Further, they only study the effect of geographical distances between spot beams in the same frequency band and geographical spaces between adjacent spot beams in different frequency bands on the overall system performance. However, they do not implement adaptivity in terms of flexible beam size and beam position in the beam plan. So far as we are aware, the authors have yet to consider the mobility aspects of non-uniformly distributed users and dynamically to change beam demand during beam pattern and footprint planning. 

In \cite{Puneeth2021}, advanced fixed multibeam patterns and footprint plans are studied to show their drawbacks to supporting non-uniformly distributed user terminals and varying traffic demands. The authors propose an adaptive multibeam pattern and footprint plan in which we design point beams with flexible sizes and positions based on the spatial clustering of users to increase the flexibility of the high-throughput satellite system. 

\paragraph{Proposed ML Solutions}
In \cite{Flor2021jo}, the use of a CNN to solve the resource management problem in SATCOMs, including beamwidth management using a suitable cost function and a realistic traffic model, is analyzed. The suggested cost function aims to minimize the error between the offered and required capacity while minimizing the number of resources used in the satellite and obtaining the adaptive beamwidth at each instant. One of the limitations of CNN in this work is the dependence on the traffic model used during training. Thus, in a real system with changes in traffic behavior not conforming to the model, the CNN will have to be retrained.

Authors also evolve the same cost function in \cite{Flor_RL2022}, analyzing the use of \ac{DRL} and \ac{RL} algorithm to manage the available resources in flexible payload architectures, including beamwidth. These algorithms are Q-Learning (QL), Deep Q-Learning (DQL), and Double Deep Q-Learning (DDQL), which are compared based on their performance, complexity, and added latency. On the other hand, this work demonstrates the superiority of a decentralized cooperative multi-agent (CMA) distribution over a single agent for multibeam scenarios.

However, implementing ML to enhance flexibility in geographical positioning has yet to be explored as another possible resource in addition to beamwidth. In that sense, it is expected that ML can enhance the gain in the cost function, a reduction in the time and complexity required because the search space is huge for conventional optimization techniques.

\subsubsection{Link Adaptation}
\paragraph{Motivation} 
\ac{ACM} techniques are among the most successful fade mitigation techniques for wireless links. The channel conditions in SATCOM links, such as rain, scintillation and delay, make it difficult for the estimation signal for feedback to choose the most suitable Modulation and Coding (MODCOD) at all times. ACM is easy for a single link without interference, where a pilot signal can measure the channel and report back to the receiver. However, the innate delay of satellite links changes the problem, especially in multi-beam satellites with many channels to estimate. 
There is an evident "dead time" between the fade event and its corresponding ACM reaction. This dead time may last between 600 and 800 ms and is caused by the two hops over the satellite and additional averaging delays. Frame errors might appear during this period if the SNR decreases. In addition, emerging SATCOM scenarios with NGSO and GEO increase the interference between systems, whereas ACM can improve communication performance by avoiding interference as far as possible. 
ACM employment includes as benefits greater distances free of errors, using smaller antennas which save on mast space, and higher availability to provide better link reliability. The usage of ACM techniques aims to improve SATCOM links' operational efficiency by increasing network capacity over the existing infrastructure – while reducing sensitivity to environmental interferences.
  
\paragraph{Description} 
ACM or Link Adaptation (LA) is a technique used in communications to match the modulation and coding schemes among other signal and protocol parameters according to the quality of the radio channel. LA dynamically and automatically changes the modulation and/or the \ac{FEC} code rate as the radio link conditions change. The receiver estimates the instantaneous SNR and signals, and the estimated \ac{CSI} is transmitted to the transmitting end through a feedback channel. The transmitting end keeps the transmit power unchanged and changes the coding and modulation mode according to the channel state. High-order MODCOD schemes are selected during good channel states, while lower MODCOD schemes are set when the channel states are bad. This means the system can transmit data at high speed under favourable channel conditions and reduce the throughput when the channel is degraded, thereby providing high spectral efficiency without sacrificing system power and \ac{BER}, maximizing the throughput obtained under momentary propagation conditions. In the uplink, the on-board ACM controller selects an appropriate modulation method and channel coding mode based on CSI sent to the satellite via the feedback channel.

\paragraph{Conventional Solutions and Issues}
Typical LA requires knowledge of the channel. This is easy for a single link without interference, where the pilot signal can measure the channel and report back to the receiver. However, the channel is stochastic, and the reported realization may differ from the actual one. In addition, the channel can substantially change during the so-called feedback delay (time between pilots being transmitted, feedback received and sent signal). The problem becomes more challenging when an interference channel is considered, or multiple channels are present, including Multiple-Input Multiple-Output (MIMO) systems. The authors in \cite{PEREZNEIRA2016141} and \cite{lagunas2016power} have considered the power adaptation for interference channels when the complete CSI is unavailable at each transmitter. Remarkably, in \cite{lagunas2016power}, different levels of coordination between transmitters are evaluated. From the point of view of emerging scenarios, an AMC scheme suitable for integrating the satellite in 5G networks is proposed in \cite{ACM5GSAT} based on 21 MODCOD schemes and different thresholds for SNR grouping.
ACM techniques for high-rate transmitters are developed on FPGA in \cite{ACMLEO} and proposed for LEO satellites. This proposal reaches a spectral efficiency of 7.068 bps/Hz with a code rate of 0.8889 in the highest rate mode.  
For the optical feeder link, time-packing is used as ACM considering low-order M-ary Pulse Amplitude Modulation (M-PAM) as shown in \cite{9221139} where time-packing enables a finer granularity on the LA capability of the optical link, enabling adjust its SE according to the moderate attenuation that thin cloud layers introduce.
\cite{7869090} uses precoding at the gateway side and SINR estimation at the terminal to address the issue of LA or ACM in a multi-user and multibeam satellite. This strategy includes a scheduler among different SNIR values and with imperfect CSI to maximize the similarity of two channels in the same frame.
The literature shows two main problems to be solved in LA with the new SATCOM paradigm. On the one hand, the wide margin that the Q/V and W frequency bands require for new mega-constellations and, on the other, the amount of information that must be estimated in multibeam satellites.

\paragraph{Proposed ML Solutions} 
An improvement of the CSI prediction in the satellite network is expected with the ML algorithms that will help realize ACM, predicting the most appropriate ACM to optimize system capacity for given channel conditions. This strategy will enable us to develop a more efficient ACM mechanism. 
The application of ML for LA has been considered in \cite{7924387} and \cite{9154263} using Bayesian methods focused on \ac{NN}, which shows performance gains over the commonly used ACM selection based on effective SINR. 
The work presented in \cite{8885616} focuses on applying ML to CSI prediction in SATCOM and using the improved prediction results to optimize system capacity. The different tested ML techniques demonstrate the improvement in system performance and the feasibility of deploying an ML-based prediction framework. The ML-based CSI prediction model provides an average capacity increase of up to 10.9$\%$ with acceptable overhead. 
The authors in \cite{9631216} propose a neural episodic control (NEC) algorithm from the DRL group as an ACM scheme for Inter-Satellite Communication Link. The proposed scheme adjusts the modulation and coding scheme region boundaries with a differentiable neural dictionary of the NEC agent, which enables the effective integration of the previous experience. The ACM proposed in \cite{9268889} utilizes an online random regression forest (ORRF) to predict time series of SNR values aiding the ACM switching decisions, introducing adaptive techniques for satellite links on Q/V-Band. The ORRF-based approach could outperform the classical approaches in terms of SE, and the parameterization of the ORRF is simpler and needs less knowledge of the channel properties than the discussed classical ACM approaches.
  


\subsubsection{Spectrum Sensing, and Interference Detection and Classification}
\paragraph{Motivation}
The usage of spectrum in current day \ac{SATCOM} follows a rigid and fixed spectrum allocation management paradigm. 
However, since the number of satellite and terrestrial services that require an ever-increasing bandwidth grows while the spectrum resource is fixed, the existing paradigm is inefficient.
A direct implication is that the available spectrum has to be shared and reused among those services. Besides interference, which is due to sharing the spectrum, there are intentional jamming sources that try to disrupt the \ac{QoS} of satellite communications systems. Both types of interference require detecting and classifying interference signals to take the appropriate measures. This will require a two-step process: 1) observing/collecting data, and 2) processing these data to decide on the current status of the spectrum.

In conventional approaches, the data-processing stage happens online and is designed by domain experts based on the specific knowledge of the signals' characteristics. On the other hand, in ML approaches, the algorithm learns the characteristics of the signals by training on a sample training data offline. Once the algorithm learns the characteristics well enough to differentiate between the different states of the spectrum, the resulting model is deployed online. 

\paragraph{Conventional Solutions and Issues}
To address the issue of spectrum sharing and the entailed interference management issue, many works in the literature focus on \ac{CR} SATCOM, where the coexisting systems are divided into primary and secondary, and the secondary system has to operate within limits specified by the \ac{QoS} of the primary system. \ac{CR} is divided into overlay CR and underlay \ac{CR} networks \cite{Zou2013}. In the former, the secondary system relies on detecting and using white spaces of the spectrum (parts of the spectrum that are not currently being used by the primary system), whereas in the latter, the secondary system relies on adapting the transmission parameters based on the maximum interference the primary system allows at its receivers to maintain the required \ac{QoS} \cite{Sharma2016}. These systems, however, require collaboration and sharing of information between the primary and secondary systems, which is usually done via a third-party (centralized) manager, which adds complexity, overhead, and cost. 
Another approach to manage interference while sharing the spectrum is interference detection and mitigation, where conventionally, a receiver analyzes the received signal based on some \emph{a priori} statistical knowledge of the desired and interference signals as well as the channels \cite{Kim2007}, and takes actions whenever interference is detected, like changing the transmission carrier frequency and subband. 
In \cite{Irfan2021}, the authors studied distributed \ac{DSS} for a system where \ac{GSO}, \ac{NGSO}, and terrestrial networks share the same spectrum in both downlink Ka-band (17.7-19.7)~GHz, and uplink Ka-band (27.5-29.5)~GHz, where terrestrial networks are the incumbent (primary) users that consist of 5G fixed wireless access and microwave links. The multiple access technique assumed for satellite-related links is Multi Frequency-Time Division Multiple Access (MF-TDMA), whereas for terrestrial networks, it is \ac{FDMA}. Two techniques were proposed: collaboration protocol, where nodes from different systems share information they capture from other nodes participating in the collaboration, and spectrum sensing, where each receiver makes decisions on spectrum allocation based on locally acquired energy maps of the spectrum by processing the raw I/Q received data. An advantage of the latter is that the receiver can detect known and unknown interference, while only interference sources shared through CP are detected in the former technique. Another advantage of decentralized spectrum sensing is that information exchange happens only between communicating entities for resource re-allocation or release, which reduces the overhead of sharing information. Although the spectrum sensing technique does not require exchanging information from entities of other systems, it requires changes in the radio modem to integrate spectrum sensing. Some satellite links, such as Global Navigation Satellite Services (GNSS) systems, require \textit{anti-jamming} capabilities. Conventionally, GNSS uses postprocessing techniques that rely on the statistical characterization of the signals to activate interference mitigation procedures since different countermeasures are appropriate to different jamming signals \cite{Qin2022}. These approaches can be effective when the jamming signals are known, and the number of jamming sources is limited. However, estimating the probability density functions for all hypotheses becomes more complex as the number of classes increases.
\paragraph{Proposed ML Solutions}
Data-driven approaches employing \ac{ML} algorithms are considered an alternative way of detecting interference \cite{Henarejos2019, Kulin2018}, where complexity is transferred offline in training the algorithm instead of the online complexity in processing the received signal in the conventional approaches. In \cite{Kulin2018}, the authors studied the problem of wireless signal identification that covers the problems of modulation recognition, and identifying the interference source, using machine learning algorithms. Three signal representations were considered: I/Q representation, amplitude/phase representation, and frequency domain representation. The authors found that amplitude/phase representation gives better accuracy for modulation recognition than the other two representations. In comparison, for the problem of interference detection, they found frequency domain representation to attain better accuracy. In \cite{Henarejos2019}, the authors considered a two-stage system based on the \ac{DL} model, wherein in the first stage, an autocorrelator is used to detect the interference signal. In the second stage, \ac{DNN} classifier is employed to classify the interference signal from three different well-known standards, namely: LTE, UMTS, and GSM, while the incumbent system is a DVB-S2 system. This means that one classifier can detect and classify the interference signals instead of different classifiers for different signals. Regarding \textit{Anti-Jamming} capabilities, in \cite{Qin2022}, the authors studied $k$-nearest neighbor (KNN) ML algorithm in detecting and classifying chirp jamming signals to GNSS systems, which is a challenging task using traditional approaches due to their variety, that requires post-processing of the signals and human-driven analysis, thus limiting their timely awareness and response jamming attacks. On the other hand, ML-based approaches can detect and classify the jamming signals automatically, which makes them better candidates in scenarios where timely awareness and response are required.


\subsubsection{Intersatellite Synchronization}

\paragraph{Motivation}
Inter-satellite synchronization is a new open challenge triggered by the boom of small satellites grouped in Distributed Satellite Systems. Currently, Distributed Satellite Systems are considered in multiple applications, from space communications to remote sensing. In the first case, the main advantages of distributed over monolithic satellite communication systems could be the increment of the system capacity and the reduction of the transmission power. Meanwhile, distributed Earth observation satellite missions can achieve higher spatial resolution requirements than single-platform architectures. However, the Distributed Satellite Systems should satisfy strict time and phase synchronization requirements to provide such performance enhancement. 

\paragraph{Description}
Synchronization algorithms are critical blocks in any communication system. They are key in the establishment of the communication link and have strong impact on the system performance. Conventional wireless communication systems implement synchronization functions in an iterative way, starting from the frame synchronization, then time synchronization and carrier synchronization. The first of these steps allows the receivers to calibrate several parameters such as accuracy of its \ac{LO} frequency, the desired timing phase and frequency of the digital samples, receiver’s gain, etc \cite{ling2017synchronization}. To this end the transmitters send pre-established signals that are known by the receivers. In addition, the time synchronization allows the receivers to sample each symbol at the perfect time while the carrier synchronization refers to the estimation and compensation of the carrier frequency
and phase offset introduced by the channel.

During the frame synchronization step, the signal detector searches for the signal energy peak by correlating a known transmitted symbol sequence with the received signal samples at different sampling times. These algorithms are known as data-aided synchronization methods. Thus, the signal peak found is used as the initial timing phase estimate for the fine-tune carrier phase and frequency synchronization. 

Timing synchronization is usually achieved by forming a feedback loop and thus it is often called \ac{TLL} or \ac{TCL}. Once the \ac{TLL} has converged, its output is synchronous to the remote transmitter timing, i.e., the remote transmitter’s channel symbol clock \cite{ling2017synchronization}. Carrier synchronization also rely on feedback loops that can be implemented in the analog or digital domain. These systems are based on phase or frequency offset detectors, loop filters and \ac{VCO} or \ac{NCO} \cite{gardner2005phaselock}. There is a huge number of data-aided and non-data-aided algorithms for frequency and phase offset detectors \cite{mengali1997synchronization}. Generally, non-data-aided methods are used when the frequency offset is comparable with the symbol rate. After the frequency offset is reduced, higher accuracy can be achieved using data-aided methods.

\paragraph{Conventional Solutions and Issues}
Conventional solutions for inter-satellite synchronization can be classified as Closed-loop or Open-loop methods based on the feedback from the receptor or target. The Closed-loop methods require a communication channel to transmit the feedback information between the external node and the Distributed Satellite System, which makes these methods more suitable for communication systems. Whereas in Open-loop, synchronization is achieved without the participation of any node other than the distributed satellites, as typical remote sensing applications require. In addition, we consider the communication between the elements of the Distributed Satellite Systems as an additional way of classification. It includes (I) synchronization algorithms based on the exchange of information among the distributed satellites, which can be done as a two-way message exchange or (II) as a broadcast, i.e., one-way communication. (III) Another option is to synchronize without any communication among the elements of the Distributed Satellite Systems by relying the control on a node out of the Distributed Satellite System. Both classifications can be superimposed or combined 

\paragraph{Proposed ML Solutions}
Even when there has not yet been any distributed satellite mission performing inter-satellite synchronization with \ac{ML} techniques, they have been used to address other problems related to synchronization in end-to-end-communication systems. For instance,  frame synchronization \cite{Wang2013learning, Wu2019DL}, sampling frequency and time offset compensation \cite{Wang2017}, carrier synchronization \cite{Lee2019}, and the characterization of phase noise \cite{Zibar2015, Tong2020} have shown promising results by using \ac{ML} techniques. 

A \ac{CNN}-based synchronization model with softmax activation function was proposed in \cite{Wu2019DL} to achieve better frame synchronization and compensate for the sampling timing errors caused by sampling time offsets. The \ac{ML} detection results showed an increment of 2~dB over the direct correlation detection in terms of correctly detecting the actual position of a frame header. Furthermore, the traditional frame synchronization techniques based on maximum likelihood and correlation present hardware implementation constraints related to frequency deviation problems which can be avoided using \ac{ML} techniques such as multi-instance learning to solve the frame synchronization problem under different frequency ranges \cite{Wang2013learning}.  

In addition, \ac{ML} techniques increase the accuracy of the amplitude and phase noise characterization of frequency references, which are essential elements for the synchronization of Distributed Satellite Systems. In this regard, a Bayesian filtering-based framework combined with expectation-maximization was used to characterize the noise of lasers in \cite{Zibar2015}. The carrier synchronization was experimentally demonstrated using the proposed framework in low \ac{SNR} scenarios.   

On the other hand, time synchronization attacks to cyber-physical systems, which mainly focus on modifying the measurements' sampling time or the time stamps, can be effectively detected by residual-based lousy data detection and conventional supervised \ac{ML}-based detectors. However, new \ac{ML}-based solutions have been proposed in \cite{Wang2017}. The "first difference aware" \ac{ML} classiﬁer presented in \cite{Wang2017} can detect two types of time synchronization attacks, namely, direct time synchronization, which only modifies some time stamps, and stealth time synchronization attack, which modiﬁes all the timestamps at a specific time. The First Difference \ac{ML} (FD\ac{ML}) technique utilizes the backward first difference of the time-series data to process the input data stream before employing a \ac{ML} method. 

In the context of integrated 5G and satellite communications utilizing \ac{OFDM}, authors in \cite{Tong2020} proposed a Cyclic-Prefix-based multi-symbol merging blind synchronization algorithm to enhance the timing accuracy. Besides, the authors proposed an improved synchronization method to accurately correct time-frequency errors in 5G SATCOM integration scenarios. Another example of \ac{ML} used in \ac{FH-FDMA} satellite communications is the challenging synchronization between \ac{DRT} and ground equipment using different hopping sequences for the uplink and downlink. In this regard, authors in \cite{Lee2019} proposed a novel \ac{ML}-based method to synchronize the Frequency Hopping signal by utilizing serial search for coarse acquisition and \ac{LSTM} network for fine acquisition. The main objective of the proposed method was to reduce the synchronization time. 

\subsubsection{Network Security-Device Authentication}
\paragraph{Motivation} 
\Acp{SCS} can be an attractive target for attackers due to their importance in global communication systems. 
Thus, a hacker can launch cyber-attacks at scale by targeting a few key components that constitute the backbone of the global \ac{SCS}. 
The risk of attacks on \ac{SCS} is further aggravated due to a wide threat surface caused by the ease of spoofing wireless devices and limited on-board  computing power, which prohibits the implementation of complex cryptographic techniques \cite{Pag:15:Web, WZY:20:Access, YAZ:22:arX}. 
One of the most common attacks in this context is spoofing where a hacker impersonates a legitimate device in the network to provide tampered or falsified information to the network users \cite{WZY:20:Access}.
These attacks put \acp{SCS} on a high risk of service outage, ransomware attacks, and even political and financial losses \cite{Ako:20:Web}. With the advent of distributed learning techniques, e.g., \ac{FL}, device authentication becomes vital to deter various attack vectors including model and data poisoning, membership inference, and reconstruction attacks by the unauthorized agents \cite{BLL:23:IF}.

\paragraph{Description}
The first line of defense against these spoofing and impersonation attacks is the device authentication procedure \cite{Cam:00:info, MK:01:WMC, Cru:96:ICSSMCN}. In traditional approaches, the device authentication and tracking may be carried out by the \ac{NCC}. Alternatively, individual devices may carry out additional mutual device authentication procedures to include another layer of security in case the \ac{NCC} is compromised \cite{CLC:09:CEE}. 

\paragraph{Conventional Solutions and Issues}
Traditional approaches for device authentication may be based on \ac{SKC}  \cite{Cru:96:ICSSMCN} or \ac{PKC} \cite{HYS:03:SOSR}. The server maintaining the secret-key tables in \ac{SKC}-based authentication schemes become an attractive target. On the other hand, high complexity of \ac{PKC}-based schemes is a major challenge for their use in low-powered devices. Some other issues in these approaches include high transmission delays and session dependence, i.e., if the key of a session is compromised the entire subsequent communication and authentication is compromised. Recent approaches for device authentication include blockchain \cite{DSC:21:Ele}, orbit-based authentication using time-difference of arrival \cite{JSS:21:ACM}, and hierarchical group key distribution \cite{LNP:22:IoT}. Some common demerits of these traditional approaches include model dependence, low reliability, and low cost efficiency due to an increasing number of network nodes.

\paragraph{Proposed ML Solutions}
In the recent years, intelligence-based approaches for authentication have emerged providing cost efficiency, high reliability, and a fairly high level of model independence \cite{FWT:19:WC}. Another major advantage of \ac{ML}-based approaches is the departure from session-based authentication to the \emph{continuous} protection throughout the duration of communication. 

In \cite{SSD:20:Sen} and \cite{SMS:19:ICL}, authors proposed a supervised \ac{ML} approach for detecting \ac{GNSS} spoofing signals. The authors used code, phase, doppler, and signal strength measurements in different frequency bands as data features sampled at 1Hz. The spoofing was performed as an intermediate timing attack aimed to affect the receiver's clock divergence by emulating a satellite clock drift. The specific algorithm utilized in this approach was \ac{SVM}, which resulted in a high accuracy of more than 98.5\% for different datasets. 

The authors of \cite{GY:20:Access} investigated similar spoofing attacks on \ac{GNSS}. They compared the performance of multiple \ac{ML} methods, namely, \ac{SVM} with radial basis function, $k$-nearest neighbors, adaptive boosting (AdaBoost), decision tree, and random forest. They found that the simple classification based on decision tree outperformed other methods in their considered scenario providing high accuracy with a very low false-alarm ratio. 

\ac{ML}-assisted antenna fingerprinting is another attractive approach for authenticating the wireless communication devices. This approach exploits the antenna manufacturing defects that are unique to each antenna and cause identifiable waveform imperfections \cite{WFK:18:EL, SNY:20:IJRFI, JRO:20:IoTM}. One major challenge in antenna fingerprinting is degradation of antenna due to different environmental factors that might alter the antenna fingerprint \cite{MER:21:IJAS}. The authors of \cite{QSG:22:MPLB} used a \ac{RNN} to capture temporal variations and provide robust antenna fingerprinting of different \ac{LEO} scenarios with a high accuracy of 99.34\% for 198 days despite antenna degradation. 

These \ac{ML}-assisted physical layer authentication methods complement the device authentication employed at higher layers. Highly reliable and model-free gate-keeping against spoofing and impersonation attacks at the physical layer greatly reduces the number of possible attacks that can be carried out at the higher layers. This makes the satellite systems more reliable and difficult to intrude by the malicious agents. 

\subsubsection{Network Security-Quantum Key}\label{QuantumKey}
\paragraph{Motivation}
The wide coverage and broadcasting nature of satellite signals make them susceptible to eavesdropping and information leakage to malicious parties \cite{JWW:15:IEEE_M_COM, LJ:20:ICONC}. Data security protocols aim to  eliminate the possibility of information leakage to eavesdroppers. Two main approaches to ensure data security are 1) \ac{PLS} and 2) data encryption in higher layers \cite{MSA:18:Nat,LFZ:20:IoT}. 

\paragraph{Description}
\ac{PLS} methods operate under the premise that the wiretap channel to the eavesdropper is inferior to that of the intended receiver. To this end, the randomness and time-varying nature of wireless channels are exploited \cite{HWZ:16:IEEE_J_WCOM, ZYL:16:IEEE_M_COM}.
New techniques are being developed that utilize multiuser \ac{MIMO} channels to ensure that eavesdropper ends up with a worse channel than the intended users due to interference \cite{SKS:19:MILCOM, SSK:21IEEE_J_IFS}. 
On the other hand, data encryption in higher layer relies on a secret key shared between transmitter and receiver to encrypt data. Data encryption in higher layers can be preferred over \ac{PLS} due to its modular nature and independence from operating/channel conditions. However, the biggest demerit of data encryption techniques is that they can be compromised if an eavesdropper has strong computational powers or a reasonably big quantum computer \cite{JWW:15:IEEE_M_COM, GE:21:Quantum}.

\paragraph{Conventional Solutions and Issues}
The advent of \ac{QKD} methods has eliminated the biggest demerit of key-based encryption by providing unconditionally secure keys that can ensure information-theoretic security \cite{BB:84:QC, QC:91:EK, HK:12:MD, SQ:05:RR, FK:14:MC, TF:12:MT, RW:22:arXiv}.

In the context of satellite communication, \ac{QKD} methods become more attractive due to a unique symbiotic relation between the two. That is, not only \ac{QKD} enables to achieve information-theoretic security of satellite communication but also \acp{SCS} have a vital role in achieving global-scale \ac{QKD} \cite{LCL:17:Nature, LCH:18:PRL}. 
To demonstrate the approach feasibility, the work in \cite{pirandola2021satellite} presents the configurable
finite-size secret key rates that protocols with continuous variable \ac{QKD} may achieve for both downlink and uplink.
\Ac{QKD} systems are very sensitive to the environmental factors, e.g., phase drift caused by internal as well as external factors, channel noise, and photon losses. Satellite channels can have high variability in these factors due to weather, satellite mobility, and imperfect internal devices. Accurate prediction and tracking of these factors is crucial to maintaining correct operation of \ac{QKD} systems. Traditional approaches for tracking the relevant parameters perform a scan before transmission, which amounts to a significant ``down-time'' for calibration of \ac{QKD} equipment \cite{WCY:18:OL, LDZ:19:PRApp}.

\paragraph{Proposed ML Solutions}
The authors of \cite{LDZ:19:PRApp} employed \ac{LSTM} network to predict relevant physical parameters in real-time and actively control \ac{QKD} devices to ensure a stable operation over the course of 10 days. The performance of resulting system was comparable to that of traditional ``scanning-and-transmitting'' schemes in terms of quantum-bit error rates and key transmission rates. Lack of down-time for scanning resulted in a high system efficiency over the course of continuous operation. The authors of \cite{WL:19:PRA} targeted the \ac{QKD} over free space on moving platforms including satellites. They employed a neural network to predict optimal parameters for operation of \ac{QKD} in variable environments. The resulting system required low computational power in its operation and was demonstrated on hardware devices including Raspberry Pi 3 and a mobile phone consuming less than 5 W. As compared to a brute-force search for the optimal parameters, the proposed approach provided up to two to four orders of magnitude speedup while providing a 95\% to 99\% of the optimal secure key rate for a given protocol. 

\ac{ML} was employed in \cite{RCL:21:IEEE_J_COML} for real-time selection of optimal \ac{QKD} protocol in a given environment characterized by dark count rate, efficiency of single photon detectors, misalignment error rate, and transmission distance. The authors employed random forest and compared it with other approaches e.g., support vector machine, K-nearest neighbors algorithm, and convolutional neural networks. Testing results showed an accuracy of over 98\% for selecting the most optimal \ac{QKD} protocol in a given scenario by random forests. 

Carrier recovery is another critical task in modern continuous-variable \ac{QKD} systems. The authors of \cite{CJZ:21:npjQI} employed an unscented Kalman filter for carrier recovery in noise over a 20 km fibre-optic link. The experimental results showed low variance and high stability in excess noise even at low pilot powers. 

These \ac{ML}-based operations and optimization of \ac{QKD} systems demonstrate a high potential of \ac{ML} in achieving a global-scale \ac{QKD} network. Considering the highly variable nature of satellite communication channels, it is clear that \ac{ML}-based solutions provide an efficient solution for system-parameter tracking and prediction, which in turn achieves optimal performance of \ac{QKD} that is an important component for realizing the goal of unconditionally secure global-scale communication over \acp{SCS}.

\subsubsection{Precoding}
\paragraph{Motivation}
Multibeam system design is based on the concept of frequency reuse, and currently, multibeam systems
typically exploit 4-color frequency reuse schemes to avoid aggressive levels of co-channel interference. In this sense, more aggressive frequency reuse schemes, such as full frequency reuse, can fully exploit the available bandwidth and achieve significantly higher performance levels. However, we need several solutions to increase system throughput when providing terabit connectivity over satellite systems. In this context, interference between adjacent beams strongly limits the overall system throughput. Consequently, it is of utmost importance to apply advanced interference mitigation techniques at the receiver, e.g., multiuser detection, or at the transmitter, i.e., precoding. 

Due to the popularity of multiuser multiple-input multiple-output (MU-MIMO) techniques in terrestrial
communications, together with the launch of the superframe structure in the DVB-S2X standard, the SATCOM community has started to evaluate and implement precoding techniques in multibeam SATCOM systems.

While the concept of precoding in satellite networks have widely studied in theoretical scientific papers, e.g. \cite{vazquez2016precoding}, an actual live-based demonstration supported by the European Space Agency (ESA) has been carried out in ESA project LiveSatPreDem (2020), validating the feasibility of such technique considering the recently amended DVB-S2X specifications to support it \cite{9771918,9373415}. Precoding is generally embedded at the gateway, thus keeping the complexity of the payload and user terminal (UT) infrastructure low.

\paragraph{Description}

The most popular precoding design in satellite communications is the low-complexity \ac{RZF} precoder, which is designed as follows:
\begin{equation}
	\textbf{W}_{\text{ZF}}= \eta \cdot \tilde{\textbf{H}}^H\left( \tilde{\textbf{H}}\tilde{\textbf{H}}^H + \alpha \textbf{I} \right)^{-1}, 
		\label{mmse_eq}
\end{equation}
where $\eta$ is a normalization factor ensuring that the output signal stays within the power limits. For a system-level power constraint, $\eta=\sqrt{P_{\text{tot}}/\text{Trace}\left\{\textbf{W}_{ZF}\textbf{W}^H_{ZF}\right\}}$ such that 
 the sum of the norm of the precoder vectors in $\textbf{W}_{ZF}$ is equal to $P_{\text{tot}}$, i.e.
 $\sum^{N}_{n=1}\left\|\textbf{w}_n\right\|^2=P_{\text{tot}}$. The regularization factor $\alpha$ is an arbitrary number, which is usually considered equal for all users and proportional to the inverse of the expected signal-to-noise ratio \cite{vazquez2016precoding}.
 
In \ac{RZF}, the precoding matrix is calculated once per frame period, and the computation of $\textbf{W}_{\text{ZF}}$ is mainly driven by the matrix inversion process.

\paragraph{Conventional Solutions and Issues}

The asymptotic complexity when assuming the Cholesky decomposition method is given by $\frac{1}{3}N^3$
flops \cite{hunger2005floating}, where $N$ is the number of beams. Even if the number of beams is high, the standard is limited to 32 unique Walsh-Hadamard (WH) sequences. Therefore, the receiver can discriminate the signals coming from the 31 nearest interfering beams, and the complexity is upper-bounded by $\frac{1}{3}32^3$. It is to be noted that such inverse operation needs to be re-calculated as soon as the scheduled users change (i.e. when the channel matrix changes). All in all, such computation is considered computationally intensive for conventional systems. Different approaches have been investigated to mitigate this challenge, e.g. precoding matrix nullification, which consists of nulling-out irrelevant coefficients.

\paragraph{Proposed ML Solutions}
While ML for multi-antenna transmission has been well 
explored in terrestrial systems \cite{9134393,8444648}, there are very few works published on the SATCOM domain. The European Space Agency (ESA) funded the 'MLSAT - Machine Learning and Artificial Intelligence for Satellite Communication' in 2020 \cite{MLSATTIA}, where the GEO satellite precoding matrix calculation was studied and evaluated using a two-step procedure involving autoencoders. The first step tries to learn the relationship between the real channel and the estimated channel matrix, i.e. understand the proper parametrizations to cover all the relevant information about the real channel matrix and the estimated channel matrix, respectively. The second step starts with the already inverted, estimated channel matrix from the simulation framework and predicts the alternative precoding matrix. Therefore, the second step does not need to learn the complex matrix inversion performed in the simulator framework but can focus on the remaining projection.
Once the individual steps are solved, the combined approach estimates the machine learning-based precoding matrix directly from the estimated channel state matrix of the terminals. According to ESA MLSAT project reports, the proposed AI-based approach showed a 3dB gain in mean SINR.

Some works like \cite{Abdu2022deepcon,VanPhuc_2022} have considered precoded GEO satellite systems and learning-based techniques but not directly to calculate the precoding matrix but to optimize the resource allocation assuming a fixed linear precoding matrix.

Boosting the precoding performance with proper user scheduling (i.e. clustering) is discussed in Section \ref{sec:sched} of this survey.

There is potential for ML techniques to improve the conventional precoding in SATCOM systems, possibly exploiting the advances made in the terrestrial domain, e.g. \cite{Lissy_2022}.

\subsubsection{Link Quality Prediction}
\paragraph{Motivation}

Satellite services are being deployed in higher frequency bands to provide higher capacities. Ka and Q/V show substantial excess attenuation in atmospheric events, especially rain. As a result, techniques for mitigating satellite link unavailabilities, such as automatic power control, gateway switching and adaptive modulation and coding schemes, have been studied in recent years. As the implementation of the mentioned fade mitigation approaches requires a non-negligible delay, link quality prediction techniques become mandatory. Indeed, for the efficient deployment of pre-emptive fade mitigation techniques, satellite systems shall accurately track the link quality.

While the terminal segment predictions require a time window similar to the propagation delay, the ground segment usually needs longer time windows for performing mission-level operations such as gateway switching. Bearing this in mind, while the terminal channel predictions can rely on former and current channel values, addressing longer time windows usually require external information such as weather predictions. Channel samples become uncorrelated with time lags similar to a few seconds.

Mobile satellite services vendors have addressed the terminal channel variations for many years as it was the main shortcoming of terminal capacity. Those systems generally rely on margins so that modulation, coding scheme, and transmitted power are selected considering a potential decay of the channel magnitude. On the contrary, ground segment systems, whose cost severely relies on high power amplifier and antenna size, focus on switching mechanisms when their fade mitigation techniques do not circumvent the channel degradation. We describe how this later operation is performed in the next paragraph.

\paragraph{Description}

As mentioned earlier, ground segment equipment is designed considering a certain channel decay tolerance and targetting a constraint cost of the teleport based on the target EIRP. To offer a 99.99\% ground location availability, teleports are located in areas with annual rain values below a certain threshold. The automatic power control can combat up to 10-12 dBs of fade. Satellite operators must rely on redundant gateways if extreme atmospheric events exceed the mentioned 12 dBs. This is traditionally coined as $N + P$ gateway diversity \cite{Fenech2014,6333101,1461558} where $N$ nominal gateways can be eventually replaced by any of the $P$ redundant ones. With this, overall mission outage is substantially reduced.

This ideal overall outage mission target cannot be attained in the system's regular operation due to the delays in performing the gateway switch. While the delay between the gateway switch decision has been made and the actual switch taking place requires a few minutes in nowadays network operation centres, after the switch, all terminals have to resynchronize again with the new ground station leading to a non-negligible additional delay. This operational delay is critical as outages with a temporal duration below this delay shall not involve a gateway switch.

Bearing this in mind, operators have to predict the gateway's link quality to reach the ideal capacity outage value conceived with the $N+P$ ground mission design. In particular, it is relevant to determine atmospheric events that yield strong fading values with a duration larger than the operational gateway switching delay.

\paragraph{Conventional Solutions and Issues}

The propagation community in different papers carefully addressed the prediction of channel values \cite{van2002short,de2008short,1095044}. Mathematical models behave well in short-term predictions and offer good results in temporal horizons below one minute. However,  the operational gateway switch requires a few minutes, so channel predictions shall consider a longer prediction window than those in the mentioned mathematical models.

The first attempt to consider several minutes of gateway prediction can be found in \cite{jeannin2019smart}. This seminal paper utilizes a Bayesian approach to consider a priori forecast weather data to attempt to predict the channel values in several minutes time windows. Prediction accuracy is linked to weather forecast accuracy as rainfall rate is the determining feature of strong satellite link fading events.


\paragraph{Proposed ML Solutions}

Due to the exponential interest in ML solutions, channel prediction and, in general, radio propagation modelling techniques have been revisited in \cite{RadProg2022-1}. The goal is to move from a model-based approach to a data-based approach which can eventually embrace high-complex scenarios while preserving a relatively low computational complexity.

This is the case of the recent works in \cite{9253575, Ventouras2019,cornejo2022method,9682090}. The goal of these works is to rely on deep learning time series prediction known approaches and apply them to channel prediction. As a general statement, the model boils down to
\begin{equation}\label{channel_prediction}
    g(\mathbf{h},\mathbf{w}; \mathbf{\theta}) = \hat{h^{\prime}},
\end{equation}
where $\hat{h^{\prime}}$ is the predicted channel value, $\mathbf{h}$ is a vector containing the staked previous channel values, $\mathbf{w}$ is the external weather features, and $\mathbf{\theta}$ is the deep neural parameters to be learnt in order to perform the prediction efficiently. Remarkably, the external weather information results to be a critical aspect in the works \cite{9253575, Ventouras2019,9682090}, which can be either a set of RADAR images or the current rainfall rate.

While \eqref{channel_prediction} shows a good performance in below few minutes time horizon prediction, addressing longer time windows could eventually require weather predictions. This is reported in \cite{deepgateway}, where external rainfall prediction techniques are utilized as a priori information for performing the channel quality prediction in long-term periods.

\subsubsection{Predistortion}
\paragraph{Motivation} 
Optimizing size and weight of the communication payload in a spacecraft is essential for reducing launch and operational costs, as well as for improving the utilization of the rather limited available space on board \cite{Shyu2000}. Conventional payload architectures usually rely on the separation among the antenna feeding segments, such as \ac{PA}, transmission line, and passive unit cell. This scheme has many drawbacks, including losses in each interconnection between segments, increased payload size, high cost, high risk of failure, and high noise figures. For all these reasons, active antennas, which centralize the active device, transmission lines, and  radiating element~\cite{gonzalez2002review} in a single device, are a promising solution for future \ac{HTS}.
\paragraph{Description} Current and future \ac{HTS} payloads need to simultaneously generate hundreds to thousands of beams over the coverage area, requiring a very large aggregated capacity (in the order of a few THz). In the context of \ac{DBF} with \ac{DRAA}, the only feasible solution for the amplification of the signals is the use of \ac{SSPA} which are inherently less efficient than the conventional \ac{TWTA} typically used in \ac{SATCOM}. 

Radiofrequency \ac{PA}s are essential building blocks in \ac{SATCOM} payloads. They are required to drive the transmitting antennas with an output power high enough to overcome all the losses and attenuation of the RF signals on their way to the receiver. In fact, the \ac{PA}s are among the most power-hungry components in communications systems. Therefore, \ac{PA}s should work with high efficiency, consuming as little energy as possible in addition to that to be delivered. This is even more critical in power-limited devices, such as battery-operated handhelds and satellite payloads. 
In this regard, \ac{RF} \ac{PA}s are most energy-efficient when operated at their highest output power, close to saturation. This gives rise to nonlinear effects upon the transmitted signals. The nonlinearities in the transmitter tend to degrade performance due to gain compression, clipping, phase distortion, intermodulation, adjacent channel interference, spectral re-growth, increased out-of-band emissions, etc, thus reducing the maximum throughput delivered by the payload. 
Particularly in digital communications, nonlinear distortion causes constellation warping and clustering, thus significantly complicating signal reception/detection. 

\paragraph{Conventional Solutions and Issues} 
Selecting the \ac{PA}s' operation point poses significant challenges for properly balancing efficiency and linearity~\cite{cripps2006rf}. 
Limiting the output signal's \ac{PEP} away from the \ac{PA}'s saturation level (i.e., to back-off the output) improves linearity at the cost of reducing the amplifier efficiency. 
This approach is particularly inconvenient when dealing with high \ac{PAPR} values, as those present in many current multicarrier-modulated signaling to achieve higher power spectrum efficiencies \cite{MCCUNE201555}. Choosing a conservative \ac{OBO} value to accommodate the \ac{PEP} in the intrinsic linear part of the amplifier's characteristic implies moving the average operating point further away into the low energy efficiency regime. In other words, the higher the back-off, the lower the \ac{PAE} of the \ac{PA}. This justifies the need for applying additional linearizing techniques to energy-efficient but highly nonlinear amplifiers.

The basic design principle behind pre-distortion is to measure the non-linearity at the PA output side and provide a complementary non-linearity at the PA input side via analog methods in the RF domain or digital in the base band domain. SATCOM systems with digital signal processing and active beamforming capabilities could employ \ac{DPD} techniques as part of the on-board signal processing. Applying \ac{DPD} will either increase the amplifier's useful output power or increase the PAE for the same linearity requirement. 
A moderate increase in processor power requirement could be traded-off against either increased \ac{PAE}, leading to a reduction of the mass of the thermal management system, or reduced beamforming and intermodulation interference at the same \ac{PAE}. 
These improvements have the potential to provide substantial power and mass reduction in different missions (e.g., MEO/GEO). Additionally, the use of \ac{DPD} techniques will allow the use of different types of waveforms, which are currently prohibitive for \ac{SATCOM}s applications, such as the \ac{OFDM} based 5G-NR standard.
\paragraph{Proposed ML Solutions}
\ac{ML} algorithms are attractive for \ac{DPD} domain due to their ability of modeling non linear behaviours. While strictly not related to \ac{SATCOM}, relevant works on \ac{ML} and \ac{DPD} has been published in literature in recent years. 
Differentiating based on the dataset form used as an input for \ac{DNN}, the work in \cite{kobal2022RNN_DPD_TMTT} proposes a novel neural network model based on a JANET architecture, \cite{zhang2019vectorDecomp_IEEEaccess} includes the phsyical features of the \ac{PA} in the input of an MLP neural network \ac{DPD}.
Paper \cite{chen2022inter_learningDPD} presents a \ac{DPD} scheme based on a dual-time delayed neural network (TDNN) learning architecture.
\cite{sun2022navigation_DPD_NN} proposes a \ac{NN} to compensates the nonlinearity given by the onboard satellite payload, considered as a whole and including \ac{PA} non linearity, and linear distortion caused by equipment such as filters and multiplexers.

While the application of \ac{ML} techniques in the field of \ac{DPD} techniques in the field of satellite communications is still limited, we expect a widespread rise in the next future.

\subsubsection{Coding}
\paragraph{Motivation} 
Coding schemes are a crucial element of modern digital communications systems. Convolutional Codes, Turbo codes, Low-Density Parity-Check (LDPC), Reed-Solomon or Polar codes are examples of \ac{FEC} algorithms used in different communications standards such as DVB or \ac{3GPP} (\ac{4G}, \ac{5G} and future \ac{6G}) for channel coding. \ac{FEC} decoding algorithms are typically the element of the receivers that require more complex computational capacity regarding resources. This computing problem is increased and aggravated in high data rate communications. Reducing the implementation complexity in communication receivers and the power consumption of FEC algorithms has become essential to evolving future communication systems to increase throughput. A demand for enormous resources in current communication receiver implementations. These resources can be the number of computations or processes to perform in each task, the computation time or the number of hardware components used in the implementation as logic gates. From the point of view of SATCOM, reducing power consumption and processing time is vital due to its repercussion on the payload size and, consequently, on the cost. Also, satellites must reduce latency and delay to integrate with terrestrial networks. Reducing the processing time favours the previous reduction or compensates for the least latency and delay innate to SATCOM's nature. All these challenges are the objective for future regenerative payloads, reducing complexity in HTS and, on the other hand, reducing the sizes of the payloads and improving the NGSO small satellites.

\paragraph{Description} Coding schemes are grouped into source and channel schemes. Source coding removes redundancy from the bitstream. Instead, channel coding adds redundancy to the bitstream; this redundancy or repeating of code is introduced to reduce SER. 

\paragraph{Conventional Solutions and Issues} 
Typical decoding relies on maximum a posteriori decoding, which consists in computing the probability that a specific bit was 0 or 1 and selecting the hypothesis with a higher probability. 
The two main drawbacks of a typical decoding approach are (i) the computational complexity and (ii) the unknown distribution of the channel noise. 
The classical Viterbi algorithm is used for decoding a bitstream that has been encoded using FEC based on a convolutional code. The Hamming distance is used as a metric for hard-decision Viterbi decoders. The squared Euclidean distance is used as a metric for soft decision decoders.
Regarding power consumption, several studies have evaluated the consumption of FPGA as hardware to support LDPC coding \cite{7360870,9059248,9136683,9294906}. In the DVB standard with a coding rate of (64800, 29160), the power consumption by LDPC is 12.92W for a throughput of $\sim$4 Gbps, which for a satellite with a maximum power consumption of 50W, the decoding process account for 25.84\% of the total of the consumption of the payload for a single reception chain \cite{7360870}. In the worst case, for small satellites like CubeSatKit, decoding at such a throughput rate is impossible since this task's consumption is larger than the power budget (2W) \cite{9059248}. In future HTS, where there could be up to thousands of parallel receiving chains on board the regenerative satellite, power consumption is increased with the number of beams up. For example, suppose an HTS satellite has a capacity of 200 beams, each being decoded by a FEC decoding chain with power consumption similar to SmallGEO. In that case, the total power consumption will be proportionally incremented by 200, $\sim$25.8\%. 
Therefore, to overcome these challenges, it is possible to consider a ML approach to learn from data.

\paragraph{Proposed ML Solutions}
The application of ML to FEC decoding is generally restricted to shortcodes due to the exponential training complexity. For instance, a message with “$k$” bits gives $2^k$ possible codewords. Fortunately, the ACM codes of DVB-S2X are generally limited, favouring such a scenario.
Two typical decoding approaches based on NN are data-driven and model-driven schemes, as shown in \cite{DL_decodingSurvey}. One of the earliest works on ML for FEC decoding is \cite{7852251}, which showed that the belief propagation decoding algorithm might be equipped with learnable multiplicative weights and trained as a NN to improve error correction performance. Recently, any academic works have been published on the topic with the objective of wireless communication receivers, e.g. \cite{phdthesis,lugosch2018learning,Haroon2013DecodingOE}.
Furthermore, \cite{9054192} presents a novel approach for decoding 5G data frames (a combined demapper and decoder) based on a combination of autoencoders and deep neural networks. The performance results of the proposed system compared with traditional implementations based on constellation demapping and LDPC decoders. The proposed approach can obtain a gain of 3 dB of SNR, which may be expanded by considering spatial-domain diversity through a MIMO approach. This scheme proposed in \cite{9054192} may be a candidate for satellite integration in the 5G network.

\subsection{Medium Layers}

\subsubsection{User Scheduling}
\label{sec:sched}
\paragraph{Motivation} Current multi-beam high throughput satellite systems use popular Orthogonal Multiple Access (OMA) schemes such as multi-frequency \ac{TDMA} (MF-TDMA) to exploit frequency flexibility. Furthermore, the traditional four-colour frequency reuse scheme is adapted to handle adjacent beam interference. The term "colour" in the four-colour frequency reuse scheme refers to the four different sets of non-overlapping channels assigned to each cell, represented by different colours. This scheme helps minimize interference between adjacent cells by using different sets of channels, allowing for efficient use of the available frequency spectrum.) However, when non-orthogonal users are scheduled together, even with the four-colour frequency reuse scheme, system performance can degrade due to user interference. This interference can arise due to the non-orthogonal use of frequency resources, which can lead to overlapping signals and reduce the quality of the received signals.
Nevertheless, the extent of performance degradation will depend on various factors, such as the number of users scheduled together, the modulation and coding schemes used, and the amount of interference in the system. Hence, the system performance is agnostic when non-orthogonal users are scheduled together. As a result, scheduling orthogonal users is crucial to reducing interference and maximizing the capacity offered.

In recent years, the full frequency reuse scheme has replaced the traditional four-colour frequency reuse scheme. Precoding has been used as an effective co-channel interference mitigation technique to enhance satellite spectral efficiency. In such an aggressive frequency reuse environment, orthogonal user scheduling is increasingly significant in precoding performance. There is a significant reduction in throughput when adjacent beams have collinear user channel vectors. Hence, a proper user scheduling strategy must be designed to select users with orthogonal channel vectors to be served simultaneously to achieve optimal precoding performance. 
Similarly to terrestrial systems \cite{1603708, 1003822}, multibeam precoding systems are sensitive to user scheduling. As a general statement, the attainable system rates increase when co-scheduled users have close-to-orthogonal channel vectors. In any case, user data flows are subject to end-to-end delays, frame encapsulation problems and user traffic prioritization, which makes this channel-vector-based scheduling challenging to be deployed in a real system.
\paragraph{Description}

A user scheduler establishes the relation between users, beams and scheduling time. In the unicast case, one user is scheduled per beam at a given scheduling time. As a result, the DVB-S2X \cite{dvb} defined XFECFRAME contains data that belongs to a single user. The scheduler's task is to schedule one user per beam for each scheduling period by considering factors such as orthogonality (better spectrum efficiency), fairness, and demand satisfaction.

As a result of the huge number of users to be served in a single beam in broadband satellite scenarios, frames are filled with data from more than one user. In this context, the resulting precoding operation becomes the so-called multigroup-multicast \cite{4443878}. This communication framework requires users embedded in the same frame to have similar channel vectors and $C/I$. This latter requirement comes from frame encapsulation restrictions that impose that all users served on the same frame will be assigned the same modulation and coding scheme.

In the multicast scenario, a group of users are scheduled per beam at a given scheduling time. As a result, multiple users share the same physical frame defined by DVB-S2X and are usually grouped based on similar SINR values. The challenge in multicast scheduling consists of two steps: the first is to group users using the same DVB XFECFRAME, and the second is to schedule these groups on time. The first step groups highly non-orthogonal users to ensure that all users sharing an XFECFRAME can decode the frame correctly. This is because the modulation and coding scheme will be selected based on the SINR of the weakest user. In the second step, the multicast scheduler schedules one group per beam at every scheduling time to make the scheduled groups' channel vectors as orthogonal as possible. Furthermore, the second step of the multicast scheduler should also consider fairness and demand satisfaction.

\paragraph{Conventional Solutions and Issues}

Scheduling users in multicast multibeam precoding is a coupled problem. Indeed, the ground station can only compute the precoding matrix once the scheduled users have been determined. To avoid iterative methods, the scheduler has to provide scheduling policies without knowing the resulting attainable rates. In this context, scheduling must involve heuristic approaches that yield to resulting scenarios where users within the same group have similar $C/I$ yet different channel vectors with co-scheduled users served at other beams to attain high data rates.

As the line-of-sight component strongly dominates satellite channels, one could resort to latitude and longitude positions to perform the scheduling. This is a substantial complexity reduction of the scheduler as it no longer requires access to reported CSI values. In \cite{8510728}, the authors propose a location-based scheduling algorithm that groups users with similar geographical locations and assigns them to the same beams. An alternative approach is suggested by the authors in \cite{6843054} who suggest using Euclidean distance to correlate channel vectors and impose channel orthogonality.

In satellite systems, cellular scheduling can be used to increase the number of users that can be supported and improve system performance. However, as the number of cells increases, the likelihood of co-channel interference also increases. From the authors of \cite{8978709}, it is also possible to avoid interference using graph theory in cellular networks to reduce intercell interferences (ICIs) using the Least Beam Collision (LBC) algorithm. This method avoids the simultaneous scheduling of two adjacent cells that might interfere with each other by refraining from scheduling them simultaneously. 

Despite geographical scheduling results into the same result as when using perfect CSI in clear sky conditions and perfect satellite feed element frequency precompensation, differences appear when a close-to-real operation is analyzed and fading is considered. In those cases, grouping users considering their Euclidean distance seems to behave well \cite{joroughi2016generalized}even when mobility is considered \cite{8353925}. Furthermore, the authors of \cite{joroughi2016generalized} propose a user grouping scheme based on random preprocessing and the previously discussed Euclidean norm.

A method to minimize the simultaneous scheduling of users in interference beams is also described in \cite {dimitrov2015radio}. This method uses partial channel state information (CSI) to avoid scheduling users in interference beams simultaneously. A study in \cite{Zhang2020Eurasip:UserSCheduling} investigates user scheduling for multicast transmissions with full frequency reuse and multicast precoding. Some authors have, however, enhanced their study by using the Euclidean norm and cosine similarity to examine channel characteristics. Also, in \cite{a6}, the authors sequentially select users with orthogonal channel vectors based on the cosine similarity metric.
Consequently, channel orthogonality and semi-orthogonal scheduling, as defined in \cite{1603708}, have been adopted widely by the satellite industry to address scheduling challenges. On a different approach, the authors of \cite{9740405} use user scheduling to achieve better demand satisfaction by considering both co-channel interference and user demands. Furthermore, considering the joint nature of precoding and user scheduling, the works in \cite{9443436,9014235,8870200} suggest an alternative solution involving suboptimal solutions that approach stationary points.

\paragraph{Proposed ML Solutions} In recent years, unsupervised machine learning techniques have been used to automatically group users into clusters with similar characteristics, which helps make the scheduling process more efficient. These clustering techniques can also take into account temporal variations in user traffic, which is particularly beneficial in the context of multibeam satellite systems. In particular, \cite{clust_Guidotti} focus on user clustering for multicast precoding in multibeam satellite systems. Two clustering algorithms have been proposed in this context: (a) fixed-size clustering, aimed at minimizing the impact of outlier users, and (b) variable-size clustering based on the K-means++ algorithm. Through numerical simulation, it was shown that the rate achieved with the proposed fixed-size clustering algorithm consistently outperforms the solutions available in the literature, and almost meets the upper bound performance. Also, variable-size clustering is more efficient than existing solutions. However, fixed-size clustering should be preferred when the number of users increases and the channel coefficient similarity increases. Nevertheless, the study does not consider the impact of non-uniform traffic requests on scheduling decisions.

Additionally, \cite{9815569} approaches the user scheduling problem in the context of clustering users as a function of the feature vector of the users, taking into account not only their geographical position but also channel information. Eventually, they evaluate three different algorithms based on UL: K-means (Km), Hierarchical clustering (Hc), and SelfOrganization (SO). However, further investigation is necessary to assess the wider implications of the research. This is because several issues remain to be addressed, such as finding the optimal number of clusters, studying other user characteristics for clustering, and evaluating its impact. In this sense, unsupervised learning techniques are ideal for exploring the structure of evaluated users.

\subsubsection{NOMA Multiple Access}
\paragraph{Motivation} The current satellite communication systems use popular \ac{OMA} schemes such as \ac{TDMA}, \ac{FDMA}, \ac{CDMA} and \ac{SDMA},  where orthogonal users share time, frequency, code and space, respectively. However, the performance of such OMA schemes is confined by the physical limitations of the resources. Conversely, \ac{NOMA} differs fundamentally from OMA in that all users share the same resources (time and frequency), but their power levels are distinct. Hence, the concept of NOMA has recently received considerable attention due to its ability to improve power-domain flexibility in managing resources and achieve higher spectral efficiency than \ac{OMA}. Furthermore, \ac{NOMA} has numerous advantages over conventional OMA, such as increased spectral efficiency, increased connectivity, reduced transmission latency and signalling costs. Despite such benefits, the resource allocation for satellite-\ac{NOMA} systems still needs to be fully explored, and many open questions exist.

\paragraph{Description} In power-domain \ac{NOMA}, by implementing superposition coding at the transmitter, different users can send their respective signals through the same time-frequency block without interfering with each other. The SIC receiver then uses successive interference cancellations to decode the different signals. 

On the transmission side, the channel conditions determine the power coefficients of users in an inversely proportional manner, i.e. transmission power for users with poor channel conditions is higher than for those with good channel conditions. 

In the receiver, since the user with the highest transmission power considers the signals of other users as noise, it immediately recovers its signal without requiring any SIC processing. Nevertheless, other users are required to perform the SIC process. As part of the SIC, each user's receiver detects the strongest signals before detecting their desired ones. This process continues until a user's signal is determined by subtracting those signals from the received signal. The final step involves each user decoding its signal by treating users with lower power coefficients as noise. 

Consequently, in NOMA, multiple users can simultaneously use a time-frequency block to improve \ac{SE} at the expense of additional complexity. This additional complexity, however, is still considered an acceptable trade-off to increase \ac{SE}.

\paragraph{Conventional Solutions and Issues}

Although \ac{NOMA} offers numerous features that could support next-generation systems, it also has some limitations that must be addressed to utilise its capabilities thoroughly. Compared to OMA, \ac{NOMA} has greater computational complexity because each user must first decode another user's signal before encoding their own signal. To decompose the overall \ac{NOMA} signal recovery into distributed low-complexity processes, in \cite{8423459}, a low-complexity receiver has been developed with iterative processing which consists of a linear \ac{MMSE} multi-user detector and a bank of single-user message-passing decoders. Other works have followed similar approaches, such as \cite{7510801} and \cite{8630078}. They examine an iterative linear MMSE multi-user detection for both symmetric and asymmetric MIMO-NOMA, and an IRA code is designed for the receiver for asymmetric and symmetric \ac{MIMO}-\ac{NOMA}. The authors of \cite{8957684} formulate a joint network stability and resource allocation optimisation problem to maximise the long-term network utility of \ac{NOMA} S-IoT downlink system. The authors consider the condition of SIC decoding and propose a practical solution under the Karush–Kuhn–Tucker (KKT) conditions, and then solve the joint resource allocation problem by introducing an optimal solution using the particle swarm optimisation (PSO) algorithm.

In NOMA, the base station should be notified of the channel gains of all users, but doing so will require a significant amount of channel state information (CSI) feedback overhead. To reduce it, \cite{8972353} propose a scheme where the CSI is acquired only if the BS estimates a high probability of successful UE pairing. \cite{9321366,NOMAimperfectCSI} study the performance of NOMA with imperfect channel state information (CSI).

\paragraph{Proposed ML Solutions} 

The authors of \cite{Andiappan2022} study the \ac{DL} aided NOMA for resource allocation, power optimization, Channel Assignment and User‑Pairing, Data rate Maximization, Signal Detection and Channel State Estimation. Using a \ac{DRL} algorithm, the authors of \cite{iot:downlinkNOMAdelayConstrained} have developed a dynamic power allocation strategy for NOMA-based S-IoT systems subject to QoS constraints to optimize power allocation factors for NOMA users. According to the authors of \cite{9685660}, a hybrid approach combining data-driven learning and model-based optimization can minimize transmission time in a NOMA-enabled satellite system by optimizing transmit power and terminal time slot scheduling. In \cite{9839197}, the authors minimize the average \ac{AoI} of a NOMA downlink by formulating a \ac{MDP} problem. Furthermore, to overcome the curse of dimensionality and non-convexity, they propose a \ac{DRL}-assisted age-optimal power allocation algorithm.

In \cite{iot:downlinkNOMALongTermPower}, \ac{DL} was applied to satellite-IoT systems to determine the optimal decoding order for NOMA's SIC process based on an approximation of the implicit mapping between queue size, channel gain, and decoding order. Also, to perform completion time minimization, \cite{8626195} employ Deep Belief Network (DBN) to solve the joint downlink resource allocation non-convex problem using end-to-end learning. Additionally, \cite{9328471} proposes the DPO combining learning and optimization to provide a feasible and near-optimal solution. Furthermore, to address the problems introduced by imperfect \ac{CSI} and channel estimation errors, the \ac{SVM} model in \cite{9053003} showed promising performance by significantly improving user pairing.

\subsubsection{Rate Splitting}
\paragraph{Motivation} Rate-splitting multiple access (RSMA) is an effective multiple access technique that includes SDMA and \ac{NOMA} as special cases. One extreme case of RSMA is the linearly precoded SDMA scheme which relies on a transmit-side interference cancellation strategy which treats interference as noise. Another extreme case of RSMA is power domain NOMA which uses a receive-side interference cancellation strategy and fully decodes interference \cite{Mao2018}.

Furthermore, RSMA can benefit from both SDMA and NOMA, and outperforms both in terms of better spectral efficiency, energy efficiency and QoS enhancements with lower computational complexity \cite{8491100}. The performance of NOMA is more effective with both underloaded and overloaded network loads, along with various channel strengths, perfect/imperfect CSI and channel directions. Thus, RSMA can be described as a computationally simpler bridge between SDMA and NOMA. Additionally, RSMA offers the advantage of controlling computational complexity and data rate by adjusting the ratio for splitting user signals into common and private portions.

\paragraph{Description} As part of RSMA, linearly precoded rate splitting (RS) occurs at the transmitter, and \ac{SIC} occurs at the receiver. The result is that part of the interference can be decoded, and the remaining interference can be treated as noise. Hence, RSMA can partially decode interference and partially treat interference as noise. While SDMA and NOMA rely entirely on either extreme, fully decoding interference and fully treating interference as noise, RSMA acts as a combination.

Specifically, the RSMA scheme divides all users' messages into two independent streams: common and private. Then, the common signals are jointly encoded to form an independent data stream. It should also be noted that private signal interference is considered noise when encoding the common part. As a result, the common stream represents the part of the desired signal expected to be used by intended and non-intended users. On the other hand, the private signals are encoded according to linear precoding schemes in which each user encodes his own signal.

In order to decode the common stream, each receiver first decodes the common stream and treats the private streams as noise. Accordingly, the common stream is decoded and removed by \ac{SIC} scheme. Later, each user decodes their private stream by treating all interference streams as noise. Thus in RSMA, a soft transition is realized between treating interference as noise and fully decoding interference. This is done by carefully adjusting the message split and the power allocation between the common and private messages. Nevertheless, it is worth noting that although the RSMA has the potential to fundamentally alter the physical and \ac{MAC} layers of wireless communication networks, it still faces numerous challenges.

\paragraph{Conventional Solutions and Issues} A comprehensive tutorial of RSMA can be found in \cite{PrimerRS} that discuss the myths associated with RSMA, along with answers to frequently asked questions about RSMA. Also, \cite{9475472} discusses open issues of RSMA in Cognitive Radio Networks (CRN), in particular, how to handle mutual interference (functions of the power allocation, rate splitting parameters or the power allocation factors). Thus, both RSMA and CRN parameters (spectrum sharing) must be solved jointly, which leads to a non-convex problem. Furthermore, open issues such as spectrum sensing, physical layer security issues, along with the impact of RSMA on the physical layer and lower MAC layers have been discussed. The authors of \cite{PrimerRS} discuss the Physical Layer Security issues and state that if encryption is used in higher layers, there will be no privacy or security issues in RSMA. 

More specifically to satellite systems, the authors of \cite{yin2020rate} discuss the Max-Min fairness (MMF) achieved by conventional linear precoding and RSMA in multigroup multicast with imperfect CSI and demonstrate the benefits of RSMA strategies in both underloaded and overloaded scenarios. Also, by combining a modified WMMSE approach with an alternating optimization algorithm, the authors in \cite{9145200} solve the MMF optimization problem and propose an RSMA approach that is highly promising for multibeam satellite communication systems to handle inter-beam interference, taking into account practical challenges such as CSI uncertainty and practical per-feed constraints. A similar approch is followed in \cite{9684855}, but for Multigateway Multibeam Satellite Systems. Security Analysis was investigated in \cite{9833359}, and a beamforming scheme based on RSMA is proposed to suppress eavesdropping.
 
\paragraph{Proposed ML Solutions} With limited knowledge of channel information and uncertainty of the communication channel, optimizing power allocation in RSMA is very challenging. In the conventional approach, the sum rate maximization of RSMA is generally achieved by the WMMSE algorithm. However, due to high computational complexity, the authors of \cite{9851793, 9837852} use a hybrid \ac{ML} technique based on deep learning called \ac{DU} with momentum accelerated \ac{PGD} algorithm and outperforms the original WMMSE algorithm in sum rate and speed. In \cite{9850358}, the power allocation problem to common and private streams is discussed. The paper proposes a highly-effective proximal policy optimization based solution, enabling the LEO satellite to learn an optimal power allocation strategy to maximize the sum rate with low computation complexity.

\subsubsection{Constellation Routing}
\paragraph{Motivation}
Satellites are usually grouped in a constellation to offer global or near-global coverage and service. Several constellations are already operating, such as O3b, Galileo, Global Positioning System (GPS) at the MEO orbit or Starlink and Iridium at the LEO orbit. Within each constellation, satellites can be connected with the neighbor satellites via inter-satellite links (ISLs) either inter-plane, i.e., the link between two satellites in a different orbit, or intra-plane, i.e., the link between two satellites in the same orbital plane. GEO system has a wide coverage and typically does not require complex routing over the satellite component. On the contrary, MEO and, especially, LEO satellites have a smaller coverage. This brings to a constellation with several satellites and many ISLs between them. As a consequence, efficient routing algorithms are of vital importance to offer global services. 
In addition, beyond 5G networks will provide reliable and ubiquitous connection while accommodating heterogeneous services, e.g. emergency services, mobile broadcasting $\&$ broadband, sensor-based networks, earth observation missions \cite{Vanelli, Yuki}.
To achieve this goal, beyond 5G satellite networks are expected to have a multi-layer (LEO-MEO-GEO) structure, that introduces several challenges in the intra-layer, making the constellation routing even more complex to handle.
In the following sections, we will describe the above mentioned challenges, together with the proposed AI/ML techniques to overcome them.

\paragraph{Description}
Routing defines a path on the substrate network, i.e., a sequence of nodes, to allow the traffic generated by a source node to reach the destination node(s). The routing algorithm is an optimization decision process based on several parameters, such as path length, resources of intermediate nodes, link available capacity and delay. The output of a routing algorithm is then translated into commands for each intermediate node. These commands are entries of the forwarding table of each device. Thanks to the forwarding table, the device knows how to forward the arrived traffic to the next hop, as decided by the routing algorithm. The routing algorithm is computed with a fixed transport network once the intermediate nodes receive the commands and traffic flows. However, when the substrate network has a highly dynamic structure, e.g. the NGSO satellite constellations, especially for low altitude orbit ones, and due to the strict power/computational constraints on-board the network nodes, i.e., satellites, the routing has to face numerous challenges. Primarily, with the increase of the nodes and links in the network, the number of available paths and their dynamic behaviour over time increases considerably. This brings more complex routing algorithms to be computed. Secondly, the heterogeneity of on-board capabilities between the GEO-MEO-LEO satellites and the difference in the link latencies additionally increases the complexity of the algorithms since the number of constraints grows. In the following subsection, we present conventional algorithms to address satellite routing and the issues highlighted for the next-generation integrated satellite-terrestrial networks.   
\paragraph{Conventional Solutions and Issues}
First, a valuable routing algorithm should guarantee each application's required QoS, e.g. latency, packet loss, and data rate. To handle this together with seamless handovers, due to the mobility of the substrate network, a temporal subdivision into time intervals is applied such that the substrate network is considered static within each time interval. Then, traditionally, the most common approach to solve the routing is the shortest-path (SP) \cite{Werner}. 
While SP was efficient with the first generation of MEO and \ac{LEO} constellations, several issues were raised with the advent of more complex and more extensive networks. Alternative conventional solutions such as the minimum hop count \cite{Stock} has also been proposed for Mega-Constellations. However, due to the high mobility of the \ac{SAGIN} nodes, i.e., satellites, flying devices and mobile users, and the high volumes of data which may easily congest the channel \cite{Razmi}, providing seamless transitions remains challenging. 
Additionally, as highlighted in \cite{Chao}, the routing problem along the \ac{SAGIN} not only deals with the link capacity constraints but also needs to consider the different caching and computing capabilities of satellites, which increases the overall computational complexity. 
\paragraph{Proposed ML Solutions}
Authors in \cite{Kato} describe a \ac{CNN} based routing algorithm to optimize the \ac{SAGIN}'s performance via traffic patterns and the remaining buffer size of GEO and MEO satellites. They propose online training, distributed over multiple satellites, due to a high volume of data, and does not interrupt the throughput while training.\\
Authors in \cite{Chao} propose a \ac{DQL} approach to overcome the computational complexity of a joint optimization routing algorithm. The DQL algorithm is described with 1) the state space, function of user-LEO angle, networking, caching and computing states, 2) the action space and 3) the reward function, defined as the efficiency of unit resource.
In \cite{Cigliano}, authors propose a DRL algorithm where the rational agent of the model learns following the replays and selects the route with the smallest \ac{RTT}. The routes chosen, although in principle crossing
more nodes, and therefore longer, in the end, have better effectiveness and a lower load on the bottleneck router.

\subsubsection{IoT Channel Access to Scheduling}
\paragraph{Motivation} 
IoT via satellite is envisioned as a viable solution to connect devices in remote places due to the ubiquitous nature of satellite coverage, and thus dramatically complements terrestrial IoT solutions. However, due to the massive number of terminals, using optimal resource access and allocation policies would result in a computational load incompatible with the processing constraint of IoT devices. On the other hand, using simple access techniques such as random access lead to under-utilization of the network resources. Hence, machine learning is an interesting solution to offer a trade-off between channel utilization and computation complexity. 
\paragraph{Description} Due to the specific nature of IoT networks, access and resource allocation protocols for IoT satellite networks must satisfy several criteria regarding efficiency, delay, complexity, and scalability. A detailed list is available in \cite[Sec.~6.3]{iot:SurveyRAforM2M}. 
The authors of \cite{iot:SurveyOnIoTTechnologies} present a survey on current technologies and standards in the field of satellite IoT technologies, as well as the challenges inherent to this type of communication.
As presented in \cite[pp.~92]{kodheli2020Survey_IEEE}, access methods can be grouped in two categories: \emph{fixed assignment} (FA) based, and \emph{random access} (RA) based. The former ensures contention-free access but requires precise synchronization, while the latter has little synchronization constraints but has to deal with packet collision probabilities. In the following paragraphs, we will separate FA and RA-based solutions with different constraints, challenges, and network and IoT device designs.

\paragraph{Conventional Solutions and Issues} 

Two leading well-known FA solutions are the \ac{NTN} narrow band \ac{IoT} (NB-IoT) \cite{iot:NB-IoT-NTN} and \ac{5G} \ac{NR} standards. Both are part of the \ac{5G} cellular access technology developed by the 3GPP. The fact that those solutions are part of a cellular standard is at the same a strength and a weakness. On the one hand, it allows seamless integration of \ac{NTN} into other cellular-based networks. On the other hand, those standards impose restrictions that do not necessarily mix well with the constraints found in NTN. One issue is the higher \ac{RTT} delay, especially in GEO links. In LEO systems, other common issues are differential doppler and high demand variability due to short orbital periods. These issues can conflict with the precise synchronization needs of \ac{OFDMA}, and lead to more complex and expensive terminal design (e.g. by including GNSS receivers). Most of the existing literature focuses on finding solutions that solve those issues with as few modifications as possible to the existing standards.

RA solutions are well suited for satellite \ac{M2M} communications as they satisfy most of the criteria inherent to this type of communication. The authors of \cite{iot:SurveyRAforM2M} provide a survey on RA solutions for satellite uplinks, focusing on \ac{M2M} communications. Originally based on the ALOHA protocol, RA methods were first used in satellite communications for capacity requests \cite{iot:RAforDAMA}. Successive improvements, such as in \cite{iot:AdvancesinRandomAccessProtocols}, have allowed new RA methods to achieve low ($<10^{-3}$) Packet Loss Probability (PLR) even at high ($\geq 1$) network loads, enabling applications in massive machine-type communications (mMTC). However, the very high number of terminals, and thus access requests, coupled with the constraints of NTN networks (e.g. high RTT), brings down the performances of those solutions in satellite IoT networks.

\paragraph{Proposed ML Solutions}
As for the conventional solutions, we will present the FA \ac{ML} solutions first.
The authors of \cite{iot:DynamicChannelAllocationRL} presented how classical DAMA with RA requests could be improved with dynamic channel allocation and \ac{DRL}, retaining the high efficiency of FA while adding the flexibility of RA. The optimization is done on the satellite side, which receives transmission requests from the terminals and schedules the data transmission up to a given number of time slots ahead. The bandwidth, the transmission completion time, and the number of scheduled transmissions for each time slot characterize the states. The reward function increases when the sum of transmission completion time decreases. The authors also present several techniques to speed up training sessions. The results showed that CA-DRL reduces by up to 87\% the delay in request satisfaction compared to first-come-first-served and pure ALOHA policies. However, no comparison is offered against more advanced techniques like CRDSA. 
This work was expanded by the same authors in \cite{iot:CA-DRLenergyefficient}, of which a description can also be found in \cite[Sec.~VI.B.a]{iot:surveyNTN5Gto6G} and \cite[Sec.~V.B.3]{iot:surveyAImodels6G}. The authors added multibeam, QoS satisfaction, and energy efficiency, extending the terminals' lifetime in remote IoT networks. The authors formulate a \ac{MDP} for allocating tasks, defined by their size and location, to channels and beams. They present how to use DRL to solve this problem. The action space is defined as all the possible channel allocations. To reduce its size, the allocation of a channel to a beam is done sequentially. The reward is a trade-off between two components: the first represents power management.
In contrast, the second represents QoS satisfaction as a service blocking rate inside the time window. The authors also construct the users' requests into an image as the input of the considered NN, which can reduce the input size and accelerate the learning process. Results show a reduction in energy consumption of up to 67\% compared to random and greedy algorithms and minimize the service blocking rate amongst all algorithms.
In both the above works, the agent is placed on the satellite, and the algorithm is supposed to be run onboard.

Regarding RA \ac{ML} solutions, the authors of \cite{iot:DeepDyna-ReinforcementLearningCongestionControl} have presented a Deep Dyna-Q Learning solution for the random access control in IoT LEO networks (slotted ALOHA). 
IoT devices connect via random access opportunities, of which the product of preambles and physical random access channels available determines the number. The authors present an optimization problem to maximize the utilization of idle RAOs while keeping the transmission delay under a given threshold. An Enhanced access barring (EAB) mechanism is used in case of congestion. The states in the \ac{MDP} are defined as the number of devices connecting to each satellite. The action set is the number of available RAO (discrete set). The reward takes values in the set \{1,0,-1\}, depending on if the utilization at the next step is respectively higher, equal, or lower than the current one. Probabilities of transition between states are expressed analytically based on the Poisson process for both contention-based and contention-free random access. The authors then present a solution to the optimization problem in the form of a Deep Dyna-Q Learning algorithm and a model-free Deep Q-learning algorithm for comparison. Both Q-learning algorithms present a 3.5-fold improvement in access efficiency over classical dynamic RAO allocation. The proposed Deep Dyna-Q Learning solution performs similarly to the model-free Deep Q-learning. It does not necessitate the feedback interactions between satellites and IoT devices needed in model-free solutions. The authors consider the agent to be placed on the satellite and consider onboard training without, however, offering an overview of the demanded computational power to run the proposed \ac{RL} algorithm.
To improve the efficiency of satellite communication systems, the NOMA technique has been applied with \ac{ML} in several works.  In \cite{9839197}, the authors minimize the average \ac{AoI} of a NOMA downlink by formulating a \ac{MDP} problem. Furthermore, to overcome the curse of dimensionality and non-convexity, they propose a \ac{DRL}-assisted age-optimal power allocation algorithm.
The authors of \cite{iot:QL-NOMA-IoT-STRN} have applied a model-free Q-learning algorithm in the context of IoT satellite terrestrial relays network (STRN) with \ac{NOMA}. This work is a combination of \cite{iot:QL-STRN} and \cite{iot:QL-NOMA-IOT}. In this scenario, the IoT devices communicate with a nearby terrestrial relay, which then transmits the aggregated data to satellites. The terrestrial relays are assumed to perform successive interference cancellation (SIC) based on the power imbalance between received signals. Each IoT device is a learning agent. The states are defined as the timeslots, while the actions are defined as the channel's choice. The reward comes from a single feedback bit transmitted by the relay to the device and is either one if the transmission is successful or minus one if it is not. The Q-learning algorithm replaces the random time slot and channel selection by learning based on the relay's feedback. Results show up to an 18\% improvement in spectral efficiency over SA-NOMA. In this context, Yan et al. \cite{iot:downlinkNOMAdelayConstrained} studied the power allocation problem for NOMA-enabled SIoT using a single-agent DQN-based DRL scheme. In their system, the agent is the satellite, whose action space is discrete, corresponding to selecting the power allocation coefficient for each NOMA user. The state space is continuous, consisting of NOMA users' average SNR, link budget, and delay-QoS requirements. At the same time, the reward is discrete, which is a function of the effective capacity of each NOMA user. Experimental results demonstrate that their proposed DRL-based power allocation scheme can produce optimal/near-optimal actions, providing superior performance to the fixed power allocation strategies and OMA scheme.
Sun et al. utilize the DL technique to optimize the \ac{SIC} decoding order for NOMA downlink system in satellite IoT networks \cite{iot:downlinkNOMALongTermPower}. This research proposes a long-term joint power allocation and rate control scheme to improve the NOMA downlink rate. Then, the Lyapunov optimization framework is adopted to convert the original problem to a series of online optimization sub-problems. The power allocation depends on the SIC decoding order, which is further affected by the queue and channel states. Due to the continuous changes, the DNN model is adopted to map from the states of queues and channels to the SIC decoding order. Moreover, the DNN is trained supervised with the data obtained by traversing all possible choices.
Literature \cite{iot:RA_ML_IoRT} presents a resource allocation for IoRT with joint time-variant channel fading process and stochastically fluctuating solar infeed process. The solution is based on SARSA based actor-critic \ac{RL} and is evaluated in average amount of downloaded IoRT data by the satellite. The authors conclude that performances depend on battery capacity.

\subsection{Upper Layers}

\subsubsection{Traffic/Congestion Prediction}
\paragraph{Motivation}
Nowadays, most existing satellite network systems operate independently. Integrating different satellite constellations is considered a promising feature of next-generation satellite networks. The connection inter different satellite constellations could improve the performance of the satellite network and create the potential for new uses \cite{de2020analysis}. Multilayer satellite networks consisting of multiple satellite constellations with different orbital altitudes have been studied as significant space network, which carries traffic to any point on the Earth \cite{Miguel2021}.

Multilayer satellite networks formed by combined networks have several advantages, e.g. the wider coverage area served by the upper layers and the comparatively shorter propagation delay provided by the lower layers \cite{SIN_2021}. From the standpoint of fair and efficient utilization of network resources, the load of satellites can be balanced by broad coverage areas of satellites with higher altitudes. Overall, due to population imbalance based on different causes such as terrain shape and weather, traffic convergence to certain satellites usually occurs in constellations of mesh satellites with lower orbits. Therefore, intelligent routing control technology and smart resource management are necessary for multilayer satellite networks to coordinate the traffic load according to QoS requirements effectively.

Monitoring changes in the incoming traffic rate is the most general way to detect network congestion that occurs when rate measurements exceed a predefined or dynamically adjusted threshold. Balancing the load is a timely procedure that, however, might not be enough to prevent network congestion effectively. 
To overcome this challenge it is necessary to develop and employ congestion prediction mechanisms for load balancing by distributing the traffic amongst available network resources. 

\paragraph{Description}
Network traffic prediction plays an important role in many areas of satellite networking. Result outputs of congestion predictors could be used for network management, network design, short and long-term resource allocation, traffic (re)-routing, and anomaly detection. Two categories of prediction methods, based on long and short-term's periods, are typically considered. The former, long-term traffic prediction, estimates future capacity requirements, enabling more effective planning decisions. The latter, predictions within minutes, even seconds, is usually linked to dynamic resource allocation. It can be used to improve \ac{QoS} mechanisms, congestion control, and optimal resource management.

\paragraph{Conventional Solutions and Issues}
The authors of \cite{nishiyama2011load} study the hybrid GEO/LEO satellite network and address the load balancing problem with QoS provision using the advantage of interlayer interconnection. They base the effectiveness of the traffic distribution scheme based on network congestion prediction using computer simulations. 

In the proposed scheme in \cite{nishiyama2011load}, two different states, i.e., normal and warning states, are defined to present the state of each LEO satellite. The normal and sign conditions imply that the corresponding LEO satellite has a high and low probability of congestion. In the normal state where no congestion event is detected, an LEO satellite attempts to catch a congestion precursor by simply monitoring the incoming traffic rate and not exhibiting any load balancing behavior. On the other hand, in the alert state where there is a threat of the cause of congestion, not only traffic monitoring is performed, and traffic distribution and information exchange for congestion prediction as it exists in congested or non-congested areas.

\paragraph{Proposed ML Solutions}
The work presented in \cite{wu2019link} proposed a link congestion prediction method using machine learning, which can operate at the upper layer of the controller. The authors propose a software-defined network system model consisting of a data layer, a control layer, and an application layer. The state information of switches and links is collected to train and test the proposed machine learning algorithm to predict the related congestion. The prediction performance of four machine learning algorithms was compared; in this case, CNN was shown to obtain the highest pressure, reaching 98.3$\%$.

\ac{NN} models are also used to predict network traffic, such as \ac{CNN}, \ac{LSTM}, and Full-Connected \ac{DNN}. These techniques mentioned above did not explicitly consider the topological information of the network. Therefore, the prediction performance on network traffic was low. Recently,  Diffusion Convolution Recurrent Neural Network (DCRNN) was proposed to capture critical topological properties of the network, which are expected to influence the patterns followed by significantly the traffic when propagating through the network \cite{li2018dcrnn_traffic}.

Authors of \cite{phenarejos2022} introduce an hybrid architecture composed by stacking \ac{CNN} and \ac{LSTM} which produce predictions of the traffic for a particular beam in a two-hours window with a resolution of $5$ minutes. In this work, the traffic data statistics provided by a european satellite operator within several months made possible to train extensively the implemented \ac{NN} to achieve accurate predictions. Despite it is not possible to detect outliers and peaks in traffic prediction, the results unveil predictions at trending level, proposing a mechanism to indicate the level of future congestion.

\subsubsection{Integrated Satellite-Terrestrial Networks}
\paragraph{Motivation}
Satellite systems were developed when reliable and secure communications were needed.
They have evolved today to cover a wide range of services, from broadband to critical communication services \cite{Hayder2021}.
Each service has its specificity in performance, architecture, or technology, which puts flexibility as a requirement for today's satellite networks.
The same constraint is also applicable in the terrestrial networks, and therefore 5G networks and beyond put flexibility and versatility at their core definition.
Since 3GPP Release 15 \cite{3gpp.23.501}, as the satellite and terrestrial networks share common objectives, satellite networks are now part of 5G and are referenced as \acp{NTN} within the standard.
Terrestrial networks expect to benefit from the wide coverage area of satellites, which does not require deploying any terrestrial infrastructure to extend their coverage.
This integration should bring more flexibility and efficiency in network deployment and management.
However, this integration is challenging as both networks have specificities and must converge technologically.
To achieve this objective, the network slicing paradigm can be used \cite{3gpp.28.530}.
A network operator can create end-to-end partitions and deploy services while guaranteeing their end-to-end \acp{KPI}.

\begin{table*}[!ht]
    \centering
    \caption{Satellite-Terrestrial Integration Modes \cite{tiomela_jou_integrated_2019}}
    \begin{tabular}{c|p{10cm}}
    \toprule
        \bfseries Integration type & \bfseries Challenges \\
        \rowcolor{Seashell1} 3GPP Access & Support of the NR waveform within the satellite \\
        \rowcolor{Seashell1} Trusted non-3GPP Access & Integration of the satellite in trusted non-3GPP Access mode within the standard \\
        \rowcolor{Seashell1} Untrusted non-3GPP Access & Support for 3GPP untrusted access mechanisms within the satellite network \\
        \midrule
        \rowcolor{Seashell2} Relay node with 3GPP Access & Support of the NR waveform within the satellite, adaptation of the relay mechanism within the satellite terminal \\
        \rowcolor{Seashell2} Relay node with Trusted non-3GPP Access & Integration of the satellite in trusted non-3GPP Access mode within the standard, adaptation of the relay mechanism within the satellite terminal \\
        \rowcolor{Seashell2} Relay node with Untrusted non-3GPP Access & Support of 3GPP untrusted access mechanisms within the satellite network, adaptation of relay mechanisms within the satellite terminal \\
        \midrule
        \rowcolor{Seashell3} Transport Network based on 3GPP System specification & Definition of a satellite system compatible with the 5G standard \\
        \rowcolor{Seashell3} Transport Network non-based on 3GPP System specification & Define an adaptation layer for the satellite network considered as a transport network \\
    \bottomrule
    \end{tabular}
    \label{table:integration-mode-5g-sat}
\end{table*}

\paragraph{Description}

A multitude of research projects has focused on the integration of satellite networks into terrestrial mobile networks.
We can cite the VITAL project, which aimed to apply the concepts of \acp{SDN} and \acp{NFV} to the satellite network \cite{vital_consortium_h2020_2015} in order to increase its flexibility and allow close integration with 4G networks.
It was followed by the SaT5G project \cite{jou_architecture_2018}, which studied the integration of satellites in 5G networks. 
The table \ref{table:integration-mode-5g-sat} summarizes the different modes of satellite integration into 5G networks: \colorbox{Seashell1}{direct access modes}, \colorbox{Seashell2}{indirect access modes as a relay node} (using gateways to connect other, smaller \acfp{UE}), and integration as a \colorbox{Seashell3}{\acp{TN}} for backhaul link services.
Following these projects, Zhu et al. \cite{zhu_integrated_2022} listed all challenges regarding the 6G-satellite integration.
Depending on the integration scheme, adaptions have to be achieved at various layers of the systems.
Interconnection and integration should take place at the orchestration plane, control plane and data plane.
This also implies comprehensively allocating and efficiently managing terrestrial and satellite components' resources.
A complete integration scheme is direct access where \acp{ST} embeds all the 5G protocol stack and connects directly to satellites also compliant with 5G \ac{RAN}.
This method is also the most complex, as satellite networks need to be fully 3GPP compliant.
Then, depending on the trusted mode, the relay node can be an intermediate solution where the \ac{ST} relays (with or without embedding the 5G protocols) all data from \acp{UE} to the \ac{RAN}.
The last and easiest mean to realize such integration is to consider the satellite network as a transport network for the 5G system and create backhaul links between 5G \acp{RAN} and \acp{CN}.
Considerable research was conducted using conventional or \ac{ML} solutions.

\paragraph{Conventional Solutions and Issues}

Ahmed et al. \cite{ahmed_-demand_2018} have worked on the integration of satellite and 5G using \ac{SDN}/\ac{NFV} and proposed a method for hybrid resource reservation across terrestrial and satellite \acp{PoP} using \ac{MILP} combined with an online algorithm to ensure a minimum end-to-end delay.
Drif et al. \cite{drif2021extensible} have analyzed the network slicing paradigm and applied it to satellite networks to enhance its flexibility and integrate it as a network slicing within 5G networks.
They have listed all the challenges related to slicing in satellites, identified interconnection points between terrestrial and satellite networks and produced a novel architecture to mutualize all satellite resources into a single physical infrastructure.
They have implemented their work on a 5G and satellite network in \cite{drif_santn_lcn_2021} and demonstrate the end-to-end service and \ac{QoS} continuity using network slicing.
Kodheli et al. have listed the challenges of using 5G direct access in \ac{LEO} mega-constellation and proposed high-level solutions in \cite{kodheli_globecom_2017} regarding the Random Access procedure.
The authors pursued their research in \cite{kodheli_random_2021}, where they have detailed all the current limitations of using mobile network protocols in \ac{GEO} \acp{NTN}.
They have proposed solutions to push these boundaries and validated them on a physical testbed using a channel emulator and a 4G modified stack of the \ac{OAI} software.

\paragraph{Proposed ML Solutions}

Lei al. \cite{lei_ai_dynamic} have presented AI-based solutions to manage network slices deployed over the integrated 5G-satellite networks.
They have listed major issues of applying \ac{AI} in such networks and proposed solutions to two distinct dynamic scenarios: predictable network variation using \ac{TEG} and unpredictable network variation (e.g., sudden change in mega-constellation topologies), which requires fast adaption mechanisms due to model drift. 
Solutions can be conventional learning (retraining a model), Transfer learning, Meta-learning, or \ac{RL}.
Rodrigues et al. \cite{rodrigues_network_2022} proposed combining network slicing with machine learning to optimize links in combined 6G-satellite networks. 
Their ranking-based framework shows an increased user acceptance ratio compared to the traditional method.
Similarly, Bisio et al. in \cite{bisio_network_2019} have developed a solution mixing queuing theory and \acf{NN} to minimize this time latencies in end-to-end 5G-satellite communications.

\subsubsection{Edge Computing and Caching}\label{Edge}
\paragraph{Motivation}
Edge computing \cite{edge} proved to be the key technology to overcome challenges such as massive and heterogeneous data transmission coming from the \ac{5G} and beyond \ac{5G} applications which have undoubtedly made the cloud computing paradigm unable to manage a such large number of computation tasks, both for the network capacity and for the power need for the cloud computing infrastructures. As emerging for mobile terrestrial networks, edge computing gained interest as relevant technology for beyond \ac{5G} satellite networks \cite{kodheli2020Survey_IEEE, Zhang, Li}. Edge computing concept is to perform data processing close to the source, which could be the key to enabling widespread adoption of \ac{ML} for satellite applications.
Satellites, due to their ubiquitous coverage and low latency, when considering LEO satellites, can strongly support terrestrial networks to apply the edge computing paradigm.
At the same time, edge caching is one of the key technological enablers in beyond 5G (B5G) networks to enhance \ac{QoE} and reduce transmission latency. 
The premise of edge caching is to prefetch popular contents at the edge caches close to end users, thus, when the users request the cached contents, they can be served immediately without transmission from the data server through the core networks.
\paragraph{Description}
Edge computing was introduced to avoid computing and storage capabilities being concentrated in centralized locations, also known as the cloud computing paradigm, and most likely far, in terms of network topology, from the network user devices. In fact, on the opposite, the edge computing paradigm runs applications and services in hosting servers/hardware close to the user devices. Edge computing is an optimization problem to decide how/where to allocate the tasks to be executed. The larger the network and the traffic demand, the higher the complexity of the problem. Accordingly, ML relies on learning techniques to support and reduce this complexity increase and achieve good solutions in a reasonable time.
Edge computing is typically used in \ac{SATCOM} in two main solutions. Computation offloading \cite{Mach}, whether total or partial, is the technique to offload computational tasks from the cloud to the local \ac{MEC} servers. This is a flexible approach, depending on the \ac{QoS} application/user requirements, e.g. tolerated latency and power constraints, the computational tasks are offloaded to servers accordingly. For instance, delay-sensitive applications are expected to be partially offloaded to servers due to time constraints.

Content caching is the second common technique for edge computing over SATCOM \cite{Wang}. The lack of spectrum availability has forced to find new solutions to minimize the traffic over information networks. Caching the contents to servers/stations close to the user, reduces the transmissions over the network and the overall latency experienced by each user. 
Edge caching in general comprises of two phases: cache placement phase and delivery phase. In the first phase, selected contents are determined and are prefetched on the caches, which usually happens during off-peak times. The second phase occurs during peak hours when users request contents. If the requested contents are available in the edge caches, they can be served immediately without being sent throught the backhaul network. Two main issues in edge caching is i) how to determine the efficiently determine the popular contents, and ii) how to optimally allocate the edge caches in the network.
\paragraph{Conventional Solutions and Issues}

Conventional caching is to exploit the long-term content popularity to determine that cached content. In general, based on the history of requested data, the most popular contents will be prefetched to the cache \cite{ref48}.
In \cite{Zhang}, several techniques are proposed to decide the location to which tasks should be offloaded. Authors propose different scenarios where tasks are either offloaded to close terrestrial servers, on-board of satellites or close to ground stations (gateways).

One of the issue/challenge, presented in \cite{Cheng}, when implementing edge computing over SATCOM, is the high dynamicity of the network. Due to fast and frequent variations, the channel features and devices availability is critically unstable and time-dependent. This might bring to delay or failures when communicating to the required servers.
In addition, as the beyond 5G SATCOM networks are expected to be very heterogeneous, especially in terms of on-board available resources, the optimization design of the offload policies should be properly analyzed, considering the variety of available on-board capabilities/resources. 
\paragraph{Proposed ML Solutions}
The majority of proposed ML solution for edge computing in the literature are developed for terrestrial networks. However, recently few contributions have emerged in the area of integrated satellite-terrestrial networks. Authors in \cite{Chao} propose a deep Q-learning approach to solve the joint optimization of routing and caching for \ac{SAGIN}. ML-based edge computing is mainly proposed for IoT scenarios, because the highly dynamic and random computing task arrival makes the edge computing deployment challenging. 
Authors in \cite{Zhou} propose a \ac{DRL} approach to offload IoT tasks with the objective of reducing the average processing delay
while considering the \ac{UAV} energy capacity (consumed by communication and computing). The \ac{RL} approach is used via a risk function which accounts for the violation of the \ac{UAV} energy availability. In fact, the main idea is to use \ac{UAV} to collect \ac{IoT} tasks, and run the \ac{DRL} to schedule them either locally on-board (UAV), in the closest base-station or to the LEO constellation. These offloading servers are complementary since UAV is very close to the IoT, the LEO constellation has a vast coverage and the terrestrial base-station has the highest computation capability. Clearly, the decision will be driven by the requirements from each task.  
A similar scenario is considered in \cite{Cheng}, where authors work on the optimization of computational task offloading for IoT-based \ac{SAGIN} networks. In this context, they propose a \ac{RL} approach to learn the timely network behaviors in order to determine the optimal offloading policy. The learning process is improved by the use of \acp{DNN}, policy gradient and actor-critic methods. The final objective is to minimize a composite function of delay, energy consumption and server usage cost. 

Centralized learning for caching is studied in \cite{ref33, ref34, ref35}. The authors of \cite{ref33} consider a SBS with limited cache capacity connected to a central controller via a limited capacity backhaul link. The time is divided into separate periods in which the combinatorial multi-armed bandit (CMAB) learning algorithm is used to learn the file popularity over the time and improve caching efficiency. The work of \cite{ref34} considers centralized content popularity learning algorithms of a network comprised of multiple Wi-Fi or WiMAX access points which have a shared cache storage and serve multiple users. The authors of \cite{ref35} consider multiple SBSs where each of them is connected to the core network using a limited capacity backhaul link. Device-to-device interactions of each user are used to estimate the file popularity matrix. This matrix is used to apply the transfer learning on each SBS to optimally cache the content according to storage limit, estimated popularity, and backhaul capacity.  
Distributed caching is studied in \cite{ref39, ref29}. The work of \cite{ref39} considers a network comprised of a central BS and multiple SBSs where each SBS serves multiple users. The authors use the coded caching for distributed file replacement in SBSs. The multi-armed bandit algorithm is used to learn the file popularity in time and reduce the traffic from the central BS to the SBSs by considering the users’ connectivity to each SBS and pre-fetching the content. The works of \cite{ref32, ref39} consider only one learner in the network which can impose a huge computation load on the learning agent. This issue is addressed in \cite{ref29} using a distributed multi-agent learning algorithm where the computation load is distributed and there is no need for the channel state information at the macrocell BS. However, authors in \cite{ref39, ref29} have not considered improving the learning convergence rate of the distributed caching by coordinating the caches through satellite connectivity. Furthermore, backhaul and user link capacities were not included in the optimization model. Theoretical analysis of time and samples to achieve a specific performance is studied in \cite{ref40}. The work of \cite{ref40} considers a network of multiple BSs and cache-enabled SBSs where users are spatially distributed according to Poisson point process. Transfer learning algorithm is used to estimate the file popularity at the BS in order to reduce the user waiting time using distributed caching. Proactive catching along with file popularity estimation has been studied in \cite{ref41, ref42}. The authors of \cite{ref41} predict user demand based on Wi-Fi access, battery status, and download bandwidth analysis. This way, the predicted data are cached before the user demand and this reduces the network congestion. The work of \cite{ref42} uses the content request delay to predict users' requests. A schedule is used to push the content into the receiver buffer and cache it to maximize the number of available user requests at the cache. 
The application of satellite communications due to high bandwidth and wide area coverage is investigated in feeding several network caches at the same time using broad/multi-cast \cite{ref43, ref44}. The work of \cite{ref43} uses the satellite broad/multi-cast to feed the caches at the user side. The authors of \cite{ref44} consider a network comprised of proxy servers with individual cache storages. If the requested content is not available in the cache, it is routed to the gateway for satellite transmission. Based on the number of requests, the file size, and the available satellite bandwidth the request will be admitted or dropped. The satellite multicasts the requested content and the global popularity. Each server uses the local and global file popularity to update the cache. While \cite{ref44} proposes satellite multicast in cache feeding, it does not consider the backhaul cost, user link, cache storage limit, handling the dropped packets at the gateway, and the interaction between the macrocell and SBSs.
Content popularity estimation, selection, and delivery are studied in \cite{ref45, ref46, ref47}. The works of \cite{ref45} studies the caching in a network where small cell networks with high capacity cache are connected to central scheduler via a limited backhaul link and serve multiple users. The central scheduler learns the file popularities by tracking users request and places the right file in the cache of the right small cell network. The authors maximize the ratio of the satisfied requests to the total user request while considering the backhaul, cache, and user wireless link capacities. The authors in \cite{ref46} consider a network comprised of a macrocell BS and multiple cache-enabled SBSs where each of them serves a group of users. The authors minimize the content delivery delay to the users by learning the content popularity at each SBS and updating the cache according to it. As a practical approach, \cite{ref47} studies the edge caching in a cellular 5G wireless network comprised of cache-enabled SBSs. Statistical machine learning is used for file popularity estimation and file caching in off-peak hours. The cached content is assumed fixed during the usage hours. The work minimizes the average backhaul load by considering the file size, file bit rate, backhaul link capacity, wireless link capacity, cached content, and user quality of experience (QoE). The authors perform big data analysis on users' data and show that it can be modelled as a Zipf-like distribution \cite{ref48}.

\subsection{Others}

\subsubsection{Integrated Sensing and Communication}\label{ISAC}
\paragraph{Motivation}
Current networks are evolving in all spectrum bands: RF, mm-wave, and terahertz to solve the shortage of spectrum. Annother promising solution is represented by multi-function systems simultaneously performing. This is the case for joint sensing and communications systems which converge spectrally through waveform design. Both radar and communications systems need wide bandwidth to provide a designated quality of service, thus resulting in competing interests in exploiting the spectrum. Hence, sharing spectral and/ or hardware resources of communications and radar is imperative for efficient spectrum utilization. This has led to the emergence of \ac{ISAC} systems which employ one of the following:  common channel access,  transmit/receive hardware units, processing, or waveforms \cite{surveyJCS}. 
\paragraph{Description}
The electromagnetic spectrum in different bands (RF, mm-wave, and terahertz) is not nearly sufficient, as the number of emerging technologies with different functions, communications, radar, positioning, navigation, spectral situational awareness, and opportunistic spectrum access, is foreseen to rise in the next years. 

Various technologies able to simultaneously performing two or more radio tasks, are currently under discussion to tackle this challenge.
One promising solution is represented by \ac{ISAC} systems \cite{9376388}, or \ac{JCS} and JCAS, as can be found in literature, where communication and sensing are performed simultaneously in a dedicated or existing infrastructure. Remote sensing conventionally refers to satellite- or aircraft-based sensor technologies to detect and classify objects usually on Earth. Remote sensing can be split into \textit{active} remote sensing (when a signal is emitted by a satellite or aircraft to the object, and its reflection is detected by the sensor) and \textit{passive} remote sensing (when the sensor detects the reflection of sunlight).
\ac{ISAC} aims to unify radar and communication systems through a combination of joint hardware, waveforms, joint signal design, and joint signal processing.
\paragraph{Conventional Solutions and Issues}
Generally, the most tracked objects in satellite video \ac{SV} used for RS are aeroplanes, ships, vehicles, and trains. However, in SVs, these objects, especially vehicles, are extremely small relative to the captured region, which raises a great challenge for general tracking tasks. Therefore, many excellent tracking methods \cite{RS3,RS4} have been developed to cope with these challenges. Early tracking methods are mainly based on movement features, and colour features, such as Kalman filter (KF) or particle filter \cite{RS_filter}. These classic filters or statistical methods, predominantly based on mathematics and physics derivations, aim at achieving relatively fast and robust tracking. However, these trackers are limited by strict parameters or single-usage scenarios. In \cite{4156613}, a model has been developed and validated to make it functional so that elevation-independent data acquisition can occur with specific considerations for LEO satellites. Some issues in conventional RS have been identified in \cite{grandRS}, such as increased spatial and temporal coverage and resolution of satellite observations or an increased information content and exploring the synergy of observations. Also, another issue present in conventional satellite RS is achieving continuity in consistent observations by satellites, and the long-term data record \cite{grandRS}. Increasing RS data's resolution in the spatial, spectral, and time dimensions is a possible solution. However, it also presents new problems with the high amount of data, such as insufficient upstream and downstream transmission bandwidth, the need for more adaptability of the remote sensing payload imaging method to the complex ground and feature changes, and time-efficiency problems with information services. 
Regarding \ac{ISAC}, literature offers already several works focused on terrestrial and aerial networks and explore promising techniques as \ac{MIMO} \cite{2020FanLiu_TCOM:JointRadarComm} or \ac{UAV} applications \cite{hu2022TCOM:UAV_ISAC,liu2022integrated}.
However, the direct application of these techniques is not directly applicable to satellite systems due to the difference in wave propagation properties. In particular, two significant differences can be highlighted, i.e., the inevitably high propagation delay and large Doppler shifts due to the long distances between the LEO satellites and the targets as well as their mobility \cite{you2022_JSAC:Beam_MassiveMIMO_ISAC}. Given these substantial challenges, research of \ac{ISAC} for \ac{SATCOM} is still in its infancy stage.
\cite{9842817} proposes a physical layer design to sense the atmosphere in the terahertz band using communication signals from LEO satellites. The work in \cite{you2022_JSAC:Beam_MassiveMIMO_ISAC} performs \ac{ISAC} in a massive \ac{MIMO} \ac{LEO} system, which can provide wide coverage for wireless \ac{ISAC}. Due to the inaccuracy of the instantaneous \ac{CSI}, authors consider statistical \ac{CSI}.
\paragraph{Proposed ML Solutions}
The above works show the promise of the potential usage of \ac{ISAC} in conjunction with \ac{LEO} satellites. \ac{ISAC}, however, is not only one of the pillars for a new air interface for \ac{6G} networks, but, if coupled with \ac{ML} can serve as foundation of a world where everything is connected, sensed and intelligent. Toward this direction, \cite{JSC1} proposes a novel approach using an auto-encoder for data-driven \ac{ISAC}. K-means-based clustering and DRL-based resource allocation (K-DRL) are proposed in \cite{JSC2}. These techniques extract pixel characteristic-based features from satellite images to improve localization accuracy. In addition, the proposed UK-medoids and DRL-based scheme (UKM-DRL) is compared with two other schemes: K-means-based clustering and DRL-based resource allocation (K-DRL) and UK-means-based clustering and DRL-based resource allocation (UK-DRL).
In table \ref{tab:Article_Categorization} we have reported also some remote sensing articles, for the sake of completeness.

\subsubsection{Cooperative Satellite Communications}
\begin{figure}[t!]
\centering
\includegraphics[height=3in, width=3in]{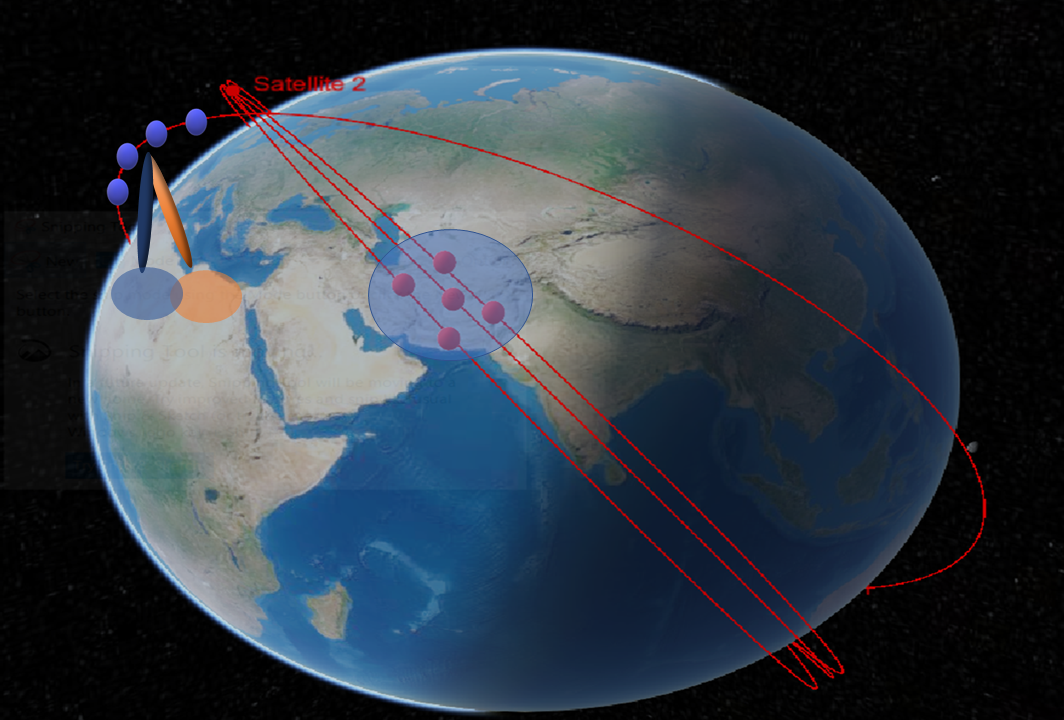}
\caption{Distributed beamforming using satellite constellation.}
\label{fig:su1}
\end{figure}

\paragraph{Motivation}
Cooperative satellite communication is integral to satellite-based applications like remote sensing, surveillance, joint communication, joint scheduling, distributed beamforming, and sensing \cite{choi2022cooperative}. Specifically for LEO satellite networks, implementing cooperative communications becomes challenging due to their high mobility. The existing cooperative satellite-based communication requires a huge amount of communication among satellites for cooperative tasks. This creates problems in establishing joint beamforming, joint scheduling, and other cooperative satellite applications in real-time \cite{dalin2020online}. In the current scenario, the LEO satellite communication networks such as SpaceX, and OneWeb are gaining popularity because of their low latency, seamless connectivity and broad coverage \cite{deng2021ultra}. However, the networking of these small satellites considers different factors for successful cooperative communication. One crucial factor is the distributed processing and synchronization of satellite networks for coordinated applications \cite{radhakrishnan2016survey}. Distributed processing refers to the decentralization of the computing processors, which are physically separated. This distributed processing is application specific. Additionally, these distributed processors can have different structural architectures. It is essential to have efficient distributed processing for effective cooperative satellite-based communication.
\paragraph{Description}  
Cooperative satellite communication comprises a wide range of applications from distributed beamforming to joint scheduling \cite{choi2022cooperative}\cite{dalin2020online}. The existing LEO satellite networks comprise monolithic antennas with different antenna elements. Deployment of such satellites is not cost-effective. Therefore, the current generation of satellite networks with SpaceX, OneWeb, and Amazon Kuiper have aimed for smaller and lighter satellite constellation networks \cite{deng2021ultra}. One of the prime applications focused on by such satellite constellations is distributed beamforming. A pictorial representation of the distributed beamforming performed by a satellite constellation is shown in Figure \ref{fig:su1}. Figure \ref{fig:su1} shows two sets of satellite swarms with different structures. One satellite set is performing the distributed beamforming.

\paragraph{Conventional Solutions and Issues}
The distributed processing in cooperative satellite communication exists in different architectures, namely a Star, a Link, Linear, Hybrid and Layered architectures \cite{palmintier2002distributed}. These existing architectures put in challenges for distributed processing in terms of power, volume, weight and size constraints of satellite platforms \cite{smartsat}. However, the main issue lies in the non-optimal utilization of the parameters. This allows employing model compression schemes via removing redundancy or adapting to a compact design with high accuracy using neural architecture search \cite{smartsat}.
\paragraph{Proposed ML Solutions}
The architectural optimization of the neural network is termed \ac{NAS} \cite{he2021automl}. The \ac{NAS} is an efficient tool for optimizing hardware architecture for distributed processing in satellites on-board processing while keeping check of power, volume, weight, and size constraints \cite{smartsat}. It works on a reward and learning model to generate high-performance architecture. The \ac{NAS} generated architectures are observed as faster, more accurate, less complex, and less expensive than conventional schemes like gradient descent, evolutionary algorithm, and reinforcement learning \cite{ghiasi2019fpn}.

\subsubsection{Image Processing}
\paragraph{Motivation}
Satellite images are becoming an essential spatial data resource for scenarios such as dynamic target tracking \cite{imag3}, disaster monitoring \cite{imag1}, and maritime surveillance \cite{imag2}, where the satellite is a crucial element. Compared with terrestrial images, the quality of satellite images is affected by additional factors, such as atmospheric disturbance, over-the-horizon imaging, the relative motion between  satellite and ground objects during observation, and scattering. In addition, to this is added the incorporation of video to SATCOM generates an enormous amount of data collected  to meet the data real-time transmission with limited bandwidth. Thus, to improve the image compression ratio or reduce the spatial sampling resolution must be performed by the satellite system. Super Resolution technique is an important solution which maps low-resolution images to high-resolution images with different scale factors. However, the above unpredictable degradation causes the reconstruction of satellite images to be quite different from terrestrial images. In addition, ground terminals' resolution varies, so it is important to reconstruct satellite images at arbitrary scales. However, due to the limitations of hardware equipment and transmission bandwidth, the images received on the ground have low resolution and weak texture. In addition, ground terminals have various resolutions and real-time playing requirements. Another problem in satellite images is the loss of high-frequency details caused by over-compression. Therefore, achieving arbitrary scale super-resolution of satellite images is essential for signal processing in SATCOM.  

\paragraph{Description}
Satellite Image Processing (SIP) consists of the images of the earth and satellites taken by artificial satellites. In broader terms, SIP is a kind of remote sensing which works on pixel resolutions to collect coherent information about the earth's surface.
Depending on the application, the images can have object land cover, clouds, or asteroids.
Firstly, the images are taken in digital form and later are processed by the computers to extract the information. SIP is applied by statistical methods to the digital images to identify the various discrete surfaces and analyse the pixel values. The key application of SIP is to detect responses to upcoming disasters or design sustainable infrastructures to monitor environmental conditions.

Four kinds of resolutions are associated with satellite images:
\begin{enumerate}     
    \item Spectral resolution measures the wavelength internal size and determines the number of wavelength intervals that the sensor measures.
    \item Radiometric resolution is defined in bits size. It gives effective bit depth and records the imaging system's brightness levels.
    \item Temporal resolution is defined as the time or days that pass between various imagery cloud periods.   
    \item Spatial resolution is defined as the pixel size of an image.
\end{enumerate}

\paragraph{Conventional Solutions and Issues}
The main application where SIP is applied is for Earth observation, which is categorized into four conventional techniques (1) Preprocessing, (2) transformation, (3) correction, and (4) classification. In pre-processing, some distortions produced on satellite images by imaging systems, sensors, and observing conditions are corrected before using the three following techniques to correct or minimize image distortions.  
In \cite{imag26}, a mixed high-order attention network is  proposed,  composed  of  a  feature  extraction network and a feature refinement network to exploit low-level feature and high-frequency detail,  respectively. 

\paragraph{Proposed ML Solutions}
For satellite images, it is essential to reconstruct the fine spatial details for subsequent applications. Many research efforts \cite{imag26,imag27,DLLimageSat,imag28} have been paid to several aspects. Some of these methods focus on improving CNN architecture. \cite{imag27} proposed a cycle convolutional neural network (Cycle-CNN), which consists of two  generative CNNs for down-sampling and SR, respectively, Cycle-CNN has great performance in re-constructing multi-spectral band images of real remote sensing satellites.
\cite{9861276} proposes a distinct residual dense network (RDN) as the feature learning module to extract hierarchical satellite image features in LR space.  Then, it is followed by an arbitrary upscale module to project LR features to  arbitrarily scale enlarged SR images. Finally, the scheme adds an edge reinforcement module to re-cover  the  edge  information  and  improve  SR  images with clearer textures.
Particularly, in \cite{9967537}, proposed a technique for spatio-temporal fusion using deep DNNs in two phases with a large amount of RS data  as the application background.  
%
As highlighted by the above works, the availability of large datasets is critical for Image processing in satellites and might represent one of the main challenges. We will present more details about the training dataset challenge in \ac{SATCOM} in Section \ref{TrainingDataAvailability}.

\subsubsection{Ground Segment Dimensioning}
\paragraph{Motivation}
Parallel to the renewal in the space segment, we find that the ground segment (GS) must also cope with the immense number of changes made in SATCOM for the space segment. This implies new challenges not yet addressed for this element of the communications system. The ground segment is that part of a SATCOM system that employs a variety of node designs and network configurations to provide and manage services delivered, on the one hand, to end-users and, on the other hand, to satellites using feeder links. Service to the GS of an NGSO system depends largely on the nature of the link, which is shorter but more dynamic than its GEO counterpart.

\paragraph{Description}
The GS dimensioning (GSD) includes determining the number of gateways (GW) needed, the antenna parameterization, the traffic model, and the switching strategies between backup GW. The location of the gateways is one key requirement and vital in NGSO constellations due to the high number of elements that make up the constellations. The location of these elements is strategic for properly functioning these constellations' new services. This strategy can be dynamic as a function of the weather conditions, considering services with greater security restrictions, such as government services and political conditions. 

\paragraph{Conventional Solutions and Issues}
\cite{groundArchitecture} compares GS architectures for constellations using feeder links in Q/V-band against those using E-band. In addition, a method is developed to determine the locations of the minimum number of ground stations that maximize the system capacity while achieving desired QoS levels. The work in \cite{ground5GNTN} analyzed the handover performance in a LEO-based Non-Terrestrial Network (LEO-NTN) via system-level simulations, focusing on ground-segment optimization by traditional methods. Two algorithms for dimensioning the optical ground station (OGS) network with MEO satellites are proposed in \cite{8887525}. These algorithms aim to minimise the network's size while ensuring the target link availabilities due to cloud blockage. A joint strategy for GSD and routing network is proposed in \cite{9149175} for satellite-terrestrial integration. CARAMUEL project \cite{CARAMUEL} presents the GSD considering ground network with Quantum Key Distribution in GEO satellites. Regarding switching strategy, \cite{9043742} proposes an optimal gateway selection in each moment using site diversity. In addition, to make the decision to locate these stations, we can make use of emerging technologies and techniques such as AI and ML. The application of this field for the optimized design in the GS that allows selecting the position of the gateways based on criteria beyond the service requirements is still in its infancy.

\paragraph{Proposed ML Solutions}
In \cite{9429227}, the authors present an overview of optimization techniques applicable to solving the GW location problem.  On the other hand, a survey on optimization methods based on ML focused on LEO satellite networks is performed \cite{s22041421}. This survey deals with GS within future autonomous transportation applications using NGSO satellites. However, it is a preliminary study using ML GSD techniques for SATCOM applications. For GEO satellite, ML-based GSD is detailed in \cite{9068955}.

\section{Hardware Solutions}\label{HardwareSolutions}

AI and ML techniques are used to optimize time and resource-critical applications nowadays. 
Satellite communication is one of those critical fields. As highlighted in Section \ref{sec:Onboard_Onground}, on-ground segment applications can be less restrictive, while onboard applications can be a real challenge from the point of view of power constraints, radiation, and fault-tolerant architectures. 
This section presents an analysis of the HW for possible AI and ML implementations for satellite communication.

\subsection{Neuromorphic Processors}
Mainstream microprocessors are unsuitable for deploying state-of-the-art deep neural networks that require many matrix and vector multiplication operations. This is mainly because on-chip memory, which holds the data for processing, is limited. In addition, operations involving large matrices require frequent access to off-chip memory and storage units, which consume a relatively large amount of power and take too long. Custom neuromorphic processors (NPs) have been built for this purpose that can solve the resulting von Neumann bottleneck. Toward this end, the hardware is specifically designed to allow tight integration of logic and memory units, and accelerator units are provided to execute frequent computational operations efficiently. Moreover, adopting spiking dynamics for the neurons also eliminates their need to perform expensive multiplication operations with high-precision real numbers represented in FP32 or another reduced-precision format \cite{Flor2022NP}. The most advanced hardware implementations to date are
\begin{itemize}
    \item TrueNorth from IBM \cite{akopyan15truenorth}
    \item Loihi and Lohi2 from Intel \cite{davies18loihi}
    \item Zeroth from Qualcomm \cite{IntroducingQualcomm}
    \item SpiNNaker and SpiNNaker 2 from University of Manchester \cite{furber20spinnaker}
    \item BrainScaleS and BrainScaleS-2 from Heidelberg University  \cite{schemmel12brainscales}
    \item NeuroGrid and Braindrop from Stanford University \cite{benjamin14neurogrid, neckar19braindrop}
    \item DYNAP devices from SynSense \cite{moradi18dynap}
    \item ODIN from Catholic University Louvain \cite{frenkel19odin}
\end{itemize}

\subsection{AI-capable Chipset/Accelarators} \label{sec:AI_chips}
So far, complex onboard AI/ML applications can only be performed with very expensive custom-designed ASICs \cite{AMDInc.2022}.
The increased performance requirements for onboard processing to support higher data rates and autonomy made the existing space CPU obsolete. New technologies, including non-qualified COTS devices, from other critical domains, are currently being explored \cite{Rodriguez2020}.
Finding a device with adequate computer capacity, proper power consumption, and compliance with standards specifications for onboard and standalone applications is one of the most critical points for the implementation \cite{ortiz2023onboard}.

The commercial AI-capable embedded chipsets market is divided into embedded GPUs and dedicated devices for hardware co-processors or standalone embedded systems. Several devices have been launched to the market, led mainly by NVIDIA, AMD, Intel, and Qualcomm \cite{2022aimarket}.

Table \ref{Tab:AI_Chipsets:_CPU} resumes nowadays embedded COTS AI-capable chipsets from the main companies\footnote{Data collected from available vendor's datasheets and published papers}:
\begin{enumerate}
    \item Intel Corp.
        \begin{itemize}
            \item Intel Movidius Myriad family \cite{LUXonis2020_Myriad}
        \end{itemize}
    \item NVIDIA Corp. \cite{Roadmap_NVIDIACorp.2022}
        \begin{itemize}
            \item NVIDIA Jetson Nano \cite{Nano_NVIDIACorp.2014}
            \item NVIDIA Jetson TX2 Family \cite{TX2_NVIDIACorp.2021a, TX2NX_NVIDIACorp.2022a}
            \item NVIDIA Jetson Xavier NX Family \cite{XavierNX_NVIDIACorp.2021}
            \item NVIDIA Jetson AGX Xavier Family \cite{XavierAGX_NVIDIACorp.2021b}
            \item NVIDIA Jetson Orin NX Family \cite{OrinNX_NVIDIACorp.2022c}
            \item NVIDIA Jetson AGX Orin Family \cite{OrinAGX_NVIDIACorp.2022}
            \item NVIDIA Jetson Orin Nano Family\cite{OrinNano_NVIDIACorp.2022a,OrinNano_NVIDIACorp.2022d}
        \end{itemize}
    \item Qualcomm Technologies, Inc. 
         \begin{itemize}
            \item Qualcomm Cloud AI 100 Family \cite{AI_100_Qualcomm2020,AI_100_DK_Qualcomm2020,AI_100_BM_Qualcomm2022}
        \end{itemize}
    \item Advanced Micro Devices, Inc.
     \begin{itemize}
            \item AMD Instinct MI200 Family \cite{MI200Series_AMDInc.2021,MI210_AMDInc.2022}
            \item  XILINX Versal ACAP AI Edge Family \cite{EDGESerie_AMDXILINXInc.2021,EDGESerie_XILINXInc.2021,EDGESeries_AMDXILINXInc.2022}
            \item XILINX Versal ACAP AI Core Family \cite{AICoreSerie_XILINXInc.2019,Versal_AMDXILINXInc.2022a}
        \end{itemize}
\end{enumerate}

AI-capable chipsets architecture includes a Central Processing Unit (CPU) and an on-chip accelerator, which define their capabilities and performance on AI and ML tasks.
The amount of CPUs and cores depends on the manufacturer and the family, varying from one to two CPUs and two to 12 cores. 
Remarkably, the multi-core ARM Cortex-R5F2 on the Versal ACAP chips and the 12 cores on Jetson AGX Orin have high-end chip specifications. The first is suitable for embedded real-time and safety-critical systems. In contrast, the vast number of cores of the second allows the use of multi-threads in sequential parallelizable software \cite{Versal_AMDXILINXInc.2022a, OrinAGX_NVIDIACorp.2022}.

In the case of the on-chip accelerators, as their architecture differs from one developer to the other, a proper comparison takes into account the performance per operation, as for Table \ref{Tab:AI_Chipsets:_CompCap}. 
Note that we have marked as \textit{NR} wherever some data are not reported in literature and marked as \textit{x} in case data are not detailed. 
As XILINX doesn't report the power consumption specifications for the Versal ACAP AI Core family, the analysis is done only for the Versal AI Core VC1902 device with the power estimation reported in \cite{AMDXILINXInc.2022_AI_inf}.

\begin{table}[th!]
\caption{AI Chipset/Accelerators off-the-shelf: Core Units}
\centering{}%
\begin{tabular}{>{\centering}p{1.5cm}>{\centering}p{1.5cm}>{\centering}m{2cm}>{\centering}m{2cm}}
\toprule 
\multirow{2}{1.5cm}{\textbf{Device}} & \multirow{2}{1.5cm}{\textbf{Provider}} & \multicolumn{2}{c}{\textbf{Core units}}\tabularnewline
\cmidrule{3-4} \cmidrule{4-4} 
 &  & \textbf{CPU} & \textbf{On-chip accelerator}\tabularnewline
\midrule
\multirow{2}{1.5cm}{Myriad Family} & \multirow{2}{1.5cm}{Intel} & \multirow{2}{2cm}{2x Leon 4 RISC} & Image/Video PA\tabularnewline
\cmidrule{4-4} 
 &  &  & SHAVE\tabularnewline
\midrule 
Jetson Nano & NVIDIA & 4x ARM Cortex-A57MP & 128-CUDA Maxwell\tabularnewline
\midrule 
\multirow{2}{1.5cm}{Jetson TX2 Family} & \multirow{2}{1.5cm}{NVIDIA} & 2x ARM64b Denver  & \multirow{2}{2cm}{256-CUDA Pascall}\tabularnewline
\cmidrule{3-3} 
 &  & 4x ARM Cortex-A57MP  & \tabularnewline
\midrule 
\multirow{4}{1.5cm}{Jetson Xavier NX Family} & \multirow{4}{1.5cm}{NVIDIA} & \multirow{4}{2cm}{6x ARM64b Carmel } & 384-CUDA Volta\tabularnewline
\cmidrule{4-4} 
 &  &  & 2x NVDLA \tabularnewline
\cmidrule{4-4} 
 &  &  & 2x PVA \tabularnewline
\cmidrule{4-4} 
 &  &  & 48 Tensor \tabularnewline
\midrule 
\multirow{4}{1.5cm}{Jetson AGX Xavier Family} & \multirow{4}{1.5cm}{NVIDIA} & \multirow{4}{2cm}{8x Carmel ARM64b } & 512-CUDA Volta\tabularnewline
\cmidrule{4-4} 
 &  &  & 2x NVDLA \tabularnewline
\cmidrule{4-4} 
 &  &  & 2x PVA \tabularnewline
\cmidrule{4-4} 
 &  &  & 64 Tensor \tabularnewline
\midrule 
\multirow{4}{1.5cm}{Jetson Orin NX Family} & \multirow{4}{1.5cm}{NVIDIA} & \multirow{4}{2cm}{6x/8x ARM64b Cortex-A78AE} & 1024-CUDA Ampere\tabularnewline
\cmidrule{4-4} 
 &  &  & 1x/2x NVDLAv2 \tabularnewline
\cmidrule{4-4} 
 &  &  & 1x PVAv2 \tabularnewline
\cmidrule{4-4} 
 &  &  & 32 Tensor \tabularnewline
\midrule 
\multirow{4}{1.5cm}{Jetson AGX Orin Family} & \multirow{4}{1.5cm}{NVIDIA} & \multirow{4}{2cm}{8x/12x ARM64b Cortex-A78AE} & 1782/2048-CUDA Ampere\tabularnewline
\cmidrule{4-4} 
 &  &  & 2x NVDLAv2 \tabularnewline
\cmidrule{4-4} 
 &  &  & 1x PVAv2 \tabularnewline
\cmidrule{4-4} 
 &  &  & 56/64 Tensor \tabularnewline
\midrule 
\multirow{2}{1.5cm}{Jetson Orin Nano Family} & \multirow{2}{1.5cm}{NVIDIA} & \multirow{2}{2cm}{6x ARM64b Cortex-A78AE} & 512/1024-CUDA Ampere\tabularnewline
\cmidrule{4-4} 
 &  &  & 16/32 Tensor \tabularnewline
\midrule 
\multirow{2}{1.5cm}{Cloud AI 100 Family} & \multirow{2}{1.5cm}{Qualcomm} & Snapdragon 865 MP & \multirow{2}{2cm}{Cloud AI 100}\tabularnewline
\cmidrule{3-3} 
 &  & Kryo 585 CPU & \tabularnewline
\midrule 
\multirow{2}{1.5cm}{Instinct MI200 Familly} & \multirow{2}{1.5cm}{AMD} & \multirow{2}{2cm}{CDNA2} & 6656/14080-Stream Proc.\tabularnewline
\cmidrule{4-4} 
 &  &  & 104/220 Core Units\tabularnewline
\midrule 
\multirow{3}{1.5cm}{Versal AI Edge Family} & \multirow{3}{1.5cm}{AMD XILINX} & 2x ARM64 Cortex-A72 & 8-304 AI Engines/ML\tabularnewline
\cmidrule{3-4} \cmidrule{4-4} 
 &  & \multirow{2}{2cm}{2x ARM Cortex-R5F2} & 90-1312 DSP Eng\tabularnewline
\cmidrule{4-4} 
 &  &  & 43k-1139k System Logic Cells\tabularnewline
\midrule 
\multirow{3}{1.5cm}{Versal AI Core Family} & \multirow{3}{1.5cm}{AMD XILINX} & 2x ARM64 Cortex-A72 & 128-400 AI Engines\tabularnewline
\cmidrule{3-4} \cmidrule{4-4} 
 &  & \multirow{2}{2cm}{2x ARM Cortex-R5F2} & 928-1968 DSP Eng\tabularnewline
\cmidrule{4-4} 
 &  &  & 504k-1968k System Logic Cells\tabularnewline
\bottomrule
\end{tabular}\label{Tab:AI_Chipsets:_CPU}
\end{table}

\begin{table}[h!]
\caption{AI Chipset/Accelerators off-the-shelf: Computer Capacity per operation}

\centering{}%
\begin{tabular}{>{\centering}p{1.5cm}>{\centering}m{1.25cm}m{1.25cm}>{\centering}m{1.25cm}>{\centering}m{1cm}}
\toprule 
\multirow{2}{1.5cm}{\textbf{Device}} & \multicolumn{3}{l}{\textbf{Computer capacity in operations}} & \multirow{2}{1cm}{Power (W)}\tabularnewline
\cmidrule{2-4} \cmidrule{3-4} \cmidrule{4-4} 
 & \textbf{I8 (Ops) } & \textbf{FP16 (FLOPs)} & \textbf{FP32 (FLOPs)} & \tabularnewline
\midrule
Myriad Family & NR & NR & NR & 1\textasciitilde 2\tabularnewline
\midrule 
Jetson Nano & NR & 512G & NR & 5-10\tabularnewline
\midrule 
\multirow{2}{1.5cm}{Jetson TX2 Family} & \multirow{2}{1.25cm}{NR} & \multirow{2}{1.25cm}{x} & \multirow{2}{1.25cm}{NR} & \multirow{2}{1cm}{7.5-20}\tabularnewline
 &  &  &  & \tabularnewline
\midrule 
\multirow{3}{1.5cm}{Jetson Xavier NX Family} & \multirow{3}{1.25cm}{21T} & \multirow{3}{1.25cm}{x} & \multirow{3}{1.25cm}{NR} & \multirow{3}{1cm}{10-20}\tabularnewline
 &  &  &  & \tabularnewline
 &  &  &  & \tabularnewline
\midrule 
\multirow{3}{1.5cm}{Jetson AGX Xavier Family} & \multirow{3}{1.25cm}{30-32T} & \multirow{3}{1.25cm}{10T} & \multirow{3}{1.25cm}{NR} & \multirow{3}{1cm}{10-40}\tabularnewline
 &  &  &  & \tabularnewline
 &  &  &  & \tabularnewline
\midrule 
\multirow{3}{1.5cm}{Jetson Orin NX Family} & 70-100T Sparse & \multirow{3}{1.25cm}{x} & \multirow{3}{1.25cm}{x} & \multirow{3}{1cm}{10-25}\tabularnewline
\cmidrule{2-2} 
 & 35-50T

Dense &  &  & \tabularnewline
\cmidrule{2-2} 
 & 20T Sparse &  &  & \tabularnewline
\midrule 
\multirow{2}{1.5cm}{Jetson AGX Orin Family} & 108-170T Sparse & 58-85T & \multirow{2}{1.25cm}{3.3-5.3T} & \multirow{2}{1cm}{15-60}\tabularnewline
\cmidrule{2-3} \cmidrule{3-3} 
 & 92-105T Sparse & 67-106T &  & \tabularnewline
 \midrule 
\multirow{2}{1.5cm}{Jetson Orin Nano Family} & 2-4T Sparse & \multirow{2}{1.25cm}{x} & \multirow{2}{1.25cm}{x} & \multirow{2}{1cm}{5-15}\tabularnewline
\cmidrule{2-2} 
 & 1-1.33T Dense &  &  & \tabularnewline
\midrule 
\multirow{2}{1.5cm}{Cloud AI 100 Family} & \multirow{2}{1.25cm}{70\textendash 400T} & \multirow{2}{1.25cm}{35-200T} & \multirow{2}{1.25cm}{x} & \multirow{2}{1cm}{15-75}\tabularnewline
 &  &  &  & \tabularnewline
\midrule 
\multirow{2}{1.5cm}{Instinct MI200 Familly} & \multirow{2}{1.25cm}{181-383T} & \multirow{2}{1.25cm}{181-383T} & 45.3-95.7T (Matrix) & \multirow{2}{1cm}{300-560}\tabularnewline
\cmidrule{4-4} 
 &  &  & 22.6 - 45.9T & \tabularnewline
\midrule 
\multirow{3}{1.5cm}{Versal AI Edge Family} & 5-202T & \multirow{3}{1.25cm}{NR} & 0.4-16.6T & \multirow{3}{1cm}{6-75}\tabularnewline
\cmidrule{2-2} \cmidrule{4-4} 
 & 0.6-9.1T  &  & \multirow{2}{1.25cm}{0.1-2.1T} & \tabularnewline
\cmidrule{2-2} 
 & 1-17T &  &  & \tabularnewline
\midrule 
\multirow{3}{1.5cm}{Versal AI Core Family (VC1902)} & 133T & \multirow{3}{1.25cm}{NR} & 8T & \multirow{3}{1cm}{\textasciitilde 87W}\tabularnewline
\cmidrule{2-2} \cmidrule{4-4} 
 & 13.6T &  & \multirow{2}{1.25cm}{3.2T} & \tabularnewline
\cmidrule{2-2} 
 & 29T  &  &  & \tabularnewline
\bottomrule
\end{tabular}\label{Tab:AI_Chipsets:_CompCap}
\end{table}

The most popular AI on-chip accelerators are embedded GPUs (SHAVE, CUDA Cores, Streams Processors) and AI-capable cores (Tensor, AI Engines, and DSP Engines).

SHAVE cores were designed by Movidius primarily for physics acceleration in mobile applications. 
Low demand for expensive physics acceleration in smartphones has forced a re-focus on image and vision processing and machine vision processing.
The union of several SHAVE cores creates a versatile parallel architecture that can adapt to other application fields such as AL and ML.

CUDA cores are NVIDIA's dedicated stream parallel processors initially released for graphics processing and 3D rendering on conventional GPUs and later ported to NVIDIA Jetson embedded Family. 
The architecture of a CUDA processor is composed of several CUDA cores as processing elements, making them a good option for AI/ML applications. 

Tensor cores are AI-dedicated processing elements released by NVIDIA since the Volta GPU microarchitecture.
They specialize in fused multiply-add (FMA) operations used extensively in neural network calculations to apply an extensive series of multiplications on weights and then add a bias. 
Tensor cores can operate on FP16, INT8, INT4, and INT1 data types executing 1024-bit FMA operations per clock per core (1024 FMA operations of INT1, 256 of INT4, 128 of INT8, and 64 of FP16) \cite{Tensor_Andersch2019}. 

 The XILINX AI Engine, a distinctive component of the Versal ACAP AI devices, is designed for intensive computing in various applications including but not limited to, machine learning and artificial Intelligence. Its architecture includes a vector processor capable of FP32 by FP16 multiply-and-accumulate (MAC) operations per clock cycle. It is integrated into an AI Engine tile with data memory and DMA engines. Each tile is connected, using AXI stream interfaces, by the four nearest tiles (except the tiles on the edges), forming a neural network and sharing the memory with the neighbours \cite{XILINXInc.2021_BeamF, XILINXInc.2020_AIE, AMDXILINXInc.2022_AI_inf}.
Versal ACAP chips, apart from the on-chip accelerators (AI and DSP Engines), include a programmable logic (PL) on-chip that can be used to design time-critical parts of the algorithm, exploiting the inherent parallelism of the hardware design.

The most supported data-types operations by the chipsets are integer bytes (\textit{INT8}), half-precision floating-point (\textit{FP16}), and single-precision floating-point (\textit{FP32}). 
Instinct MI200 \cite{MI200Series_AMDInc.2021, MI210_AMDInc.2022} and Cloud AI 100 \cite{AI_100_Qualcomm2020, AI_100_DK_Qualcomm2020, AI_100_BM_Qualcomm2022} families have the greatest operation rates in half-precision floating-point, but their form factor and power consumption make them not suitable for embedded systems applications.
NVIDIA's new-generation AI processors family (Orin NX and AGX Orin \cite{OrinNX_NVIDIACorp.2022c, OrinAGX_NVIDIACorp.2022}) promises to achieve good operations per second (\textit{OPs}) rate, as good as the Versal AI chip, increasing the power consumption compared to previous NVIDIA families. 

\begin{figure*}[th!]
\begin{centering}
\includegraphics[width=16cm]{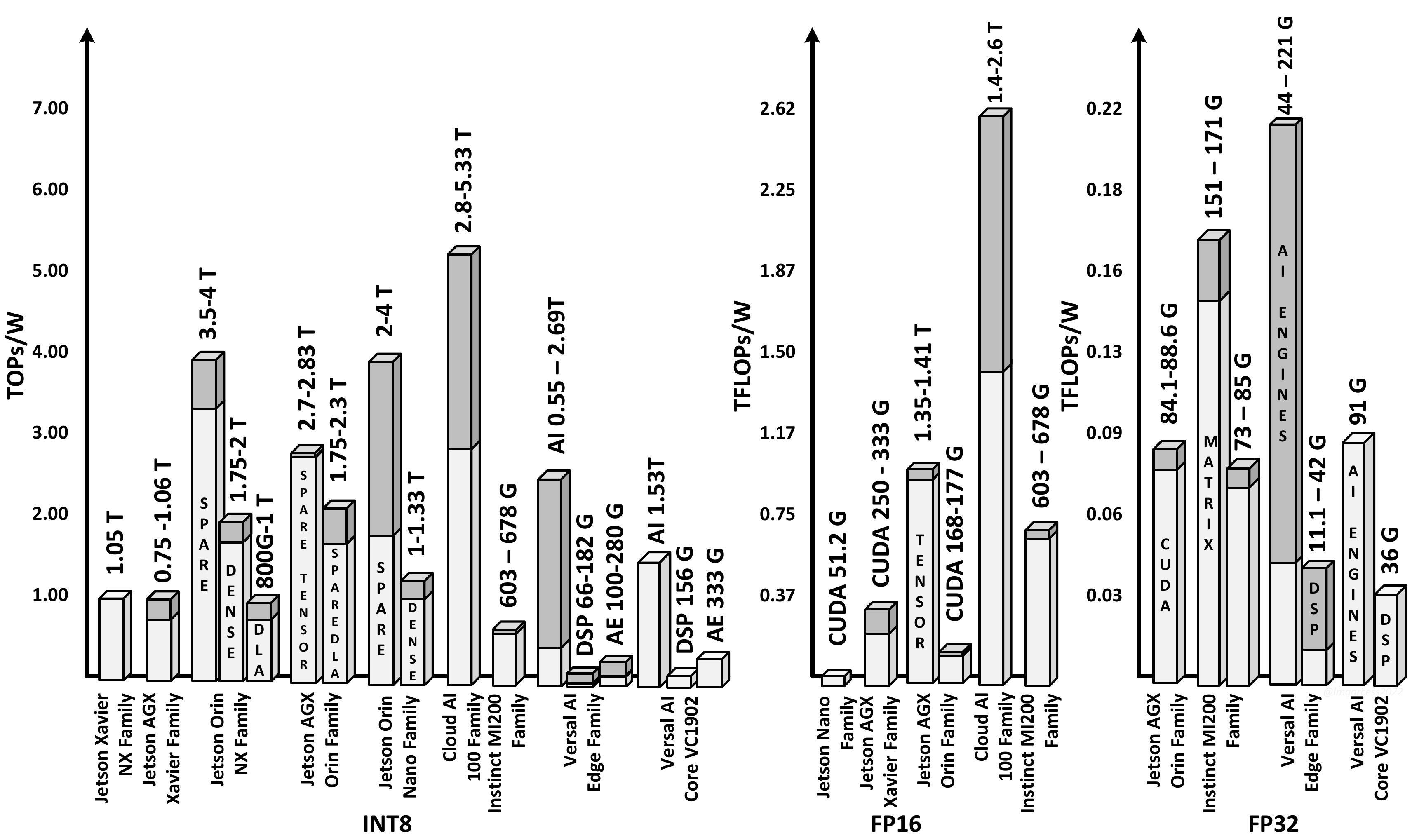}
\par\end{centering}
\caption{AI Chipset/Accelerators off-the-shelf: Computer Capacity per operation}
\label{Fig:AI_Comparison}
\end{figure*}

It is essential to analyze the computer capacity per Watt to select the proper chipset for the application. 
Figure \ref{Fig:AI_Comparison} resumes this information for the most common integer bytes operators and floating point. 
Cloud AI 100 family has better performance on byte (INT8) operations (between 2.8 and 5.33 TOPs/W), followed by the Orin NX Family (3.5 and 4 TOPs/W) and the incoming Orin Nano Family (2 and 4 TOPs/W).  
However, Versal's AI Engines on Edge Family gets a wide range of performance (between 0.55 and 2.69 TOPs/W), similar to the AGX Orin Falimy (2.7 and 2.83 TOPs/W). 
On Versal's AI Engine on AI Core Family, although its reported computer capacity per operation is outstanding, the power consumption reported in \cite{AMDXILINXInc.2022_AI_inf} for the VC1902 chips is too high, especially for onboard processing on space missions.  
With half-precision floating point (FP16), Cloud AI 100 gets between 1.4 and 2.6 TFLOPs/W, followed by Orin AGX family (1.35 – 1.41 TFLOPs/W), while Versal AI's data is not reported for this kind of operations. 
For single-precision floating-point (FP32), Versal's AI Engines of Edge Family achieve the best performance (up to 221 GFLOPs/W), followed by the Instinct MI200 Family (between 151 and 171 GFLOPs/W on Matrix operations), while the Jetson AGX Family using CUDA cores (84.1 and 88.6 GFLOPs/W) reports slightly less performance than the Versal AI Core VC1902 AI Engines (up to 91 GFLOPs/W).  
This comparison proves that Versal AI Edge Family has a remarkable and wide range of performance per Watt, being a good solution for AI and ML applications for onboard Satellite Communications. 
Analyzing the  Versal AI Core family for similar applications requires more accurate studies of chip performance.   

Finally, Table \ref{Tab:AI_Chipsets:_Size_Avail} details the chip dimension and availability.
Proper chip size comparison takes much work to perform with the available datasheets. Some sellers, like Intel and XILINX, report the die size. Others, like NVIDIA, report the available PCB, including auxiliary chips and interfaces, and Qualcomm and AMD report the final product.   
Myriad Family reports the smallest size varies from 8$\times$9.5 mm$^2$ for Intel's Movidius Myriad 2, actually discontinued, to 14$\times$14 mm$^2$ for the Myriad X.
NVIDIA's family reports size from 45$\times$69.6 mm$^2$ to 87$\times$ 100 mm$^2$. 
NVIDIA's roadmap announced the new Orin Family, planned to be released between December 2022 and January 2023, depending on the chip availability. This family was recently updated with the Jetson Orin Nano, a device for edge applications with lower power consumption \cite{Roadmap_NVIDIACorp.2022, OrinNano_NVIDIACorp.2022a, OrinNano_NVIDIACorp.2022d}. 
Qualcomm AI 100 and AMD Instinct MI200 families report more significant devices with PCIe interfaces, more suitable for systems co-processors and accelerators, not for standalone and on-board applications.
On Versal's AI families, the chip size varies from 23$\times$23 mm$^2$ to 45$\times$ 45 mm$^2$. However, only one development board has been released, the VCK190 powered by the Versal AI Core VC1902, which is too big and power demanding for on-board applications (190$\times$241.3 mm$^2$).
\begin{table}[h!]
\caption{AI Chipset/Accelerators off-the-shelf: Dimension and availability}

\centering{}%
\begin{tabular}{>{\centering}m{1.5cm}>{\centering}m{2cm}>{\centering}m{2cm}>{\centering}m{1.5cm}}
\toprule 
\textbf{Device} & \textbf{Form Factor} & \textbf{Status} & \textbf{Comments}\tabularnewline
\midrule
Myriad Family & 8\textasciitilde 14$\times$9.5\textasciitilde 14 $mm^{2}$ & Released & \tabularnewline
\midrule 
Jetson Nano & 80$\times$100 $mm^{2}$ & Released & \tabularnewline
\midrule 
Jetson TX2 Family & 45\textasciitilde 50$\times$69.6\textasciitilde 87 $mm^{2}$ & Released & \tabularnewline
\midrule 
Jetson Xavier NX Family & 45$\times$69.6 $mm^{2}$ & Released & \tabularnewline
\midrule 
Jetson AGX Xavier Family & 87 $\times$ 100 $mm^{2}$ & Released & \tabularnewline
\midrule
Jetson Orin NX Family & 45 $\times$ 69.6 $mm^{2}$ & In development & Planed for Jan 2023\tabularnewline
\midrule 
Jetson AGX Orin Family & 87 $\times$ 100 $mm^{2}$ & In development & Planed for Dec 2022\tabularnewline
\midrule 
Jetson Orin Nano Family & 45 $\times$ 69.6 $mm^{2}$ & In development & Planed for Jan 2023\tabularnewline
\midrule
Cloud AI 100 Family & 46\textasciitilde 68.9$\times$110\textasciitilde 169.5 $mm^{2}$ & Released & PCIe Interface\tabularnewline
\midrule 
Instinct MI200 Familly & 111$\times$267 $mm^{2}$ & Released & PCIe Interface\tabularnewline
\midrule 
Versal AI Edge Family & 23\textasciitilde 45$\times$23\textasciitilde 45 $mm^{2}$ & Released & No development board available\tabularnewline
\midrule 
Versal AI Core Family & 35\textasciitilde 45$\times$35\textasciitilde 45 $mm^{2}$ & Released & VCK190 190$\times$241.3 $mm^{2}$\tabularnewline
\bottomrule
\end{tabular}\label{Tab:AI_Chipsets:_Size_Avail}
\end{table}

\subsection{Hardware Analysis-Solutions Comparison}
Although machine learning and artificial intelligence are emerging fields for computing in-the-edge, implementing those algorithms for space applications is challenging for the scientific community due to the duration of the missions, the radiation effects on the chips, the temperature span and the performance-to-power ratio.
The use of COTS Processors for onboard space missions requires radiation-hardness-by-design (RHBD) or radiation-tolerance (RT) of the chip, fault-detection support and correction codes, single-point thermal load, long-term support and availability \cite{Steenari2021,Ginosar2021}. 

Although none of the COTS AI-capable devices is considered radiation-hardened-by-design new studies are emerging for using COTS chipset on onboard AI applications. 
This is the case with the GPU for Space project (GPU4S) that uses embedded GPUs for space workloads \cite{Kosmidis2021}. 
The authors focus on four premises for achieving the results: the parallelization capabilities of the software, the comparison with the most available COTS device in the market at this moment, and the use of a benchmarking tool created as one of the milestones of the project (GPU4S Bench) reported in \cite{Rodriguez2019,Steenari2021_2}. 
The test was performed for the ARM Mali G-72, NVIDIA Xavier NX, NVIDIA TX2, and AMD embedded Ryzen V1605B using complex software operations for space workloads applications. 
The NVIDIA Xavier reported better performance for the matrix multiplication benchmark than the rest of the platforms. Regarding energy efficiency, the best choice has been either the NVIDIA Xavier NX or the TX2.

As part of the same project, in \cite{Rodriguez2020}, Rodriguez \emph{et al.} present a case study for an onboard infrared detector algorithm implementation using an embedded GPU.
The algorithm is based on image matrix operations, targeting an NVIDIA Jetson Xavier's embedded GPU, using the CUDA cores for the algorithm acceleration and the Carmel CPU for offload computation.
The CPU implementation, for a standard 2k $\times$ 2k pixels image, reports performances of 98 times compared with the same algorithm on the LEON2 processor and 15.5 times on the PowerPC 750, while the GPU implementation reports performances of 806 and 128 times, respectively.
The authors confirm that embedded GPUs can be an option to consider in future space missions \cite{Rodriguez2020}.

In \cite{Kosmidis2019,Kosmidis2020}, the same authors continue the analysis of the existing space application domains and survey the COTS and soft-IP embedded GPU domain to assess which meets the required computational power and identify the challenges which need to be addressed for their adoption in space.

Steenari \emph{et al.} present in \cite{Steenari2021} a survey of high-performance processors and FPGAs for onboard processing and ML applications.
 The study includes as test device types the single and multicore processors, DPS and many-cores processors, embedded GPUs and FPGAs, while the considered data focus on the peak performance and size, qualification status, radiation support and ML tools availability. 
The authors conclude a significant gap in processing capabilities between RHBD/RT and COTS devices, as space processors need to catch up to commercial devices in terms of process technology and performance. However, the available COTS processors and accelerators have not been provided with RT-qualified packages. 
They also affirm that using high-performance COTS processors could cover future high-performance requirements for the onboard mission. However, they could compromise mission lifetime, availability and reliability as the COTS devices usually need longer-time support.

In \cite{Marques2021}, Marques et al. analyze the feasibility of ML applications for onboard space weather detection on COTS devices.
The neural network was trained to distinguish between background and radiation particles. 
The authors expose a set of HW devices and development tools for ML and AI.
The HW tested includes FPGA devices like Zynq-7000 SoC, Zynq Ultrascale+, Kintex Ultrascale and Versal ACAP; processors like the onboard space qualified processor (GR740/Leon4) and generic ARM Cortex A53 available in the Zynq US+ MPSoC; and SBC like Unibap iX5 and Myriad X.

Comparing the implementation results for FPGA-based platforms, Versal ACAP AI Core shows the best performance for inference on Coronal Mass Ejection (CME) Detection and Particle Detection. In contrast, Kintex Ultrascale KU040 show the worst results \cite{Marques2021}.
In the case of microprocessors, although the results are not as good as previous platforms, the processing time of a non-optimized implementation on a single-core LEON4 is already sufficient for onboard detection with better latency than detection on the ground after downlinking. At the same time, the CNN model for the ARM Cortex A53 MPSoC needs 5.8 times more time than the FPGA implementation on the same platform.
In the case of the SBC platforms, the best results are achieved using the embedded GPU (Unibap iX5 GPU) for CME Detection inference. At the same time, Myriad X gets three times better results in Particle detection Inference. 

This analysis demonstrates that using dedicated HW architecture for ML and AI, like the AI Engines or embedded GPU, makes a difference in increasing the performance of implementing these types of applications.

In \cite{AMDXILINXInc.2021_5G_BF, XILINXInc.2021_BeamF}, XILINX presents an application note for the implementation of massive MIMO beamforming algorithms for 5G New Radio. 
As the beamforming requires significative compute density and advanced high-speed connectivity to meet 5G throughput and low latency, they propose using AI Engines available on the Versal AI Core VC1902 to increase the performance per Watt of the implementation.
The authors claim an increase of 2.14 times the performance per Watt compared with a similar implementation over an Intel Agilex FPGA.

In \cite{AMDXILINXInc.2022_AI_inf}, XILINX published a solution brief for AI inference with a Versal AI Core VC1902 for vision and video processing.
The exploit of the Versal AI architecture results in an adaptable accelerator that exceeds the performance, latency and power efficiency of traditional FPGAs and GPUs for AI/ML Workloads. 
The authors ensure the delivery of 2.7 times the performance per Watt over competing with Intel Agilex FPGA, using 87 Watts of power, estimated using the XILINX Power Estimator (XPE) Tool. 
Even when this implementation proves this device's capability to accomplish highly complex operations on ML and AI applications, the power consumption is too high for onboard satellite applications.

Working on onboard device requirements,  AMD announced the release of the first Space-Grade Versal Adaptative SoCs enabling onboard AI processing in space \cite{AMDInc.2022, Marques2021, AMDXILINXInc.2022_XQR}. 
The actual information about this device pointed out that the XQR Versal AI Core XQRVC1902, to be available in early 2023, will be based on the XILINX Versal AI Core VC1902, detailed in Section \ref{sec:AI_chips},  including Class B qualification, radiation-tolerance and a 45 $\times$ 45 mm$^{2}$ packing \cite{AMDXILINXInc.2022_XQR}. 
There is no information yet about the possible power consumption, its predecessor's principal inconvenience, and the chip's computer capacity.

\section{Challenges and Open Discussion}\label{Challenges}
While the literature covered in this survey provides an interesting overview of the main technical challenges for specific use-cases, there are still a number of points that need to be further investigated to unleash the potential of ML into the satcom industry.

The main point of discussion resides on \textit{where} the ML is deployed. If we opt for on-board deployment, there are many requirements in terms of power consumption (which depends on the tasks to be executed), the mass incurred by this additional block and the additional interfaces to interact with this block. So far, most flying satellites operate in a simple relaying-mode, with no data processing on-board except the frequency conversion and amplification. If the ML chip is deployed on-board, the specific input data and the required storage may add constraints to the payload design. At this point, we could not identify any previous work where a trade-off in terms of ML potential performance benefit versus added cost and complexity on the satellite device. To be noted that most of the HW components in a payload are duplicated to add redundancy to the satellite in case of failure. The latter should probably apply to ML-components as well if the play a key role in the operations of the satellite.

When the ML deployment is done on-ground, the requirements can be relaxed a bit in the sense that the consumed power and mass are not as critical as for a flying satellite. Similarly, the access to different data indicators on-ground is much afforable. 

In the below sub-sections, we have aimed to separately address different challenges that we believe may act as leading reasearch directions in future ML for SatCom works.

\subsection{Implementation Challenges}\label{ImplementationChallenges}
 Implementation challenges in applying ML in the satellite communications industry are numerous and complex. One of the main challenges is determining the best place to deploy ML: should it be deployed on the ground or onboard the satellite?

If on-board deployment is chosen, limitations in terms of power consumption, mass, and the need for additional interfaces to interface with the ML block must be considered. In addition, data storage and specific data input may add constraints to the payload design. In addition, implementing ML components may require redundancy, further increasing satellite complexity and development costs \cite{ortiz2023onboard}.

On the other hand, if the ML is deployed on the ground, power and mass consumption requirements may be somewhat relaxed. Access to different data indicators on the ground is also more affordable. However, the challenges of data transmission back and forth between the satellite and the ground control center must be taken into account.

There are challenges in selecting the most appropriate ML algorithms for each specific task, the availability of training data, the extensive training length, validating and verifying the model to ensure its reliability, and implementing adequate security systems to protect the ML and the data it handles.

Implementation challenges also include the hardware required for Machine Learning implementation in the satellite communications industry.

On the one hand, high-performance hardware is required to run complex ML algorithms. This may include graphics processing units (GPUs) or tensor processing units (TPUs) that are specifically used for Machine Learning computation acceleration. It is important to consider the radiation tolerance of the hardware used. SatCom systems operate in extremely hostile environments, including cosmic radiation, solar radiation, and solar particle radiation events. Therefore, hardware capable of withstanding these harsh environments and ensuring the long-term reliability and performance of the system is required.

Another challenge is the integration of the ML hardware into the satellite payload. The mass, power consumption and complexity of the hardware must be carefully considered to ensure that the complete system is robust enough for use in space.

In that regard, it is important to keep in mind that hardware technology continues to evolve rapidly and that hardware component costs can be significant for the development and implementation of ML-based NTN systems. Therefore, careful evaluation of long-term costs and benefits is required to ensure the economic viability of ML-based systems.

\subsubsection{Training datasets Availability}\label{TrainingDataAvailability}
The availability of real-world datasets is one of the most crucial preconditions for evaluating any proposed \ac{ML} based method. So far, the cost and access difficulty of highly representative datasets has hindered the development in \ac{SATCOM} of \ac{ML} techniques.
Indeed, while some learning approaches can be trained with small simulated datasets, advanced techniques such as \ac{DL} and \ac{DRL} requires a large dataset to converge and reach the desired performance. It is time consuming and costly to generate large dataset for \ac{SATCOM} using simulators.
In addition, even assuming large datasets can be produced via simulations, their accuracy remains a key open discussion point.
This review paper has shown that the development of the \ac{ML} research does not go hand in hand with the development and availability of training datasets.
Most of the datasets available opensource are datasets to train \ac{ML} models on satellite imagery \cite{cornebise2022open, van1807spacenet, schmitt2021there} and it is pretty rare to find open high-resolution datasets for the other use-cases presented in Section \ref{UseCases}.
This makes it a tremendous challenge to compute the accuracy and scalability of the proposed \ac{ML} models in the real world.

\subsubsection{Data Dimensionality Reduction}
Using large data dimensions during the learning and testing increases computational complexity and slows the decision-making process. A promising solution to this problem is mapping the data from high dimension space to a lower one while preserving the most relevant and important features. Specifically, data dimensionality reduction is the process of scaling down a set of data with high dimensions into data with lower dimensions, ensuring that it concisely conveys similar information without affecting the original information.
Data dimensionality reduction removes irrelevant, noisy, and redundant data to obtain acceptable accuracy. 
In satellite communication systems, high-dimensional datasets are naturally aggregated from heterogeneous data sources and massive data streams, which bring about many processing challenges. Thereby, it is essential to find optimal techniques or solutions for dimensionality reduction while catering for the effect of reducing high-dimensional data without compromising the data value. Selecting an appropriate data dimension reduction technique will expedite the processing time and reduce the efforts required to extract insightful information.

The data representation space size can be reduced via feature extraction or selection techniques. The feature extraction transforms the initial data attributes into a new one formed by the linear or non-linear combination of the initial attributes. Contrarily, the feature selection method identifies the most relevant attributes according to some given criteria. In this context, different dimensionality reduction techniques are available in the literature to eliminate irrelevant and redundant features, and the work in \cite{AYESHA202044} has reviewed the existing dimensionality reduction techniques and their suitability for different types of data and application areas with highlighting the issues that may affect the accuracy and relevance of the obtained results. Furthermore, an unsupervised deep-learning framework named local deep-feature alignment (LDFA) for dimension reduction has been developed in \cite{Zhang2018} to extract the essential features. In addition, two-dimensionality reduction techniques, Linear Discriminant Analysis (LDA) and Principal Component Analysis (PCA) are investigated in \cite{Reddy2020} on some popular Machine Learning algorithms. 

In this direction, the work in \cite{Gahar2019} has proposed using the Random Forest imputation method to extract useful information and reduce the search space to facilitate data exploration. 
Moreover, a combination of convolutional deep belief network (CDBN) with autoencoding feature algorithm is applied for data dimensionality reduction and layer-wise learning \cite{Abdullah2023}. Specifically, CDBN is a hierarchical generative model that employs probabilistic max-pooling to extract high-level features by replacing the restricted Boltzmann machine (RBM) in the deep belief networks used for the individual layers by convolutional RBMs. Thus, this technique has been proved as an efficient algorithm for dimensionality reduction, collaborative filtering, and faster feature learning \cite{Lee2009}. Likewise, Auto-Encoder and PCA approaches are used for dimensionality reduction in \cite{electronics8030322} for designing intrusion detection systems with machine learning. Reference \cite{rs12081261} has considered multivariate statistical methods like independent component analysis (ICA) and minimum noise fraction (MNF)  as dimensionality reduction techniques for processing satellite data. 

\subsubsection{On-board deployment and Radiation Tolerance}
The deployment of ML onboard satellites faces a significant challenge in terms of radiation tolerance. Satellites are exposed to radiation from space that can cause electronic component failures. Radiation can affect the performance and lifetime of the hardware; therefore, it is necessary to design hardware components resistant to radiation \cite{9288809}.

ML onboard satellites are especially susceptible to radiation effects because the electronic components used for ML implementation are susceptible to radiation effects. Processing input data and implementing ML models require a large amount of computation, often performed by custom integrated circuits. Radiation exposure can cause errors in these circuits, resulting in hardware and software failures\cite{9288809}.

Therefore, it is essential to design radiation-tolerant hardware for satellite-borne ML implementation. This may include the use of integrated circuits specific to aerospace applications, as well as design techniques to reduce radiation susceptibility. In addition, it is important to perform rigorous testing of hardware components to ensure that they are radiation resistant and capable of operating in the space environment

\subsection{AI Accelerators}
Successful deployment of machine learning algorithms in SatCom systems presents unique challenges. One of these is limited computational resources in space environments, where efficient and fast data processing is needed. AI acceleration technology, which uses specialized hardware to perform specific machine learning computations, has been developed to address this challenge \cite{li2020survey}.

AI accelerators can optimize the data processing and energy efficiency of SatCom systems, enabling greater data processing and transmission capacity on satellites. Accelerators can also improve the accuracy and speed of machine learning models, enabling better real-time decision-making \cite{li2020survey}.

However, implementing AI accelerators in SatCom systems also presents challenges despite the potential benefits. Integrating specialized hardware into existing systems can be costly and require significant changes to satellite infrastructure. In addition, careful design and rigorous validation are required to ensure the reliability and safety of SatCom systems using these advanced technologies.

\subsubsection{Neuromorphic Computing}
In the future, deploying ML for satellite communications will set new challenges for computer processors, which will have to support large workloads as efficiently as possible in harsh environmental conditions. Legacy computing technologies, including CPUs and GPUs, which are based on a semiconductor architecture, still may need to be suited for some ML workloads, particularly for unstructured time-sensitive signal processing. Besides, legacy technologies are expected to hit a "digital wall" in 2025, potentially forcing a paradigm shift in terms of computing technologies \cite{berggren2020roadmap}.

From this context, bio-inspired neuromorphic processors are a promising alternative to work as efficient coprocessors in ML applications where temporal signals are involved and/or require continuous adaptation in SATCOM systems. As opposed to conventional processors that operate in batch mode, i.e., collecting many samples before processing them, NPs can process in streaming mode. Owing to their energy efficiency and continuous on-board adaptability, NPs represent an excellent opportunity to open up the potential benefits of AI solutions for SATCOM systems. Moreover, NP implementation can take advantage of non-volatile memory devices based on technologies beyond CMOS \cite{Flor2022NP}.

\subsection{Future AI Frontiers/Visions for Satellite Communication}
\subsubsection{Federated Learining}
\ac{SATCOM}'s deployment of systems globally has led to an increasing demand for ML solutions operating in distributed systems. In this regard, the federated ML approach has emerged as a promising solution for deploying machine learning models in global \ac{IoT} systems such as SatCom \cite{s22176450}.

The federated ML approach refers to training ML models on multiple devices, each of which performs a part of the training process on its local data. Instead of sending data to a centralized server for training, the data is maintained on local devices, and only model updates are sent to the server. This collaborative learning approach can provide a scalable, decentralized solution for training models on large data sets without requiring data transfer between devices.

However, the federated ML approach also presents some technical challenges. First, local devices and the central server must communicate reliably to ensure data integrity and model updates. In addition, federated ML models need to be able to handle heterogeneous data and devices with different processing capabilities.

Another major challenge is ensuring data privacy and security. Because data is held on local devices, robust security and privacy measures must be implemented to protect user and organizational data.

\subsubsection{ML and Blockchain}
Security and resilience are features claimed by the new constellations and satellite systems, especially for earth observation applications or Government communications such as those provided by the GOVSATCOM project. The typical long distances in SATCOM lead to significant communication delays among the network nodes, compromising the data's integrity and reliability. Establishing prioritized, secure, and efficient command and data communication is essential for space and ground systems. Using \ac{BC} assures a secure automated network infrastructure for many applications in SATCOM, including autonomous satellite constellation control through satellite-to-satellite communication, prioritized and stable data and command communication between space and ground stations, among others. Similarly, \ac{BC} networks can benefit from SATCOM thanks to their broad geographic reach and broadcast capabilities.
The \ac{BC}-based network nodes participate by receiving and sending data across the distributed network architecture. The data maintain a unique structure transferred by transactions \cite{BCdefinition}. These data can be anything, such as personally identifiable data, satellite command, or key/value paired asset data stored on blocks and linked together in a chain. Any alteration in a transaction detects an introduction or threat in the communication, directly invalidating the transmission.
The \ac{BC} application to benefit SATCOM has been proposed for different use cases such as interference detection in \cite{BC1}, spectrum management in \cite{BC2}, satellite manufacturing supply chain in \cite{BC3}, and Space Situational Awareness in \cite{BC4}. Different contributions of \ac{BC} in the space industry are identified in \cite{9480642}. Some of them are (1) identity authentication and exchanged data integrity for LEO satellites which include its dynamic topology and regular link switching, and (2) reducing complexity across a range of business, operational, and security applications using smart contracts by BC networks. In \cite{9172516}, from the point of view of communication, a solution to reduce end-to-end latencies in SATCOM is proposed based on a BC-based reputation framework and routing protocol.
ML in blockchain applications applied to SATCOM is currently a crucial challenge where only a few preliminary works are presented in the literature. Notably, Space-Air-Ground Integrated Networks (\ac{SAGIN}) project in \cite{9918062} presented a federated reinforcement learning approach to trusted traffic offloading using BC. Conventional previous proposals are subject to increasing their benefits by adding ML techniques.

\subsubsection{Tiny Machine Learning for Satellite}
TinyML deploys \ac{ML} algorithms onto low-cost, low-power and limited resource devices, storing \ac{NN} models directly within memory and directly running inference on onboard sensors' output. 
TinyML aims to create algorithms and software capable of performing efficiently on intelligent or limited resources edge devices.
This approach avoids communicating with the cloud to transmit data for external processing.  

TinyML aims to improve the efficiency of \ac{DL} \ac{AI} systems by requiring far less computation power, a less costly hardware platform and improving energy efficiency. \cite{prakash2023tinyml} shows that TinyML is capable of using 
a fraction of the compute resources needed for traditional ML systems. 
Finally, while the heterogeneity and limited resources of MCU devices present new challenges for on-device training, model updating, and deployment, recent research and the development of MLframeworks such as TensorFlow Lite for Microcontrollers have increased the accessibility of TinyML.
Due to its nature, TinyML is one of the fastest growing areas of \ac{DL} and might become a disruptive application in satellite communication.

\subsubsection{Quantum Computing and AI}
In this survey we have shown (Section \ref{QuantumKey}) the importance of quantum communication for \ac{SATCOM} in its role of ultra-secure communication. 
Beyond this, quantum computing holds a great potential for AI developers looking to create more advanced AI systems that can solve problems faster and more efficiently. Specifically, quantum computing can make AI training process faster and more accurate as it allows utilizing larger datasets to train AI models to be more accurate and better at decision-making. Although quantum computing is still in its early development stages, there are several imminent benefits AI can gain from quantum computers, including but not limited to:
\begin{itemize}
    \item Increased computational power: Quantum computers have the capability of solving complex optimization problems exponentially faster than classical computers. This could open up new possibilities for AI development such as enabling more complex machine learning algorithms to be trained in less time. For instance, Quantum computers can help AI developers to optimize complex functions more efficiently by leveraging quantum annealing and other optimization techniques that are not available in classical computing \cite{Junaid2023}.
    \item Improved algorithm development: Quantum computing can help AI developers improve existing algorithms or develop new ones that can take advantage of the unique capabilities of quantum computing. For example, quantum machine learning algorithms can leverage quantum entanglement and superposition to achieve better accuracy and faster training.
    \item Enhanced data analysis: Quantum computing can help AI developers to analyze large datasets more efficiently and identify patterns that are difficult to detect using classical computing. This can lead to improved data-driven decision-making and insights.
    \item Quantum computers can provide a means of safeguarding the confidential data used by AI systems from hacking and other forms of cybercrime. Additionally, the parallel processing capabilities of quantum computing can be used to develop countermeasures against cyberattacks. 
\end{itemize}

Furthermore, a major challenge in the operation of the CubeSats and small satellites within NTNs in lower altitudes is the rather low information processing capabilities of the onboard processors. Consequently, developing and utilizing AI models in these entities can be beyond the capability of a single satellite processor. Alternatively, quantum technologies along with space-based quantum clouds can be employed in such scenarios in order to offload the computational burden from small satellites. Thus, a space quantum network can be structured and interconnected via FSO links, which will benefit from the high computational capacity of the quantum servers with certainly enhanced security performance to efficiently use AI and ML techniques \cite{QNTN_2023}.


\subsubsection{AI and Optimization Methods}
This paper has shown that \ac{ML} and \ac{AI} have tremendous potential in future \ac{SATCOM} networks. Thus, while it is clear that the next generation of satellites will rely on \ac{AI}, it is unclear how \ac{AI} should be integrated into the architecture of satellite networks.
Most of the data-driven approaches surveyed in this paper (see table \ref{tab:Article_Categorization}) are based on \ac{NN} and \ac{DRL} techniques that require a large amount of data and are complex to process.
Therefore, it is convenient to investigate that to mitigate the computational complexity, latency, and power consumption. 

Hybrid \ac{ML} models allow incorporating expert knowledge in the learning approach. 
The external knowledge can be incorporated as an initialization procedure \cite{zappone2019wireless} or leveraged in the learning approach, as in deep unfolding \cite{shlezinger2023model}.
In the former, the training set for the \ac{DNN} can be generated offline, alleviating the satellite of the burden of dealing with the high complexity of the model. The model can be thus optimized online with few empirical data. The intuition behind this approach is that, despite generally not being perfectly accurate, the initial \ac{DNN} contains some critical feature of the satellite network, and less data is needed to reach the desired accuracy compared to the case of no-training being performed.
In the latter,  \acp{NN} impact the complexity and computational requirement of the \ac{ML} model on the satellite. For instance, deep unfolding aims to map an iterative algorithm into a learnable \ac{NN} and to determine the number of nodes by the number of iterations of the algorithm \cite{Lissy_2022}. Deep unfolding offers the possibility of initializing a \ac{NN} and refining it using empirical data. The first attempts to utilize Deep Unfolding for \ac{SATCOM} applications have been mentioned in the Rate Splitting Use Case with the mentioned works \cite{9851793, 9837852}.
To reduce the complexity of the power allocation and timeslot-terminal assignment in satellite NOMA systems, we have shown in this paper that the work in \cite{9685660} makes use of a hybrid approach to low the complexity of the problem without, however, proposing an end to end \ac{NN} that may not be feasible.
 
While an extensive description of hybrid ML models is beyond the scope of this paper, we believe incorporating expert knowledge in \ac{ML} models will receive increasing attention in the next future to overcome the conventional NN complexity and the relative challenges presented in Section \ref{ImplementationChallenges}.

\subsection{Security Aspects}
We have shown in this survey how big data is the core of many \ac{SATCOM} research activities. In fact, Image Processing and Remote Sensing \ac{AI} developments rely on the data collected by satellites and drones and use it to train models for different outputs and use-cases.
For this reason, \ac{AI}/\ac{ML} tools used to handle and analyze diverse sources of data are closely linked with several ethical issues, not only the well known issue of privacy. Together with it, explainability, bias also ethical issues are becoming increasingly important as a result of data collection, the approaches used for analysis, the manner and purpose for which the results from such analyzed are used \cite{kochupillai2022earth}.

Security aspects are crucial in the satellite communications industry and the use of ML introduces new challenges. One of the biggest risks is the potential manipulation of training data and exploitation of vulnerabilities in the trained ML model, which could adversely affect the quality of transmitted data and the integrity of the SatCom system. In addition, the use of real-time ML systems also introduces the possibility of real-time attacks, such as malicious data injection into the ML model.

An additional major challenge is a need to ensure the privacy and security of the data that is used to train the ML models. In particular, the collected and processed data may be privacy sensitive, such as location or enterprise customer data. Appropriate measures must be taken to protect this data and ensure it is only used for legitimate purposes.

The security of the hardware and software used to implement the ML models is also an important consideration. SatCom systems must be designed and built with security in mind, including protection against physical and logical attacks. The exploitation of vulnerabilities in the hardware or software used to implement the ML could have serious consequences for the integrity of the SatCom system.

\subsection{Cost Optimization}
Cost optimization is a critical factor in the satellite communications industry, and the deployment of artificial intelligence technologies is no exception. When considering the deployment of ML systems, the costs of the necessary components and the resources required to train and maintain the ML models must be considered. In addition, the cost of putting a satellite into orbit must also be considered, and cost-benefit optimization of ML systems must be sought.

The cost of deploying a Gbps in orbit can be extremely high \cite{ortiz2017method}. The weight of the satellite components and associated power consumption can also be costly regarding satellite payload capacity and battery life. Therefore, more efficient and cost-effective design solutions, such as using low-power data processing technologies and optimizing ML algorithms to minimize the resources required, should be explored.

Cost optimization must also consider the scalability of ML systems. As more and more satellites are deployed, scalable and cost-effective solutions must be found to train and maintain ML models on all satellites. This may include using distributed computing technologies and implementing automation solutions to reduce maintenance and training costs.

\bibliographystyle{IEEEtran}
\bibliography{IEEEabrv, References_manual.bib}
\end{document}